\documentclass[11pt,a4paper,onecolumn,usenatbib,referee]{mnras} 
\usepackage{graphicx}
\usepackage{color}
\usepackage{amssymb}
\usepackage{amsmath}
\usepackage{mathabx}
\usepackage{stmaryrd}
\usepackage{multirow}
\usepackage{multicol}
\usepackage{amscd}
\usepackage{mfpic}
\usepackage{verbatim}
\usepackage{hyperref}
\usepackage{pstricks,pst-node,pst-text,pst-3d}
\usepackage{ftnxtra}
\usepackage{fnpos} 

\definecolor{cream}{rgb}{.97, .95, .88}
\definecolor{darkcream}{rgb}{1., .88, .5}
\definecolor{lightpink}{rgb}{0.98, 0.88, 0.87}
\definecolor{lightwhite}{rgb}{1., 0.98, 0.95}
\definecolor{lightsalmon}{rgb}{1., 0.95, 0.90}
\definecolor{lightviolet}{rgb}{0.9, 0.8, 0.9}
\definecolor{lightgray}{rgb}{.96, .96, .96}  
\definecolor{lgray}{rgb}{.75, .75, .75}
\definecolor{LemonChiffon}{rgb}{0.95, 1., 0.7}
\definecolor{lightolivegreen}{rgb}{0.84, 0.89, 0.25}
\definecolor{lightgreen}{rgb}{.664, 1., .52}
\definecolor{llgreen}{rgb}{.900, .983, .960}
\definecolor{tristle}{rgb}{0.87, 0.67, 0.77} 
\definecolor{pink}{rgb}{0.95, 0.45, 0.75}
\definecolor{magenta}{rgb}{1., 0, 1.}
\definecolor{violet}{rgb}{0.9, 0.20, 0.85}
\definecolor{darkolivegreen}{rgb}{0.55, 0.65, 0.35}
\definecolor{maroon}{rgb}{0.7, 0.26, 0.56}
\definecolor{lightmaroon}{rgb}{0.85, 0.38, 0.58}
\definecolor{darkmaroon}{rgb}{0.604, 0.169, 0.451}
\definecolor{ddarkmaroon}{rgb}{0.2, 0.03125, 0.150}
\definecolor{mediumorchid}{rgb}{0.8, 0.33, 0.83}
\definecolor{mediumorchidd}{rgb}{1., 0.33, 0.63}
\definecolor{darkgreen}{rgb}{0.1, 0.6, 0.13}
\definecolor{lightyellow}{rgb}{1., 1., 0.82}
\definecolor{turquoise}{rgb}{0.042, 0.586, 0.512}
\definecolor{turquoisel}{rgb}{0.66, 0.94, 0.83}
\definecolor{darkturquoise}{rgb}{0.21, 0.55, 0.50}
\definecolor{coral}{rgb}{1., 0.6, 0.21}
\definecolor{lightorange}{rgb}{1., 0.88, 0.75}
\definecolor{orangered}{rgb}{1., 0.5, 0.}
\definecolor{orange}{rgb}{1., 0.65, 0.1}
\definecolor{orangel}{rgb}{1., .85, .3}
\definecolor{darkorange}{rgb}{0.875, 0.4, 0.204}
\definecolor{ddarkorange}{rgb}{.675, .218, .05}
\definecolor{bluesky}{rgb}{0.48, 0.53, 1.}
\definecolor{gold}{rgb}{1., 0.85, 0.25}
\definecolor{goldd}{rgb}{0.95, 0.75, 0.05}
\definecolor{darkviolet}{rgb}{0.54, 0.04, 0.84}
\definecolor{ddarkviolet}{rgb}{.382, .063, .657}
\definecolor{lightblue}{rgb}{0.30, 0.86, 0.89}
\definecolor{LightBlue}{rgb}{0.68, 0.85, 0.9}
\definecolor{lblue}{rgb}{0.78, 0.90, 0.95}
\definecolor{darkblue}{rgb}{.105, .308, .707}
\definecolor{lightmaroon}{rgb}{0.85, 0.38, 0.58}
\definecolor{darkmaroon}{rgb}{0.604, 0.169, 0.451}
\definecolor{darkpink}{rgb}{0.879, 0.020, 0.766}
\definecolor{ddarkpink}{rgb}{0.738, 0.195, 0.406}
\definecolor{grey}{rgb}{0.717, 0.717, 0.717}
\definecolor{lightgrey}{rgb}{0.800, 0.800, 0.800}
\definecolor{brown}{rgb}{0.740, 0.323, 0.182}
\definecolor{redbrown}{rgb}{.575, .158, .05}
\definecolor{darkbrown}{rgb}{0.34, 0.25, 0.05}
\definecolor{orangebrown}{rgb}{0.433, 0.262, 0.06}
\definecolor{pinkl}{rgb}{1., 0.788, 0.918}
\definecolor{salmon}{rgb}{1., 0.66, 0.5}
\definecolor{lightbrown}{rgb}{0.703, 0.508, 0.121}

\def\etal{{\it et al.}}

\def\Name#1#2 {{#2} {#1, }}

\def\Journal#1#2#3#4{{#1} {\bf #2}, (#3) #4}
\def\cir#1{{\GCN} #1}
\def\rep#1{{\GCR} #1}
\def\AA{\em A.\& A.}

\def\APJ{\em ApJ.}
\def\APJL{\em ApJ.Lett.}

\def\EAS{\em EAS Publications Series}

\def\GCN{\em GCN Circ.}
\def\GCR{\em GCN Rep.}

\def\JCA{\em J. Cosmol. Astrop. Phys.}

\def\MRA{\em MNRAS}

\def\NAT{\em Nature}
\def\NATA{\em Nature Astro.}

\def\PRD{{\em Phys. Rev.} D}
\def\PRL{\em Phys. Rev. Lett.}

\def\SCI{\em Science}

\def\SSC{\em Space Sci.}

\def\be{\begin{equation}}
\def\ee{\end{equation}}
\def\bea{\begin{eqnarray}}
\def\eea{\end{eqnarray}}
\def\bes{\begin{equation*}}
\def\ees{\end{equation*}}
\def\beas{\begin{eqnarray*}}
\def\eeas{\end{eqnarray*}}

\title{Properties of jet and surrounding material of GW/GRB 170817A}

\author[Houri Ziaeepour]{Houri~Ziaeepour$^{1,2}$\thanks{Email: houriziaeepour@gmail.com} \\
$^1$Institut UTINAM, CNRS UMR 6213, Observatoire de Besan\c{c}on, Universit\'e de Franche Compt\'e, 41 bis ave. de l'Observatoire, \\BP 1615, 25010 Besan\c{c}on, France \\
$^2$Mullard Space Science Laboratory, Holmbury St Mary, Dorking, Surrey RH5 6NT, UK}

\date{Accepted XXX. Received YYY; in original form ZZZ} 
\pubyear{2019} 

\begin{document}


\label{firstpage}  
\pagerange{\pageref{firstpage}--\pageref{lastpage}} 
\maketitle

\begin{abstract}
{We use published data in radio, optical and X-ray bands to analyze and model afterglows of 
GW/GRB 170817A. Our analysis is based on a phenomenological gamma-ray burst generator model which we 
previously used to study the prompt gamma-ray emission of this important transient. We find a 
multi-component model and a few of its variants that are consistent with broad band $\sim 1$ year 
observations of afterglows, once the contribution of kilonova in optical/IR band is taken into 
account. Considering beaming and off-axis view of relativistic outflows, we interpret the components 
of the model as approximately presenting the profile of a relativistic structured jet with a rapidly 
declining Lorentz factor from our line of sight, where it had a Lorentz factor of 
$\mathcal {O}(100)$, to outer boundaries, where it became a mildly relativistic cocoon with a 
relative velocity to light of $\sim 0.4-0.97$. Properties of the ultra-relativistic core of the jet 
obtained here are consistent with conclusions from analysis of the prompt gamma-ray emission. 
In particular, our results show that after prompt internal shocks the remnant of the jet retains in 
some extent its internal collimation and coherence. Slow rise of the afterglows can be associated to 
low density of circum-burst material and low column density of the jet. The long distance of 
external shocks from the merger, which could have been in part responsible for extensive thinning of 
the jet through expansion and energy dissipation before occurrence of external shocks is responsible 
for the peak of emission being $\gtrsim 110$~days after the merger. We discuss implications of these 
observations for origin and properties of circum-burst material around Binary Neutron Stars (BNS). 
This analysis confirms our previous results showing that an outflow with a Lorentz factor of 
$\sim 2-5$ cannot explain observed afterglows without an additional X-ray source or significant 
absorption of optical/IR photons.}
\end{abstract}



\section{Introduction} \label{sec:intro}
There is no general consensus about physics behind the unusual afterglows of the short GRB 170817A - 
the electromagnetic counterpart of the first detected Gravitational Wave (GW) from merger of a 
BNS~\citep{gw170817multimess,gw170817fermimulti}. This burst can be singled out by the faintness of 
its prompt gamma-ray and early X-ray afterglow, and its later brightening leading to the detection 
of a X-ray counterpart only after $\sim T+10$~days~\citep{gw170817xray,gw170817cxc2day,gw170817swiftnustar}, 
where $T$ is the trigger time of the Fermi-GBM~\citep{fermi} and Integral-SPI-ACS~\citep{integral} 
about $2$~sec after the chirp of the GW from the merger~\citep{gw170817fermi,gw170817integral}. 

The initial interpretation of these observations was off-axis view of an otherwise ordinary short 
GRB~\citep{gw170817latexraystructjet,gw170817xray,grboffaxprobab0,gw170817earlyradio1}. Another 
popular explanation was emission from break out of a cocoon - a mildly relativistic ejecta with a 
Lorentz factor $\Gamma \sim 2-3$~\citep{gw170817cocoon,gw170817cocoon0,gw170817earlyradio,gw170817cocoonsimul,gw170817cocoonevid}. 
However, the decline of flux in all 3 observed energy bands, i.e. 
radio~\citep{gw170817lateradio,lateradio,gw170817latexoptradio1,gw170817lateradio,gw170817lateradio300}, 
optical~\citep{gw170817lateopt,gw170817lateopt160,gw170817lateopthstir}, 
and X-ray~\citep{gw170817latexary,gw170817chandraxray260,gw170817chandraxray260a,gw170817latedecline,gw170817xraycxc260,gw170817chandraxray358} 
after $\gtrsim T+200$ days is much earlier than the prediction of a side viewed 
jet~\citep{grboffaxis} (see also earlier versions of~\citep{gw170817latexraystructjet}). 
Predictions of jet break out~\citep{gw170817cocoon,gw170817cocoon0} and 
cocoon~\citep{gw170817lateradio} models for 6~GHz band are compared with data taken up to 
$\sim T+110$~days in~\citep{gw170817latexary}. We should emphasize that predictions of the break 
time are model dependent and some off-axis models such as those studied 
by~\citep{grbjetoffaxis,structuredjet} find a peak time $\sim \mathcal{O}(200)$~days, i.e 
consistent with peak time of GRB 170817A afterglows. But, these works present general models and 
are not adjusted to special properties of GW/GRB 170817A. 

Gradually it became clear that the presence of a highly relativistic component in the outflow at 
late times is inevitable. For instance, using relativistic hydrodynamic simulations 
of~\citep{mhdjetsimul,mhdjetsimul0} for determining characteristics of the BNS merger outflow and 
asymptotic formulation of external shocks and synchrotron emission by~\citep{emission1} for fitting 
the afterglows,~\citep{gw170817latexary,gw170817aglowlorentz} found that 
an outflow with a narrow relativistic core having 
a Lorentz factor of $\sim 100$ and a sheath/side lob with a Lorentz factor of $\Gamma \sim 3-10$, 
where our line of sight passes through with an angle of $\sim 20^\circ$, can explain observations up 
to $\lesssim 140$~days. Following the detection of superluminal motion of the radio afterglow 
due to an oblique viewing angle,~\citep{gw170817lateradiosuprlum,gw170817lateradiosuprlum0} concluded 
$\Gamma \sim 4$ for the outflow at $\sim 230$~days and estimated an initial Lorentz factor of 
$\sim 10$ for the jet at the time of prompt gamma-ray. They also estimated an off-axis angle of 
$\theta_v \sim 20^\circ$ for the 
line of sight. In a further work~\citep{gw170817lateradio300} ruled out the cocoon/jet break out 
model suggested by some of these authors in their previous works and fit the spectrum with a 
phenomenological non-linear 2-component broken power-law expression~\citep{synchradiofit} in contrast 
to a simple power-law used e.g. in~\citep{gw170817lateradio,gw170817latexary,gw170817latebroad}. 
Finally, using all the observations in radio, optical/IR and X-ray during the first year after the 
merger event,~\citep{gw170817lateopthstir} found that the data can be fit by a 2-component jet model 
consisting of an ultra-relativistic component with $\Gamma \gtrsim 100$ and a relativistic component 
with $\Gamma \sim 5$. 

In~\citep{hourigw170817ag} we used a phenomenological shock and synchrotron emission model to show 
that an outflow with a Lorentz factor of $\sim 2-3$ underestimates X-ray flux. The same 
phenomenological formulation was used in~\citep{hourigw170817} to model the prompt gamma-ray emission 
of GW/GRB 170817A. The range of Lorentz factor studied in~\citep{hourigw170817ag} is the same as those 
employed in the early literature on this burst - specially those associating the unusual afterglows 
to a mildly relativistic cocoon\footnote{In the literature ``mildly relativistic'' indicates a large 
range of Lorentz factors from $\Gamma - 1 \sim \mathcal{O} (0.1)$ to $\Gamma \sim \mathcal{O} (1)$.}. 
The reason for this choice was the assumption that at $t > T+10$~days, i.e. well after internal 
shocks, the weak ultra-relativistic jet responsible for the faint prompt gamma-ray burst had 
dissipated its energy and its Lorentz factor had to be much smaller than $\Gamma \sim 10-100$ 
concluded in~\citep{hourigw170817}. In line with the same argument, the observed deficiency of X-ray 
with respect to optical and radio concluded from simulations with aforementioned low Lorentz factor 
implied either an additional source of X-ray - for instance a contribution from the decay of 
radioactive elements produced by the kilonova - or a significant absorption of optical photons. 

In the present work we use the same phenomenological model as the one employed 
in~\citep{hourigw170817,hourigw170817ag}, but we drop the assumption of a dissipated jet. We show 
that a multi-component model, including both ultra-relativistic and mildly relativistic components, 
and a kilonova, can explain all the data. The components of the model approximately present angular 
profiles of density and Lorentz factor of the polar ejecta from the BNS merger and its evolution. 
We use this model and properties of circum-burst material to investigate reasons behind the late 
brightening of the afterglows.

In Sec. \ref{sec:ag} we describe the model and compare it with afterglow models of GW/GRB 170817A 
in the literature. Interpretation of the 3-component model is discussed in Sec. \ref{sec:interpret}. 
Parameters of the phenomenological model are summarized in Appendix \ref{app:def}. Degeneracies of 
these parameters are discussed in details in~\citep{hourigrbmag,hourigw170817} and are not repeated 
here. Nonetheless, to investigate whether conclusions made in Sec. \ref{sec:interpret} can be 
significantly impacted by them, in Appendix \ref{app:paramvar} we present light curves of several 
variants of our best model and discuss their properties. Our results are summarized in 
Sec. \ref{sec:outline}. 

\section{Afterglow model} \label{sec:ag}
The phenomenological model of relativistic shocks and synchrotron-self-Compton 
emission of~\citep{hourigrb,hourigrbmag} used in the present work is reviewed 
in~\citep{hourigw170817,hourigw170817ag} and we do not repeat it here. Nonetheless, 
for the sake of self-sufficiency definition of parameters of the model are given in 
Table \ref{tab:paramdef}. Some details about how kinematics of the shock and 
dynamics of the emissions are modelled and related are given in Sec. \ref{sec:riseslope}

It is important to remind that there is a significant difference 
between our approach, in which a synthetic burst is generated for a set of input parameters 
characterizing the jet and its surrounding, and modelling of afterglows according to asymptotic 
power-law behaviour of light curves and spectra based on the original calculations of general 
aspects of synchrotron emission from external shocks by~\citep{emission1}. Notably, in our 
approach simulated bursts explicitly depend on distance and column density of the jet and 
thereby give an assessment of these quantities. Although analysis of GW/GRB170817A afterglows e.g. 
by~\citep{gw170817cxc2day,gw170817latexary,gw170817latexoptradio1,gw170817latefasttail} are 
based on the jet characteristics obtained from Magneto-Hydro-Dynamics (MHD) or relativistic 
hydrodynamics simulations, synchrotron emission is calculated according to the asymptotic formulation 
of~\citep{emission1}, which estimates power-low behaviour of afterglows in a given energy band by 
comparing emission's frequency with characteristic and cooling frequencies of accelerated electrons. 
However, backreaction of shocks and energy dissipation are not explicitly taken into account. 
The model used here is also phenomenological and an approximation. In particular, some of important 
processes and quantities, which cannot be easily formulated from first principles, are presented 
by parametrization and their initial values are chosen by hand. Nonetheless, the model takes into 
account in a systematic manner backreaction and evolution of physical properties important for the 
synchrotron/self-Compton emission. Moreover, the model is applicable to both 
internal~\citep{hourigrbmag,hourigw170817} and external shocks, and thereby allows to compare and 
to verify consistency of parameters obtained from the two types of emission in the same framework.

Due to large number of parameters in the model and CPU time necessary for each simulation it is not 
possible to perform a systematic search for the best fit to data\footnote{To give an idea about 
calculation time, on a 3.06 GHz Intel Duo CPU T9900 processor a simulation with 3 time intervals 
- regimes - takes about 20-30 minutes without calculating inverse Compton and about 2 times longer 
with inverse Compton. If only 5 parameters are changed on a lattice with 5 nodes for each, total 
calculation time without inverse Compton would be about $10^5$ minutes or about 130 days. In practice 
more than 5 parameters should be adjusted to find best models and the time necessary for a systematic 
search would be much longer than above estimation.}. Nonetheless, despite their 
apparently arbitrariness, physically acceptable values of parameters are not completely random. The 
distance of external shocks from central source is determined by wind nebula surrounding progenitor 
neutron stars and its termination shock. In pulsars wind nebula extends to 
$\sim 10^{15}-10^{17}$~cm~\citep{nstarbowshock,nssheath,nstarmatt}. But its dependence on the properties 
of neutron stars and their evolution is not well understood. Our simulations show that an initial 
distance of $\sim 10^{16}$cm, which is in the logarithmic middle of the range given here leads to 
acceptable fit to the GW/GRB 170817A data. 

The density of circum-burst material on which the jet/outflow is shocked has a lower limit 
corresponding to the ISM density of $\lesssim 0.04$~cm$^{-3}$ in the host, concluded from the absence 
of significant neutral hydrogen in NGC 4993~\citep{gw170817earlyradio}. The spectrum of accelerated 
electrons in the shock is also fairly constrained to $\sim 2$ by Particle In Cell (PIC) 
simulations~\citep{fermiaccspec}. We do not consider any external magnetic field in the simulations 
presented here. The remaining parameters define the geometry of ejecta and surrounding material, and 
are adjusted by trial and error to fit the data as good as possible.

We find that a 3-component model consisting of a diluted ultra-relativistic jet with 
$\Gamma \sim \mathcal{O}(100)$, a relativistic outflow with $\Gamma \sim \mathcal{O}(10)$ and a 
mildly relativistic outflow/cocoon with $\Gamma -1 \sim \mathcal{O}(0.1)$ can explain  
observations in radio and X-ray bands and satisfies upper limit constraint imposed on optical/IR 
data, see below for more details. From now on we call these components C1, C2, and C3, respectively. 
Parameters of the model for these components are listed in Table \ref{tab:param}. We remind that as 
the parameter space was not systematically searched, values of parameters for the best model given 
here should be treated as order of magnitude estimations. For this reason we do not provide any 
uncertainty for them. Moreover, to see how variation of parameters can affect light curves and 
spectrum, and whether there is large degeneracies in the parameter space, which may invalidate 
interpretation of the model and its comparison with data, in Appendix \ref{app:paramvar} we present 
light curves of several variant models for each of the above components and compare them with the model 
presented in Table \ref{tab:param}. Notably, the model presented in Fig. \ref{fig:paramvarc3}-a 
for component C3 has a Lorentz factor of $4$, that is similar to estimation from apparent superluminal 
motion of radio counterpart~\citep{gw170817lateradiosuprlum}, and fits radio data up to 
$\sim T+200$~days roughly as good as C3 in Table \ref{tab:param}. However, at later times the latter 
provides a better fit. See Sec. \ref{sec:jetangle} for further discussion of this subject.

\subsection {Data used for comparison with models} \label{sec:data}
For comparing models with observations we use published data from various sources:
 
Radio data is in $5-6$~GHz radio band taken from~\citep{gw170817latedecline} except for 
the last two points which are taken from~\citep{lateradio} and~\citep{gw170817lateradio300}. The 
radio data in~\citep{gw170817latedecline} is reproduced from observations 
of~\citep{gw170817earlyradio,gw170817lateradio}. 

Optical/IR observations in $r$ and $i$ bands or energetically close bands i.e. $R$, HST $F606W$ 
and HST $F814W$. For $< T+10$ days the data is taken from~\citep{gw170817optdes}. The $R$ magnitude 
at $\sim T+10$ is taken from~\cite{gw170817rprocess}, $r$ band magnitude at $\sim T+110$ is 
from~\citep{gw170817lateopt}, HST $F606W$ at $\sim T+134$ from \citep{gw170817latedecline} 
(originally from~\citep{gw170817latexary}), HST $F814W$ magnitude at $\sim T+160$ 
from~\citep{gw170817lateopt160} and HST $F606W$ magnitude at later times 
from~\citep{gw170817lateopthstir}. Magnitude $m$ is changed to flux density in ph/sec/cm$^2$ using 
$F (\text{ph/sec/cm}^2) = 10^{-0.4 (m + 48.6) + 23} (\Delta \lambda / \lambda) \times 1.509 \times 10^3$.

X-ray data is in $0.3 - 8$~keV for Chandra observations and $0.3 - 10$~keV for data taken by the 
Swift-XRT and by the XMM-Newton. For the epoch before $\sim T+9$ days only 2 upper limit for X-ray 
flux is available: an upper limit for time interval $\lesssim T + 10$~days from the Neil Gehrels 
Swift-XRT~\citep{gw170817swiftnustar} and an upper limit at $\sim T + 2.2$ from Chandra 
observatory~\citep{gw170817xray}. The X-ray counterpart was detected and followed up later and 
we use data from Chandra observations at $\sim T+9$ days~\citep{gw170817cxc2day}, 
$\sim T+16$ days~\citep{gw170817xray}, $\sim T+110$~\citep{gw170817latexary}, XMM-Newton 
observation at $\sim T+134$ days~\citep{gw170817latedecline}, Chandra observations at 
$\sim T+260$ days~\citep{,gw170817xraycxc260,gw170817chandraxray260a} and at 
$\sim T+359$ days~\citep{gw170817chandraxray358}.

Despite heterogeneity of the data we did not attempt to homogenized it for not to add further 
uncertainties. This fact and mismatch between observations and simulated bands should be taken 
into account when models are compared with data. 

After the detection of the prompt Gamma-ray spikes by the Fermi-GBM~\citep{gw170817fermi} and the 
Integral-IBIS~\citep{gw170817integral} GW/GRB 170817 is not detected in high energy electromagnetic 
bands. Only an upper limit on any extended emission in $15-50$~keV band, based on averaging of 
background counts per 16 days during from $T \gtrsim 2$~sec to $\sim T + 1$~year, is obtained from 
Swift Survey data~\citep{batgammaupper}.

\subsection{Comparison with observations} \label{sec:compdata}
\begin{center}
\begin{figure}
\begin{tabular}{p{9cm}p{9cm}}
{\bf a)} & {\bf b)} \\
\includegraphics[width=9cm]{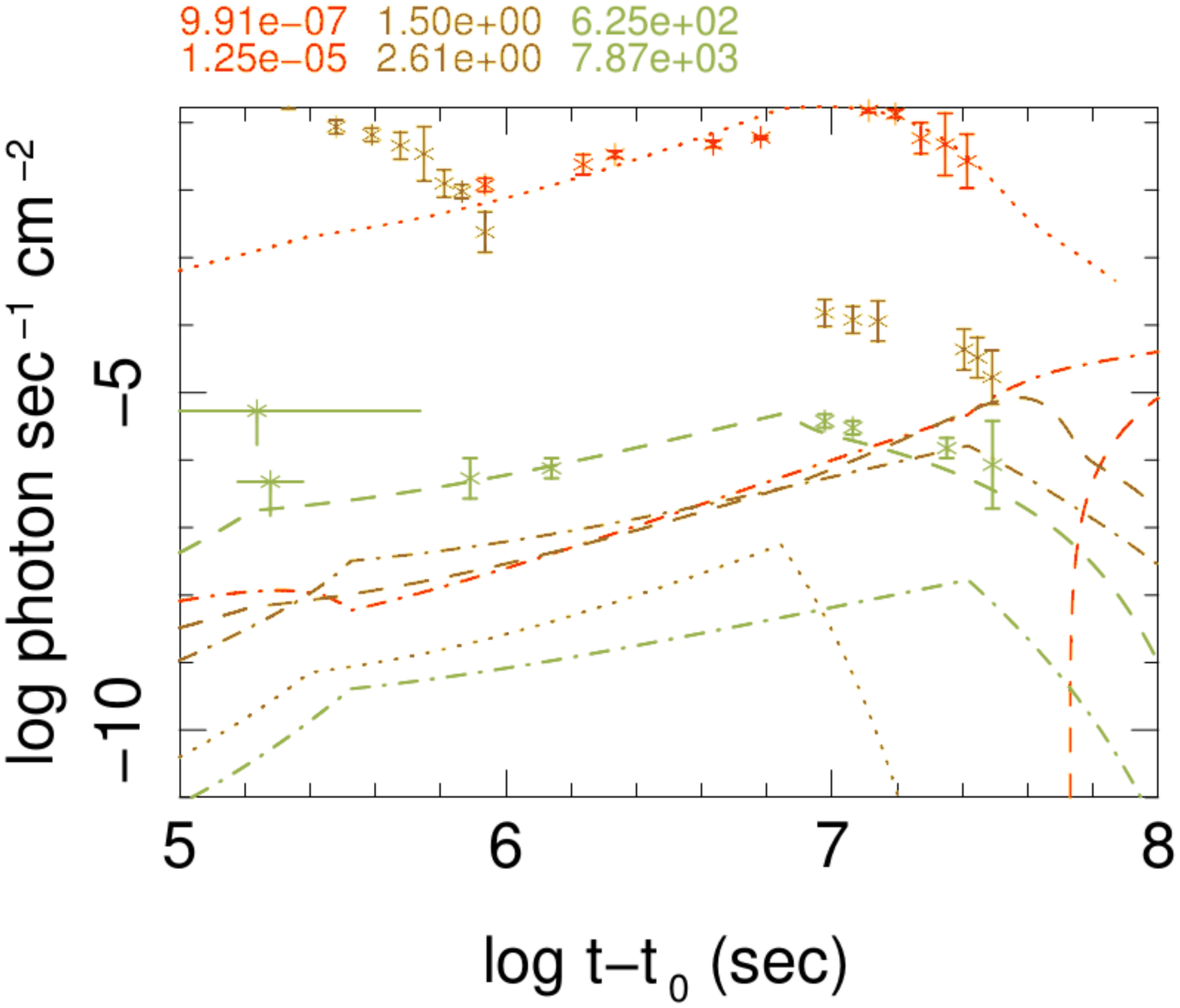} &
\hspace{-1.5cm}\includegraphics[width=9cm]{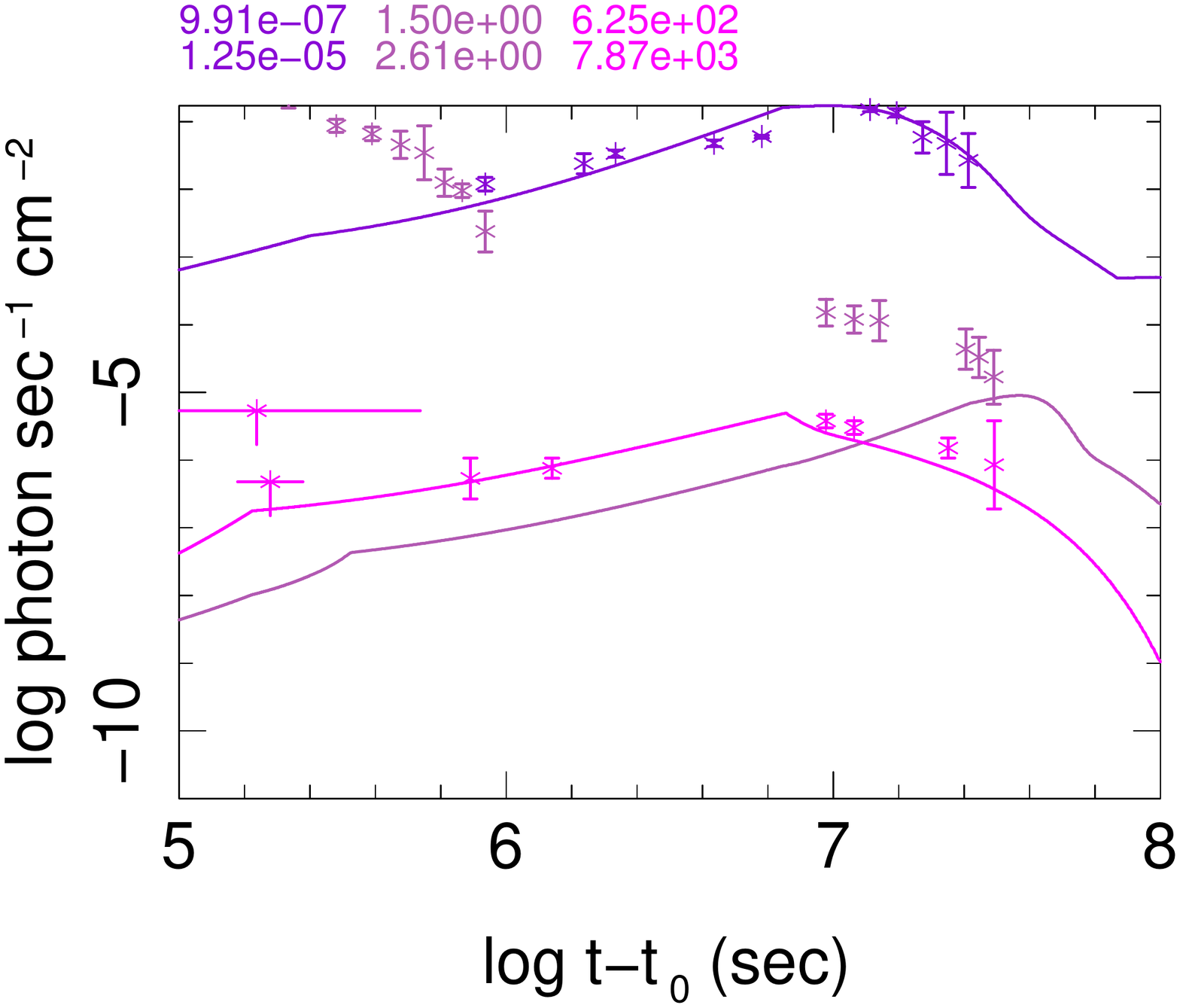}  
\end{tabular}
\caption{{\bf a)}: Radio, optical/IR, and X-ray light curves of simulated 3-component. Left: Light 
curves of the 3 components: ultra-relativistic (C1) (dash lines), relativistic (C2) (dash-dot), 
mildly relativistic (C3) (dotted lines). The energy range for each band is written on the top of 
each plot in the same colour/gray scale as the curves. Stars and upper limits present the data 
described in Sec. \ref{sec:data}. {\bf b)}: Sum of the light curves of the 3 components: radio 
(magenta/light grey), optical (purple/medium grey), X-ray (dark purple/dark grey).
\label{fig:lc}}
\end{figure}
\end{center}

Fig. \ref{fig:lc}-a shows light curves of each component of the model and Fig. \ref{fig:lc}-b the 
total light curves in each band. We notice that X-ray and radio light curves have much better fit to 
data than simulated optical light curve. However, we know that optical data, specially at early 
times, is dominated by kilonova 
emission~\citep{gw170817bluekilonova,gw170817bluekilonovamod,gw170817bluekilonovapol,gw170817kilonovaspeed} 
and cannot be modeled with shock/synchrotron emission. Thus, presentation of early optical data in 
Fig. \ref{fig:lc} and other light curve plots in this work is for the sake of completeness. Although 
a priori kilonova contribution can be modeled and removed, the residual would be model dependent. For 
this reason, we show optical data as it is observed, but use it as an upper limit for the GRB contribution 
which should be respected by any model of GRB 170817A afterglows. As the model presented in 
Fig. \ref{fig:lc} (and some of its variants discussed in Appendix \ref{app:paramvar}) fit well radio and X-ray data, we 
presume that their prediction for optical emission should be reliable. Under this assumption, these 
models show that after $t_{kn} \sim T+200$ to $\sim T+300$~days - depending on the 
model - kilonova contribution in optical/IR emission was not anymore significant and the afterglow 
was dominated by synchrotron emission from external shocks of the relativistic outflow\footnote{Note 
that in addition to model dependence one has to take into account the heterogeneity of 
the optical/IR data and mismatch with the simulated energy band. For these reasons the time of 
dominance of synchrotron with respect to kilonova contribution given in the text should be considered 
as nominal rather than exact.}. This estimation for the time of kilonova fainting $t_{kn}$ is consistent 
but somehow larger than that by~\citep{gw170817kilonovafaint}, who find $t_{kn} \gtrsim 100$ days. 
However, due to degeneracies in the parameter space of their kilonova emission 
model~\citep{gw170817kilonovafaint} do not fit it to data. Moreover, synchrotron light curves used 
for estimating GRB contribution are one possibility between many, see the discussion of 
degeneracies in Appendix \ref{app:paramvar}. Other works on the evolution of the kilonova emission, 
e.g.~\citep{gw170817optkilonovath,gw170817bluekilonova,gw170817kilonovamassexc,gw170817kilonovafaint0} 
are limited to initial tens of days after merger and cannot be compared with late optical data. 
Thus, we conclude that at present a satisfactory model for late optical emission of kilonova is not 
available.

\begin{center}
\begin{figure}
\begin{tabular}{p{8cm}p{7cm}p{5cm}}
{\bf a)} & \hspace{-1.5cm} {\bf b)} & \hspace{-3.5cm} {\bf c)}\\
\includegraphics[width=8cm]{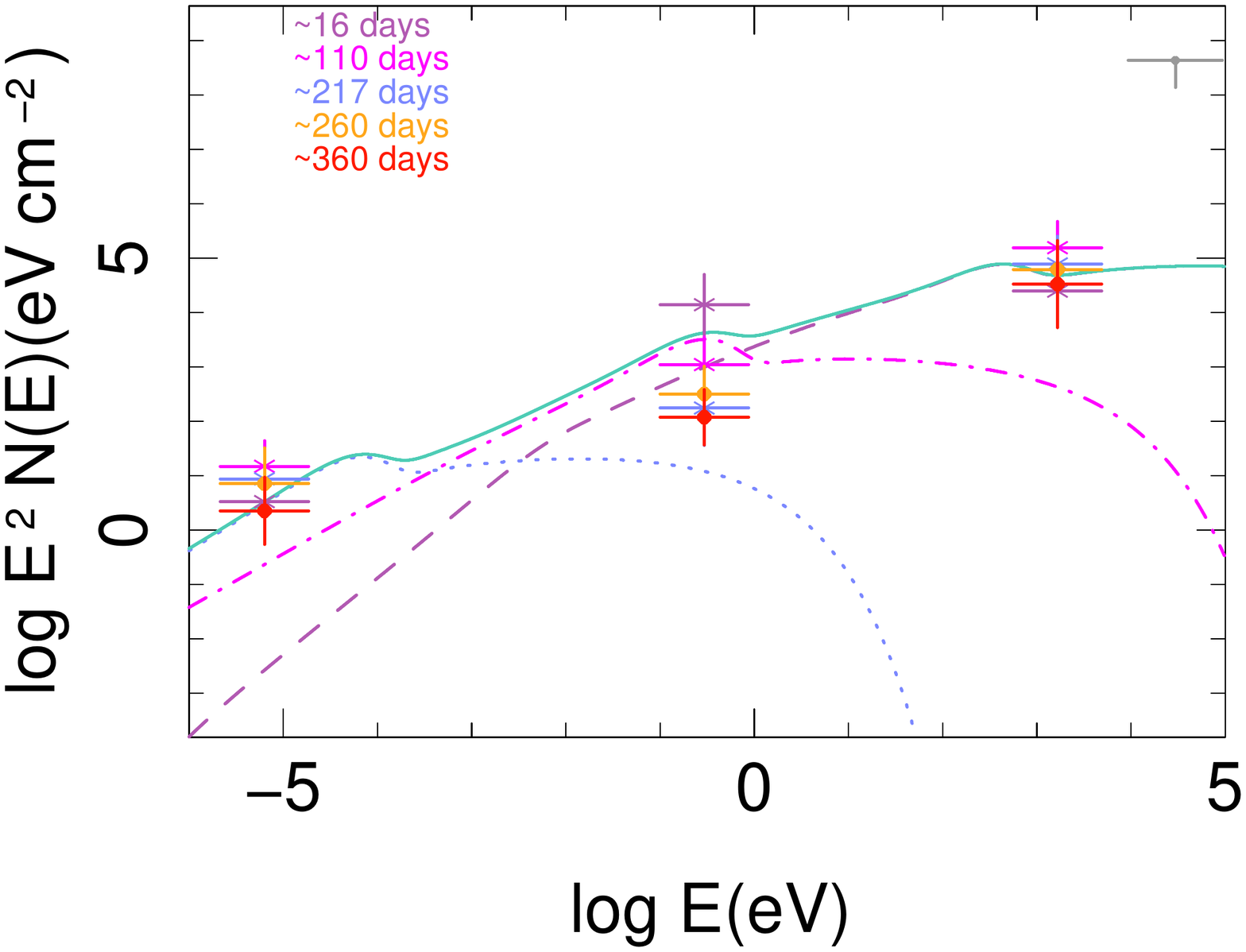} & 
\hspace{-1.5cm}\includegraphics[width=8cm]{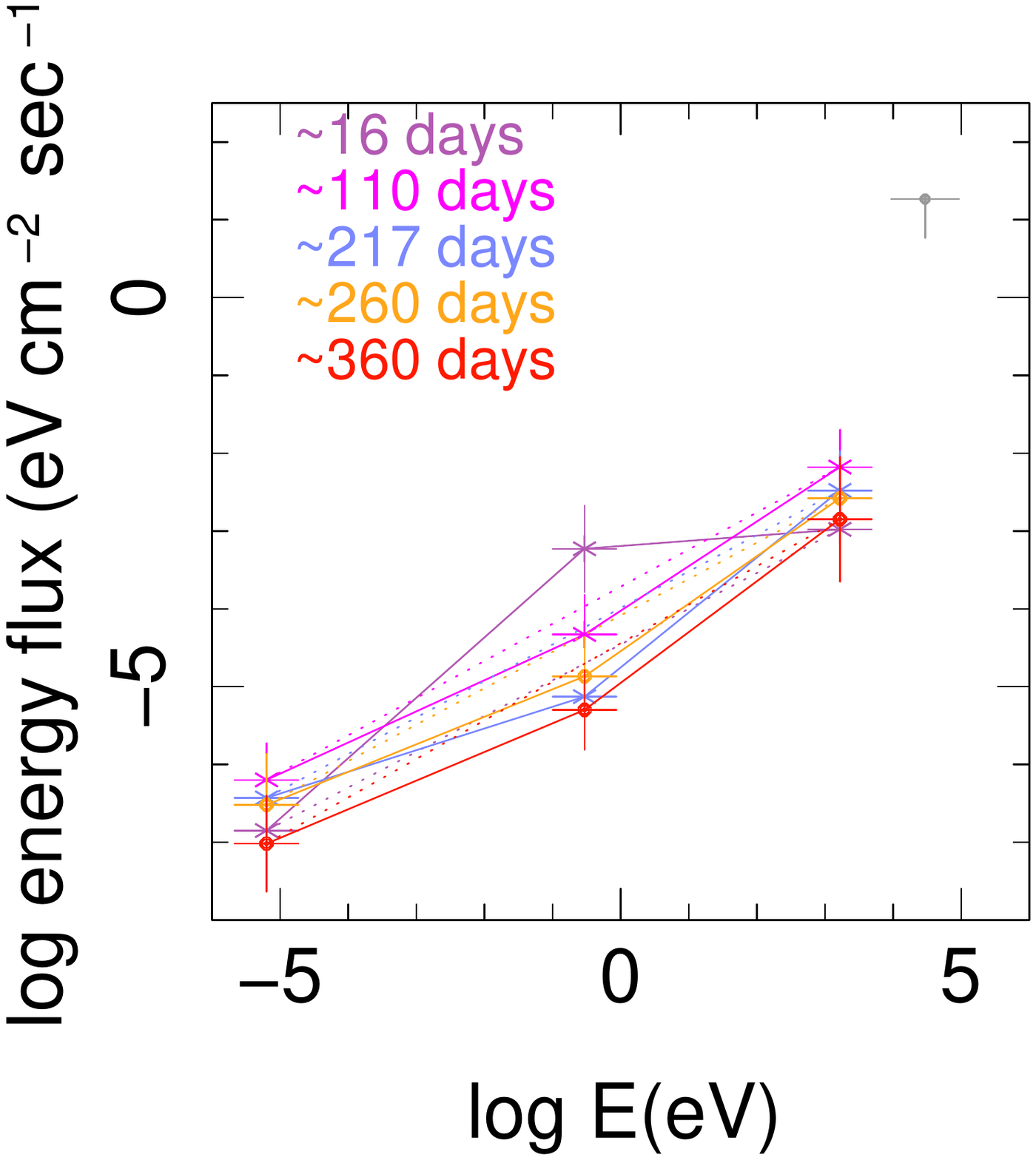} & 
\hspace{-3.5cm}\includegraphics[width=8cm]{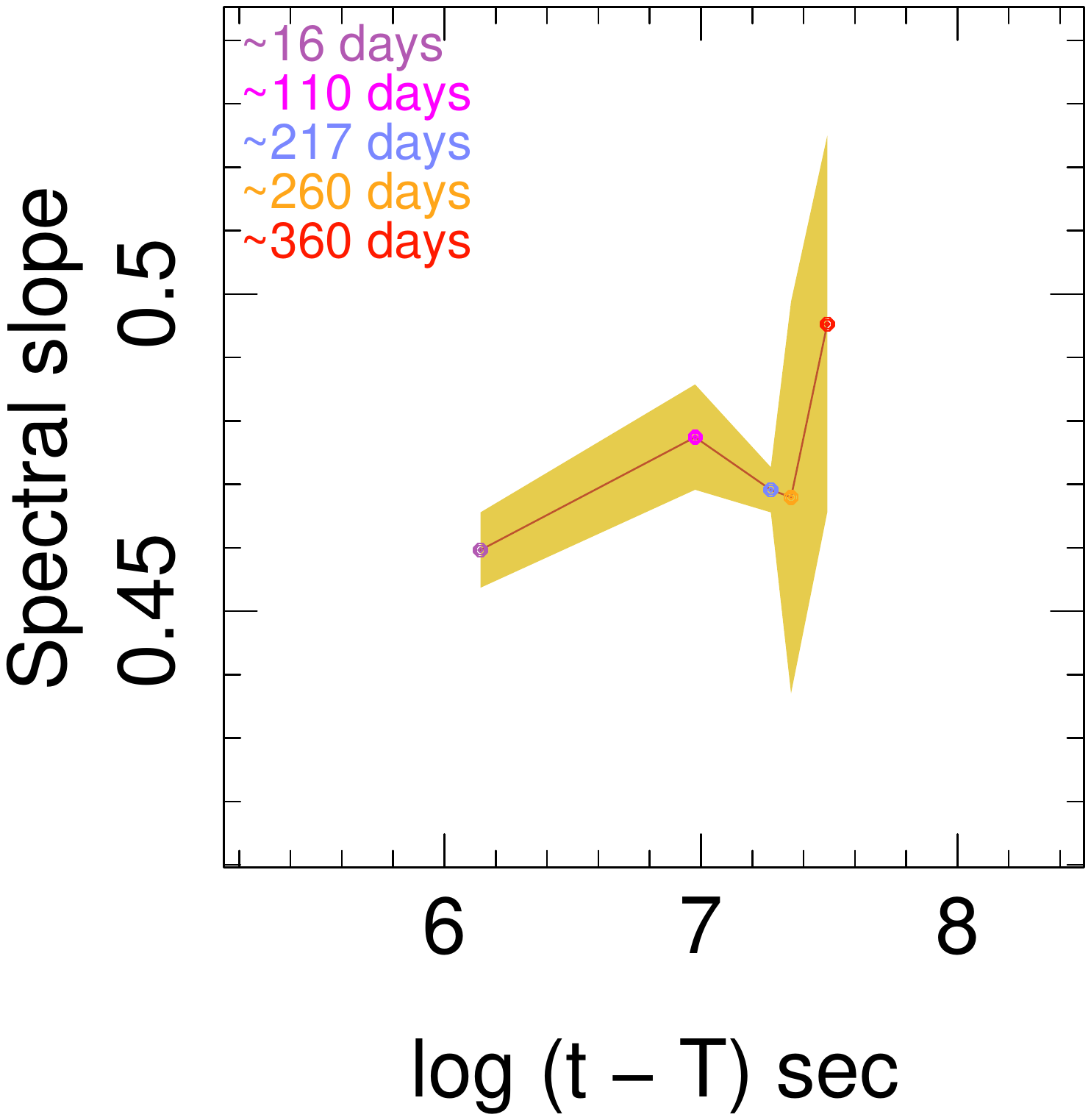} 
\end{tabular}
\caption{{\bf a)}: Spectra of components and their sum: ultra-relativistic (C1) (dash line), 
relativistic (C2) (dash-dot), mildly relativistic (C3) (dotted line), sum of 3 components 
(full line). Crosses present observations at different times in radio, optical/IR, and X-ray, 
optical and radio bands. When data for a time interval was not available an interpolation has been 
used. The width of crosses presents the width of the corresponding filter and are much larger than 
observational uncertainties. The upper limit at $E \sim 15-50$~keV is from the Swift-BAT survey 
data~\citep{batgammaupper}, see Sec. \ref{sec:data} for more details. To generate a pseudo-spectrum 
from flux measurements, we have normalized data such that X-ray at $T+217$~days become equal to 
maximum of simulated spectrum in the simulated energy interval. {\bf b)}: Spectrum of energy flux. 
The lines connecting the data points are added to facilitate the illustration of spectral variation. 
Similar to {\bf a)} plot width of crosses presents filler width. {\bf c)}: Evolution of the slope 
of pseudo-spectrum using only radio and X-ray data, i.e slope of dotted lines in {\bf b)}. The shaded 
region is uncertainty of calculated slopes. \label{fig:spect}}
\end{figure}
\end{center}

Another assessment of the performance of the model can be made by comparing simulated spectrum with 
the photometric spectrum reconstructed from observations shown in Fig. \ref{fig:spect}-a. It shows a 
good a consistency between simulated spectrum and the data. As expected, amplitude of optical emission 
at $\sim T+ \mathcal{O}(10)$~days is higher than the model. Moreover, due to the dominant contribution 
of kilonova emission the pseudo-spectrum of energy flux shown in Fig. \ref{fig:spect}-b at this epoch 
is significantly different from those of later times. In this plot if we neglect optical data and 
use only radio and X-ray (dotted lines in Fig. \ref{fig:spect}-b), spectra of all observation epochs 
have similar behaviour. Of course, a 2-point pseudo-spectrum is a very crude presentation of 
the broad-band spectrum. Nonetheless, it is not affected by kilonova emission. Fig. \ref{fig:spect}-c) 
shows evolution of spectral slope, which is similar to afterglows of other GRBs, namely softer during 
earliest observations around $T+\mathcal{O}(10)$~days, gradually becomes harder until the peak of 
emissions around $T+110$~days, and finally softens at later times. The last data point in 
Fig. \ref{fig:spect}-c is obtained from radio and X-ray data with largest uncertainties 
(see Fig. \ref{fig:lc}-b) and apparent increase of slope and hardening of spectrum during last 
observation is very uncertain.

\begin{table}
\begin{center}
\caption{Parameter set of simulated models. \label{tab:param}}
\end{center}
{\scriptsize
\begin{center}
\begin{tabular}{p{5mm}p{5mm}p{5mm}p{12mm}p{8mm}p{10mm}p{5mm}p{5mm}p{5mm}p{5mm}p{6mm}p{5mm}p{5mm}p{5mm}p{8mm}p{8mm}|}
\hline
Comp. & mod. & $\gamma'_0$ & $r_0$ (cm) & $\frac{\Delta r_0}{r_0}$ & $(\frac{r}{r_0})_{max}$ & $p$ & $\gamma_{cut}$ & $\kappa$ & $\delta$ & $\epsilon_B$ & $\alpha_B$ & $\epsilon_e Y_e$ & $\alpha_e$ & $N'$ (cm$^{-3}$) & $n'_c$ (cm$^{-2}$) \\
\hline 
\multirow{4}{5mm}{Ultra. rel. (C1)} 
 & 1 & 130 & $10^{16}$ & $10^{-7}$ & 1.5 & 1.8 & 100 & -0.5 & 0.5 & $0.08$ & -1 & 0.1 & -1 & $0.04$ & $5 \times 10^{22}$ \\
 & 2 & -   &   -      &  -    & 15  &  -  & 100 & 0.3 & 0.1 &  -     &  0 &  -   &  0 &   -    &    -   \\
 & 2 & -   &   -      &  -    & 20  &  -  & 100 &  0.4  & 0.05 &  -     &  1 &  -   &  1 &   -    &    -   \\
\hline
\multirow{4}{5mm}{Rel. (C2)} 
 & 1 & 5 & $10^{16}$ & $10^{-6}$ & 2 & 2.1 & 100 & -0.5 & 1 & $0.08$ & -1 & 0.1 & -1 & $0.04$ & $10^{23}$ \\
 & 2 & -   &   -      &  -    & 40  &  -  & 100 & 0.4 & 0.1 &  -     &  0 &  -   &  0 &   -    &    -   \\
 & 2 & -   &   -      &  -    & 100  &  -  & 100 &  0.5  & 1 &  -     &  1 &  -   &  1 &   -    &    -   \\
\hline
\multirow{4}{5mm}{Mildly rel. (C3)} 
 & 1 & 1.06 & $1.5 \times 10^{16}$ & $10^{-2}$ & 1.5 & 1.8 & 100 & -0.5 & 1 & $0.08$ & -1 & 0.02 & -1 & $0.008$ & $10^{24}$ \\
 & 2 & -   &   -      &  -    & 10  &  -  & 100 & 0. & 0.1 &  -     &  0 &  -   &  0 &   -    &    -   \\
 & 2 & -   &   -      &  -    & 10  &  -  & 100 &  1  & 1 &  -     &  1 &  -   &  1 &   -    &    -   \\
\hline
\end{tabular}
\end{center}
}
\begin{description}
\item {$\star$} Each data line corresponds to one simulated regime, during which quantities 
listed here remain constant or evolve dynamically according to fixed rules. A full simulation 
of a burst usually includes multiple regimes (at least two). 
\item {$\star$} Horizontal black lines separate time intervals (regimes) of independent 
simulations identified by the label shown in the first column.
\item {$\star$} A dash as value for a parameter presents one of the following cases: it is 
irrelevant for the model; it is evolved from its initial value according to an evolution 
equations described in~\citep{hourigrb,hourigrbmag}; it is kept constant during all regimes.
\end{description}
\end{table}

\subsection{Comparison with other analyses} \label{sec:comp}
As we described in the Introduction, some other authors have analyzed late afterglows of 
GW/GRB 170817A by using multi-component models including an ultra-relativistic jet. For 
instance,~\citep{gw170817latexary,gw170817latexoptradio1} consider a structured jet with a top-hat 
ultra-relativistic component with $\Gamma \sim 100$ in the inner $\theta \lesssim 9^\circ$, where 
$\theta$ is angle with respect to symmetry axis of outflow, and a relativistic flow which its Lorentz 
factor decreases in the angular interval $10^\circ \lesssim \theta \lesssim 60^\circ $ and has a mean 
$\Gamma \sim 10$. The authors consider a line of sight angle $\theta_v \sim 20^\circ$ from the jet axis. 
This model is very similar to the model described in the latest version 
of~\citep{gw170817latexraystructjet}. But Lorentz factor and energy profile of the jet in the two 
works are different. \citep{gw170817lateopthstir} consider two profiles for the jet, one similar to 
2-component model of~\citep{gw170817latexary,gw170817latexoptradio1}. The other model assumes a jet 
with Gaussian energy and Lorentz factor profiles. Both models find a small central/core angle of 
$\lesssim 5^\circ$. In their 2-component model high and low Lorentz factors are $\sim 100$ and $\sim 5$, 
respectively. However, in the Gaussian model on-axis Lorentz factor can be as large as $\gtrsim 900$. 
\citep{gw170817lateradio300} model light curves of afterglows with a phenomenological function, which 
effectively has two coupled components. A priori they can be related to their Lorentz factor and 
density profiles. However, the complicated and non-linear form of this model make any physical 
interpretation difficult.

A question arises here: Is the presence of an ultra-relativistic component in our models and those 
cited in the previous paragraph inevitable ? The existence of such a component means that our line 
of sight passed through the ultra-relativistic core of the jet. However, several authors find 
that the data can be fit with synchrotron emission of a jet with a Lorentz factor $\lesssim 10$ in 
our direction. For instance,~\citep{gw170817aglowlorentz} find several variants of such models and 
show that they fit the data up to $\lesssim 150$~days. However, none of their models fit earliest 
X-ray data and aside from radio bands their data set include only a couple of points in X-ray and 
optical/IR. The model of~\citep{gw170817latexary,gw170817latexoptradio1} reviewed in the previous 
paragraph has an ultra-relativistic core with $\Gamma \sim 100$ up to $\theta \sim 9^\circ$, but the 
viewing angle of $\lesssim 20^\circ$ found by these analyses means that the line of sight passes 
through a region with a Lorentz factor $\ll 100$ - they do not indicate the exact value of Lorentz 
factor on the line of sight. Similar to~\citep{gw170817aglowlorentz} their model under estimate 
early X-ray data. To these examples one should add other models, namely 
cocoon~\citep{gw170817lateradiosuprlum,gw170817lateradio300,gw170817cocoonsimul} and cocoon 
break out~\citep{gw170817cocoon, gw170817cocoon0}, which have much smaller Lorentz factors. None 
of these models is consistent with late X-ray data and over-estimate its flux. Their consistency 
with optical data is also uncertain. Interestingly, with publication of later data, a general trend 
in the literature toward models with larger Lorentz factors is discernible - specially in 
works using observations up to $\sim 200$ days and beyond. For example, in earlier versions 
of~\citep{gw170817latexraystructjet} a Lorentz factor of $\lesssim 10$ was associated to the jet. But 
in the final version of this work, which includes more data, the value of Lorentz factor is increased 
to $\mathcal{O}(100)$. ~\citep{gw170817lateopthstir} who use observations from earliest time to 
roughly one year after merger find a Lorentz factor $\gtrsim \mathcal{O}(100)$. In conclusion, it 
is unlikely that the necessity of an ultra-relativistic Lorentz factor along our line of sight found 
here be due to degeneracies or quirk of our models.

In addition to differences in Lorentz factor on the line of sight, predictions for time of kilonova 
decay $t_{kn}$ from modelling of GRB 170817A in the literature are not the same and are somehow 
shorter than $t_{kn} \sim 200 - 300$ days obtained from the model of Table \ref{tab:param} and its 
variants in Appendix \ref {app:paramvar}. Specifically, GRB afterglow models 
of~\citep{gw170817lateopthstir,gw170817optfit,gw170817kilonovadec} seem to fit optical data at 
$\gtrsim T+100$ days. A priori this means that a $t_{kn} \lesssim 100$ days can be concluded from 
these works. However, they do not show early optical/IR data points and do not explicitly 
discuss the contribution of kilonova in the optical flux. In fact, it seems that they adjust the 
models to fit optical data after $\sim 100$ days. In this case, the time of kilonova decay in their 
models is a prior rather than a posterior conclusion derived from data. By contrast, as described 
earlier, in selection of plausible models here optical data was only considered as an upper limit 
and no attempt was made to fit simulated light curves to it.

\section{Interpretation} \label{sec:interpret}
\subsection{Origin of ultra-relativistic component} 
In what concerns the Lorentz factor of the ultra-relativistic component of our model and similar 
models in the literature, they are consistent with analysis of~\citep{hourigw170817}. Indeed, 
modelling of the prompt gamma-ray emission of GW/GRB 170817A shows that the best estimation for 
the Lorentz factor of the jet at the end of main gamma-ray spike is $\gtrsim 100$. 

A priori it is not expected that a relativistic jet preserves its Lorentz factor during propagation 
from the site of internal shocks, which in the case of GW/GRB 170817A was at 
$r_i \equiv r_0 \sim \mathcal{O}(1) \times 10^{10}$~cm from merger~\citep{hourigw170817}, up to its 
collision with the ISM or circum-burst material at 
$r_e \equiv r_0 \sim \mathcal{O}(1) \times 10^{16}$~cm, in which afterglows are generated. However, 
weaker late internal shocks, cooling of shocked material, and 
shocks on circum-merger material between $r_i$ and $r_e$ can dissipate kinetic energy of the jet and 
decelerate it. Indeed, weak gamma-ray and X-ray spikes and continuous emission are observed in both 
long~\citep{xrtafterglow} and short bursts~\citep{grbshorttailbepposax,grbshorttailbat} and are 
interpreted as emission from internal shock of side lobes - high-latitude regions - of the 
jet~\citep{grbtailsidejet} or weaker internal shocks~\citep{xrtafterglow,hourigrbmag}. 

Shocks on the material close to the BNS can be simulated by considering short distance external shocks 
similar to models presented in Fig. \ref{fig:paramvarc1}-d, -e, \& -f, which have 
$r_e = r_0 = 10^{13}$~cm or $r_e = r_0 = 10^{15}$~cm. Because the value of parameter $\kappa$, which 
parametrizes variation of ISM/circum-burst material density is the same as C1 model (and its 
variants), it can be considered as extension of the same distribution. Thus these simulations 
presents interaction of the ultra-relativistic jet with material near the merger. An important 
properties of these simulation is that their emissions peak at early times. Thus, along with weak 
late internal shocks, they can a priori produce a plateau regime around $ < T+10^5$~sec, which is 
observed in some short GRBs such as GRB 070724A, GRB170127B, and GRB 111020A. The latter GRB is a 
an interesting case because it occurred at redshift $0.02$~\citep{grb111020aredshift} and was only 
a few times brighter than GRB 170817A. Unfortunately, for the latter burst no data is available in 
this time interval and these emissions cannot be investigated\footnote{If such data existed, 
amplitude of models in Fig. \ref{fig:paramvarc1}-d \& e could be adjusted to the data.}.

Simulations presented in this work and those reviewed in Sec. \ref{sec:comp} are the evidence that 
despite energy loss, a fraction of ultra-relativistic jet survives without significant 
dissipation\footnote{One of shortcomings of the phenomenological model 
of~\citep{hourigrb,hourigrbmag} is that it is deterministic and cannot give a probability or 
distribution for dissipated and non-dissipated particles in the jet. \label{foot:shortcom}}. Comparing 
the column density of ultra-relativistic component C1 in Table \ref{tab:param} (and its variant in 
Fig. \ref{fig:paramvarc1}-b) with the initial column density of the jet before internal shocks, 
which according to analysis of~\citep{hourigw170817} was $10^{25}$~$cm^{-2}$, shows that it is 
$\sim 200$ folds less than what it was pre-prompt emission. Assuming that dissipation of the second 
fainter and softer prompt spike
corresponds to relativistic component C2 listed in Table \ref{tab:param} and its variants discussed in 
Sec. \ref{sec:c3var}, we observe that its column density had become $\sim 5-50$ folds lower, i.e. 
reduced from $5 \times 10^{23}$~$cm^{-2}$ to $10^{23}$~$cm^{-2}$ in C2 in Table \ref{tab:param} or 
to $10^{22}$~$cm^{-2}$ for models in which Lorentz factor $\gamma'_0 = \Gamma = 30$, i.e. the same as 
the second prompt gamma-ray peak. Evidently, there is no proof for any relation between C2 model 
and the second spike in the prompt gamma-ray because time resolution of late observations does not 
allow to discriminate between afterglows of different density shells. Nonetheless, it is reasonable 
to presume that late time slower component include contribution from slower and more easily 
dissipatable part of the prompt outflow. Alternatively, rather than considering these models as 
competitive, they may be considered as finer decomposition of jet at late times, where the model 
with $\Gamma = 30$ presents a highly reduced remnant of second peak and C2 in Table \ref{tab:param} 
presents dissipated and laterally expanded remnant of both prompt shells. In this case, column 
densities and/or thickness of active regions of these components must be slightly smaller than what 
is given in Table \ref{tab:param} and Fig. \ref {fig:paramvarc2} to make the total emission 
consistent with data, see Sec. \ref{sec:c3var} for more details. In any case, these variant models 
and their parameter space are consistent with our warning in Sec. \ref{sec:ag} that values of 
parameters must be considered as order of magnitude estimations. Another important conclusion from 
variant models presented in Appendix \ref{app:paramvar} is that it does not seem possible to explain 
observed data with a 1-component or even 2-component model\footnote{The necessity of a 
multi-component model is the clear evidence that the phenomenological formulation used here is too 
simplistic. Notably, it does not take into account anisotropic density and Lorentz factor of 
particles inside the jet. Asymptotic formulation of~\citep{{emission1}} has also this problem.}.

If the jet were an adiabatically expanding cone, its column density had to decline by a factor of 
$\sim (r_i/r_e)^{-2} \sim 10^{12}$ where $r_i$ and $r_e$ are distance from center when 
internal and external shocks occurred, respectively. The much smaller dilation factors we 
find for the 3-component model means that the material inside the jet had preserved in some extent 
its internal coherence and collimation - most probably through imprinted electric and magnetic fields 
in the plasma - and its expansion was not completely adiabatic and free. Alternatively, accretion of 
material between $r_i$ and $r_e$ could a priori compensate reduction of column density due to an 
adiabatic expansion. Assuming a power-law variation of jet's column density by accretion per unit fly 
length, i.e. $dn'_a (r) / dr = (\alpha n'_a (r_0)/ r_0)(r/r_0)^{\alpha - 1}$ and neglecting loss of 
kinetic energy to radiation, from conservation of momentum we find: 
$n'_c (r) = n'_0 (r/r_0)^{-2} + n'_a (r_0) [(r/r_0)^\alpha -1]$ and 
$\beta (r) = n'_0 \beta (r_0) / [n'_0 + n'_a (r_0) (r/r0)^2 ((r/r_0)^\alpha -1)]$. For $r/r_0 = 10^6$ and 
$n'_c (r) / n'_0 \sim 200$, $\beta (r) \sim 2 \times 10^{-10} \beta(r_0)$. This means that the jet would 
be practically stopped. Therefore, small reduction of column density could not be due to accretion of 
circum-burst material. On the other hand, much smaller column density of C1 with respect to the prompt 
jet demonstrates that energy dissipation during prompt shocks due to the accretion of material close 
to the merger and/or internal interactions and loss of internal coherence had indeed occurred and 
led to lateral expansion, and thereby dilation of the jet's core and reduction of its Lorentz 
factor, specially in its outer boundaries. Therefore, our assumption about dissipation of the 
ultra-relativistic jet in~\citep{hourigw170817ag} was justified. Furthermore, thanks to long 
and extended follow up of this transient, for the first time we are clearly detecting the tiny 
remnant of the ultra-relativistic core of the jet in a short GRB and can distinguish its signature 
from slower part of the polar outflow of the BNS merger.

\subsection{Jet profile and viewing angle} \label{sec:jetangle}
Since the observation of unusually faint prompt gamma-ray of GRB 170817A, understanding the 
underlying physics has been the subject of significant debate. Specifically, the viewing angle of 
the ejecta had a central role in the proposed explanations.

Observation of gravitational waves led to an estimation for orbital inclination angle of 
$ 18^\circ \lesssim \theta_{in} \lesssim 27^\circ$~\citep{gw170817decline}. Moreover, superluminal 
motion of radio afterglow with an apparent speed of $\beta_{app} = 4.1 \pm 0.5$ is 
observed by~\citep{gw170817lateradiosuprlum,gw170817lateradiosuprlum0} and in combination with 
information from light curves and standard shock and synchrotron model and jet 
simulation~\citep{gw170817lateradiosuprlum} estimated an off-axis angle of $\sim 20^\circ \pm 5^\circ$ 
for the line of sight. The above value of $\beta_{app}$ constrains 
Lorentz factor of the source to $\Gamma \gtrsim 4$. In this case, C3 component in 
Table \ref{tab:param} is not consistent with the observed superluminal motion. Nonetheless, a variant 
of C3 with $\Gamma = 4$ and smaller ISM/circum-burst density and active region thickness presented 
in Fig. \ref{fig:paramvarc3}-a satisfies superluminal constraint. It fits radio data as good as C3 
in Table \ref{tab:param} up to $\sim T+230$~days, which coincides with the second observation epoch 
of~\citep{gw170817lateradiosuprlum}. However, at later times the model with smaller Lorentz factor 
is a better fit to the data. 

On the other hand, some other issues must be considered. 
Formulation of the phenomenological model developed in~\citep{hourigrb,hourigrbmag} assumes 
$\Gamma \gg 1$. Therefore, uncertainty of C3 simulation with $\Gamma$ close to 1 and its parameters 
should be larger than other components. Moreover, estimation of the error on $\beta_{app}$ measurement 
might have been too optimistic. Fig. 2 (and Fig. 1 of extended data) 
of~\citep{gw170817lateradiosuprlum} shows both error ellipses and synthesized beam shape, and 
explicitly says that the source was not resolved. Giving the fact that the size of the image 
displacement is comparable to beam size, the size of 1-sigma uncertainty ellipse seems too small. 
In addition, lateral expansion of the cocoon between two observations used for determining 
$\beta_{app}$, which were separated by $\sim 150$~days, might have changed the position of image 
centroid~\citep{jetbeaming,jetlateralexpan}. It is expected that due to scattering and energy 
dissipation in the core of the jet, the cocoon becomes slower, denser and expands 
laterally~\citep{{jetpicsimul}}. Therefore, its effective off-axis angle with respect to line of 
sight might have changed between two observations, leading to larger apparent displacements. Thus, 
although $\beta_{app}$ was certainly nonzero, it could be much smaller than published value, and 
thereby consistent with a smaller Lorentz factor. An alternative possibility is to consider both 
C3 in Fig. \ref{fig:paramvarc3}-a and C3 in Table \ref{tab:param}. In this case the former presents 
the state of out flow at angles further than C2 and the latter its state at even larger angles. 
Considering the uncertainty of data and parameters, such a 4-component model is consistent with 
radio data. In any case, most of our conclusions remain valid for both $\Gamma \sim 1.1$ and 
$\Gamma \sim 4$ for the component C3, and both 3 and 4 component models. For this reason we only 
consider 3-component model and discuss differences in interpretations for low and high Lorentz 
factor C3 whenever necessary. 

As a final evidence for plausibility of a mildly relativistic component with small Lorentz factor, 
we remind direct observation of a cocoon in the long GRB 171205A and its associated supernova 
SN2017uk~\citep{grb171205asn2017uk}, which measures $\beta \sim 0.3$ corresponding to 
$\Gamma \sim 1.05$, i.e. similar to C3 component in Table \ref{tab:param}.

Giving the fact that synchrotron emission is highly directional and emission from a relativistic 
source is beamed, off-axis view of the jet has significant consequences for observations. For 
instance, assuming a uniform Lorentz factor across the emission surface, a far observer receives 
oblique radiation only from a cone with half angle $\theta \sim \arcsin (1/\Gamma)$. Therefore, 
if the jet of GW/GRB 170817 was structured, radiation received by an off-axis observer would be 
from jet's slower wings rather than faster core, and thereby dominated by photons with lower 
energies because they had been produced by weaker and less boosted shocks in the jet's high 
latitude (boundary) region. Defining maximum visibility angle $\theta_{max} \equiv \arcsin (1/\Gamma)$, 
its value for components C1, C2, and C3 of the model listed in Table \ref{tab:param} are 
$\theta_{max_1} \sim 0.5^\circ$, $\theta_{max_2} \sim 11.5^\circ$ and $\theta_{max_3} \sim 65^\circ$ or 
$14.5^\circ$ for C3 in \ref{tab:param} and its alternative in Fig. \ref{fig:paramvarc3}-a), 
respectively.

Fig. \ref{fig:lc} shows that each of the components of the model dominates the total emission in 
one of the three energy bands with observational data, and higher the Lorentz factor of the component, 
higher is the energy of its dominant emission among the three components
\footnote{For the time being, the simulation code used in this work uses an analytical expression 
for determining synchrotron/self-Compton flux. In the simulation code terms depending on higher 
order of $\theta$, angle between emitting element and line of sight, are neglected. This is a good 
approximation when $\Gamma \gg 1$. Under this approximation $\theta$ dependence is only through a 
$(cos (\theta) + \beta) \leq (cos (\theta) + 1)$ factor, which must be integrated between 
$\theta_0 \geq -\theta_{max}$ and $\theta_1 \leq \theta_{max}$. However, angular size of emitting 
surface may be smaller than $2 \theta_{max}$. In this case integration over maximum visible angle 
over-estimate the flux. But the difference would be at most a factor of few and comparable to other 
uncertainties of the model. Indeed, for this and other simplifications and approximations applied 
to the model that we should consider parameters as order of magnitude estimations.}
Taking into account the observed inclination of the BNS orbit and superluminal motion of radio 
counterpart, we interpret the 3-component model as a structured jet, in which each component 
approximately presents characteristics of the jet and its shocks on the surrounding material from 
our line of sight up to outer boundary of the outflow. The value of $\theta_{max}$ for the components 
are consistent with this interpretation\footnote{Although formulation of 
relativistic shocks and synchrotron emission in~\citep{hourigrb,hourigrbmag} is more systematic 
than standard approach, it remains very much simplified and phenomenological. For instance, it 
considers a uniform column density and Lorentz factor for the jet. However, GRMHD simulations 
show that these assumptions are not true and column density is a function of both Lorentz factor 
and angle. It is why three components are necessary to explain the data. Therefore, this division 
and characteristics of components must be considered as an effective and simplified description.}. 
Accordingly, their Lorentz factor presents azimuthal variation of velocity inside the polar 
outflow of the merger up to a $\cos \theta$ factor, that is 
$\Gamma_{simul,i} = \Gamma_i (\theta_i) \cos \theta_i$, where $\theta_i$ is angle between 
centroid of component $i$ of the model and the line of sight. The value of $\theta_{max}$ for the 
components constrains $\theta_i \lesssim \theta_{max}$ and thereby the effect of projection on the 
estimation of parameters. Specifically, for C1 it is negligible and for C2 is about $10\%$. If we 
consider the model with $\Gamma = 4$ for C3, the effect of off-axis is again about $10\%$. But for 
C3 model in Table \ref{tab:param} it can be as large as $42\%$.

\begin{figure}
\begin{center}
\includegraphics[width=9cm]{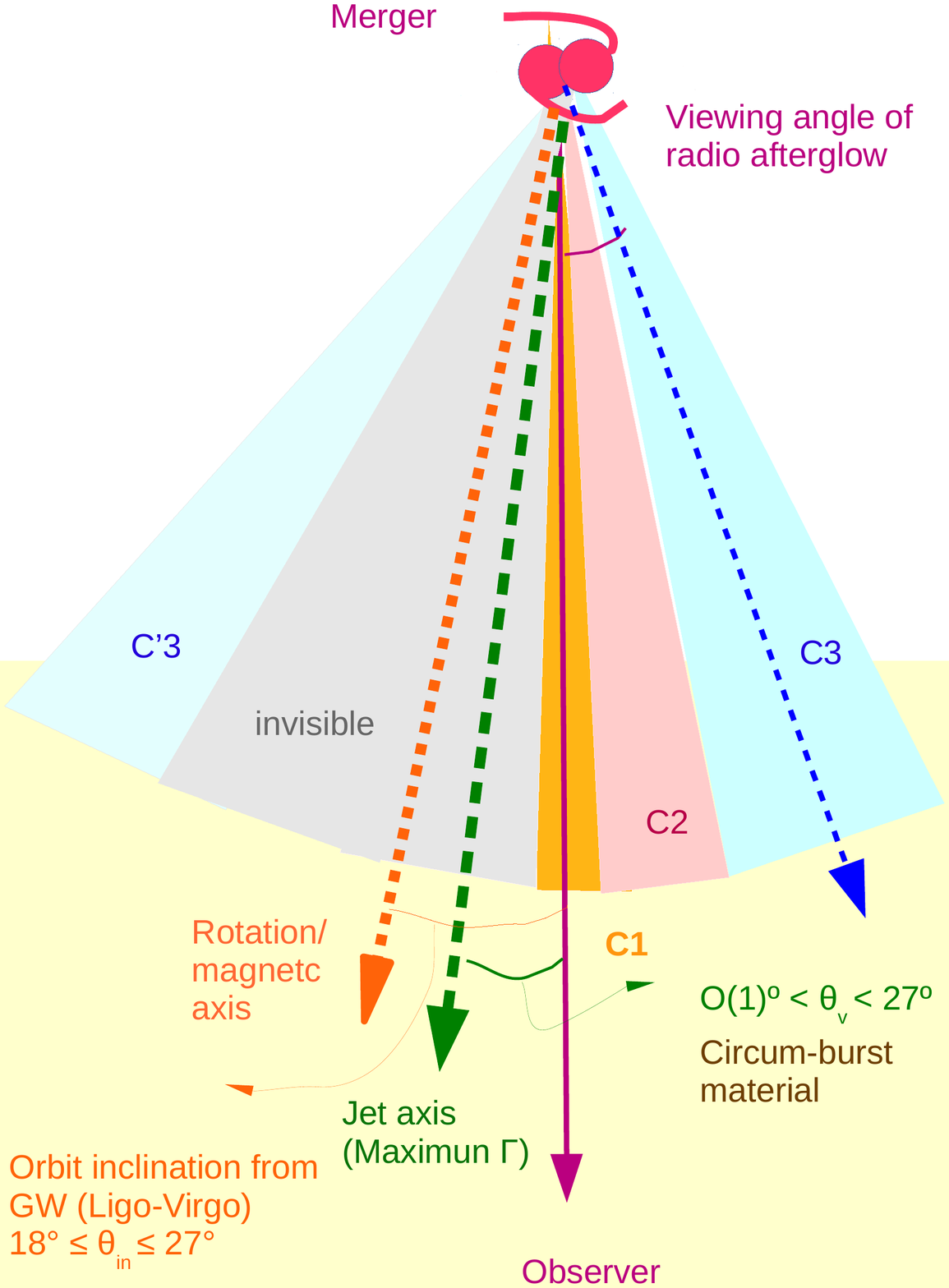}
\caption{Schematic description of polar outflow of merger at the time of its encounter with 
circum-burst material. C1, C2, C3 refer to components of the simulated model. Grey shaded 
region on the opposite side of the jet with respect to observer's line of sight is approximately 
invisible because of its large Lorentz factor and off-axis angle. Nonetheless, C'3 region 
which has even larger off-axis may be visible if its Lorentz factor is sufficiently low. Therefore, 
there can be a contribution to component C3 of the model from emission of this region. 
In any case, due to its large off-axis contribution of C3' would be subdominant. For the 
sake of simplicity here we have assumed that magnetic field direction and rotation axis 
coincide. This may not be true. \label{fig:angles}}
\end{center}
\end{figure}

Fig. \ref{fig:angles} shows a schematic presentation of the structured jet, components of the model, 
and their positions with respect to our line of sight and symmetry axis of the outflow, assumed to 
have the highest Lorentz factor. Simulations of particle acceleration by transfer of Poynting to 
kinetic energy in the polar outflow of mergers show that the direction of maximum acceleration somehow 
deviates from the rotation and magnetic field axes~\citep{grbjetsimul}. Thus, using the estimation 
of orbit inclination, the angle between our line of sight and direction of maximum Lorentz factor 
$\theta_v$ can be constrained as $\mathcal{O}(1)^\circ \lesssim \theta_v \lesssim 27^\circ$, 
independent of jet model. This interval is also consistent with estimations of viewing angle 
by~\citep{gw170817latexary,gw170817latexoptradio1,gw170817lateradiosuprlum}\footnote{In 
Fig. \ref{fig:angles} jet axis and the line of sight are assumed to be in the same projected side 
with respect to the rotation axis. Thus $\theta_v < \theta_{in}$. More generally, considering a cone 
around the line of sight with rotation axis on its surface, any line passing inside of its 
cross-section will have a smaller angle with line of sight than the rotation axis. Therefore, for 
$\theta_v > \theta_{in}$ the jet axis must be outside the cone. In this case the line of sight 
would be even further from jet axis than what is depicted in Fig. \ref{fig:angles}, and closer 
to non-relativistic or mildly-relativistic cocoon. But, as discussed in the Introduction, such setup 
is not able to explain observations.}

The knowledge of orbital inclination is not enough to constrain Lorentz factor profile up to its 
invisible core of the jet. Nonetheless, the sharp reduction of Lorentz factor from C1 which must be 
approximately along the line sight to C3 at $\lesssim 65^\circ$ or $\lesssim 14^\circ$ - depending 
on which C3 model is used - means that the visible part of the jet has an exponential 
$\Gamma \sim \Gamma_{max} \exp(-(\theta + \theta_v)/\theta_\tau)$ or Gaussian 
$\Gamma \sim \Gamma_{max} \exp(-(\theta + \theta_v)^2/2 \sigma^2_\theta)$ profile, where $\theta$ is 
off-axis with respect to the line of sight, and $\theta_v$ is the angle between jet's symmetry axis 
and line of sight. In absence of any knowledge about centroid of components, we must find the range 
of parameters which satisfy constraints on $\theta_i$, and on parameters of the profile, 
namely $\theta_v > 0$, $\sigma^2_\theta > 0$, and $\theta_\tau > 0$, respectively for Gaussian and 
exponential profiles. 

\begin{center}
\begin{figure}
\begin{tabular}{p{8cm}p{8cm}}
\includegraphics[width=8cm,angle=90]{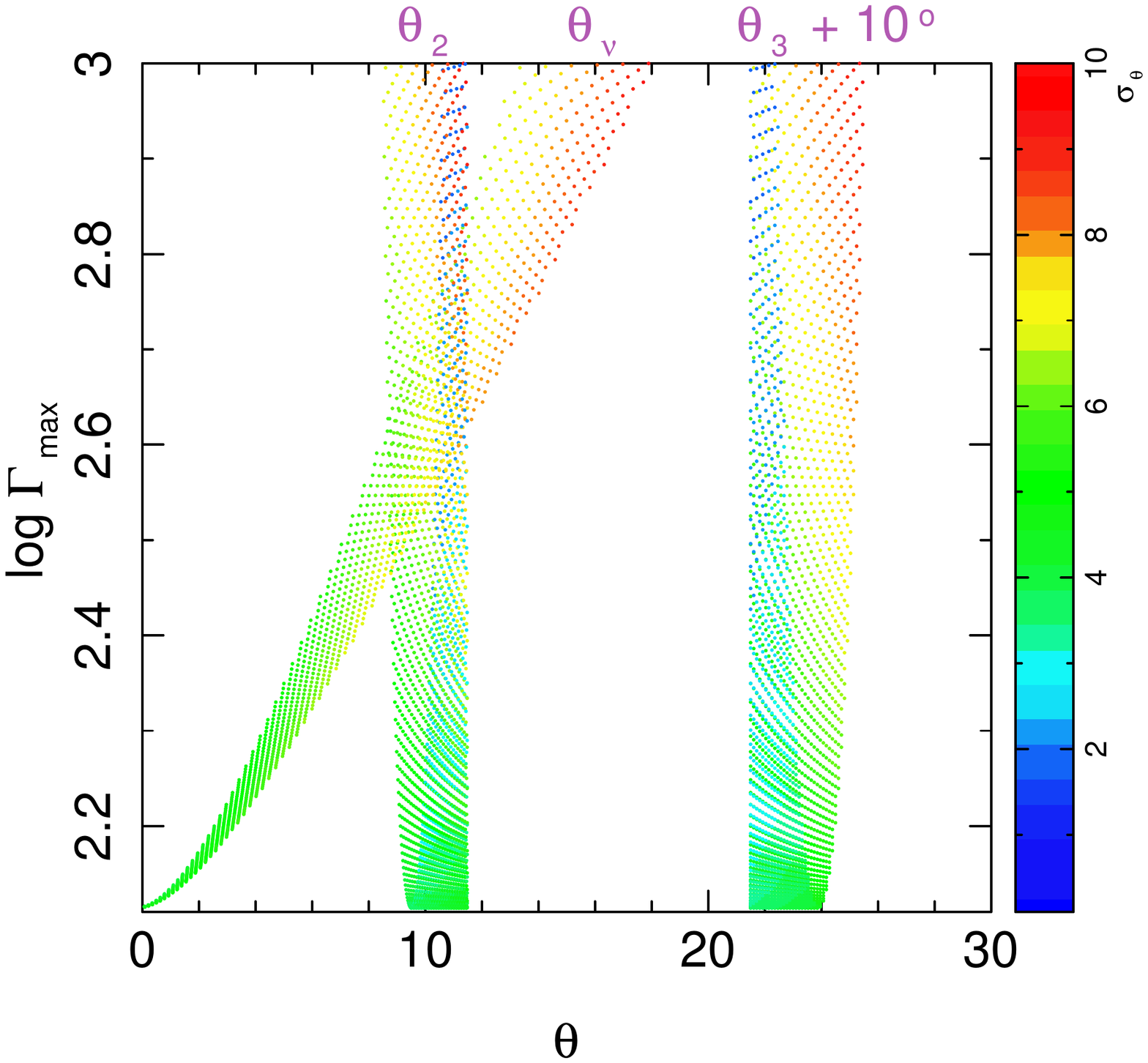} & 
\includegraphics[width=8cm,angle=90]{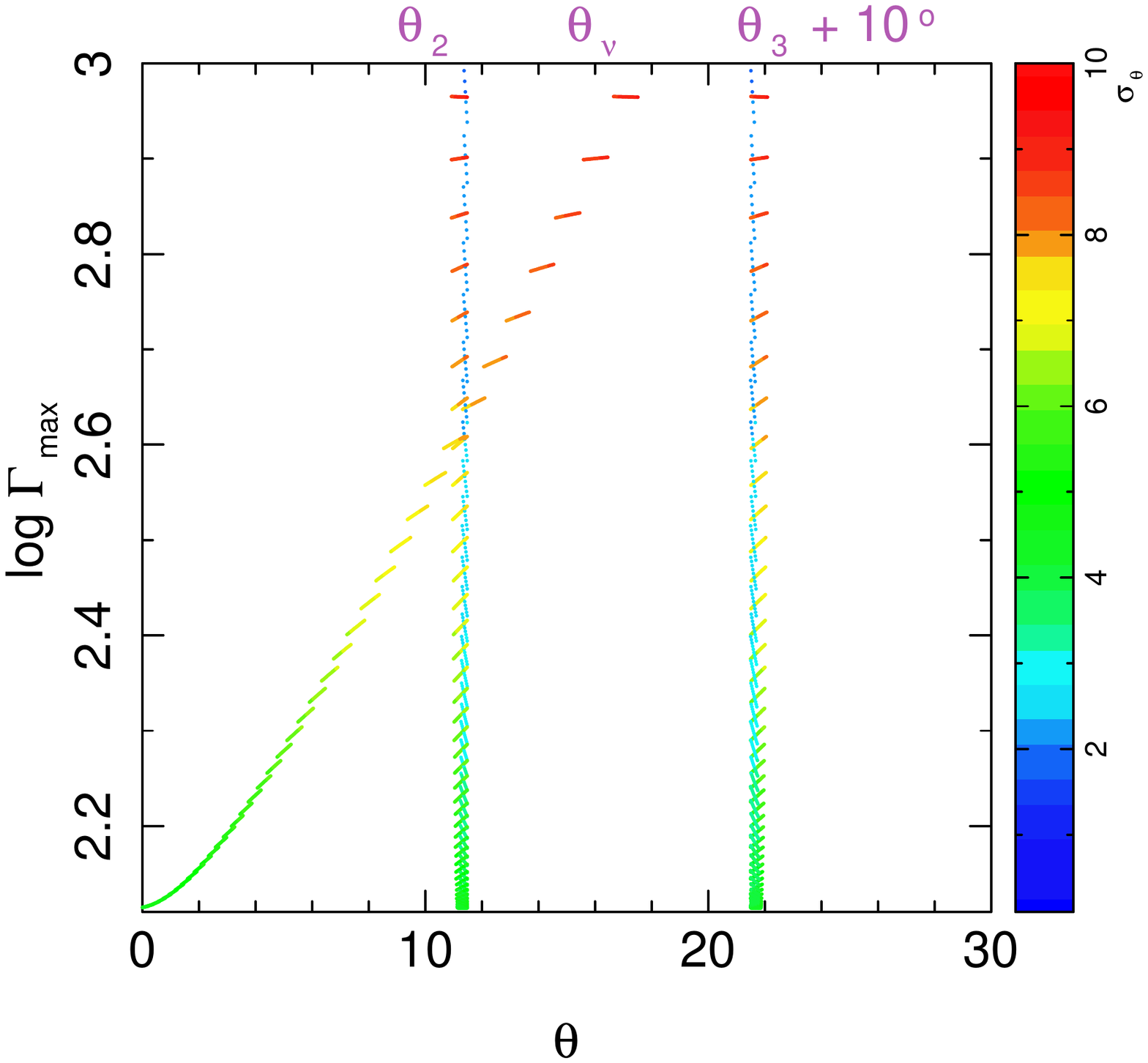} 
\end{tabular}
\caption{Parameter space of Gaussian profiles satisfying $\theta_v > 0$, $\sigma^2_\theta > 0$, and 
having Lorentz factors obtained from simulations at centroid of each component. Left: Model of 
Table \ref{tab:param}; Right: Model of Fig. \ref{fig:paramvarc3}-a. To make these plots 
$\theta_1 = 0.25^\circ$ is assumed and centroid angles of C2 and C3, i.e. $\theta_2$ and $\theta_3$ are 
selected in the range $\theta_{max_1} \leq \theta_2 \leq \theta_{max_2}$ and 
$\theta_{max_2} \leq \theta_3 \leq \theta_{max_3}$  and a solution for $\theta_v$, $\Gamma_{max}$, and 
$\sigma_\theta$ is obtained and added to $\theta_v-\log \Gamma_{max}$ plane only if the 
solution satisfies above constraints (parabolic-like region). Approximately orthogonal regions 
correspond to $\theta_2-\log \Gamma_{max}$ (closer to origin) and $\theta_3-\log \Gamma_{max}$. The 
latter is shifted horizontally by $10^\circ$ for clarity. The color code presents $\sigma_\theta$. 
Horizontal axis is a general $\theta$ variable and labels on the top axis show angle to which each 
track corresponds. \label{fig:anglegamma}}
\end{figure}
\end{center}

Fig. \ref{fig:anglegamma} shows the consistency region for a Gaussian profile for both the model of 
Table \ref{tab:param} and its C3 variant shown in Fig. \ref{fig:paramvarc3}-a. Assuming 
$\theta_1 \sim 0.25^\circ$, for the model in Table \ref{tab:param} we find: 
$\theta_2 \sim 9^\circ-11^\circ$, $\theta_3 \sim 12^\circ-15^\circ$ for $\Gamma_{max} < 1000$. 
Interestingly, this figure shows that the allowed values for viewing angle $\theta_v$ are strongly 
correlated with $\Gamma_{max}$ and are restricted to 
$5^\circ \lesssim \theta_v \lesssim 7^\circ$ for $\Gamma_{max} \sim 250$ and 
$14^\circ \lesssim \theta_v \lesssim 18^\circ$ for $\Gamma_{max} \sim 1000$. For second model the 
range of allowed angles are even more restricted: $\theta_v \sim 7.5^\circ-8^\circ$, for 
$\Gamma_{max} \sim 250$ and $\theta_v \sim 8^\circ - 15^\circ$, for $\Gamma_{max} \sim 1000$, 
$\theta_2 \sim 10.5^\circ$, $\theta_3 \sim 11.5^\circ$, for $\Gamma_{max} < 1000$.  
Considering the estimation of orbit inclination, the range of values for viewing angle obtained 
here confirms simulations of particle accelerations~\citep{grbjetsimul} that show a deviation 
between magnetic field - assumed to be the same as rotation axis - and maximum of acceleration. 
In the case of GW/GRB 170817A the deviation was $\sim [7^\circ-15^\circ]$.

It is straightforward to see that in the case of exponential distribution $\theta_v$ and $\Gamma_{max}$ 
are fully degenerate. This degeneracy induces a relation between Lorentz factor and centroid of 
three components of the model, i.e. $(\theta_2 - \theta_1) / (\theta_3 - \theta_2)) = 
\ln (\Gamma_1 / \Gamma_2) / \ln (\Gamma_2 / \Gamma_3)$. Using this relation and imposing 
$\Gamma_{max} \sim 500$ and $\theta_1 = [0.25^\circ - 0.5^\circ]$, we find 
$\theta_v = [0.06^\circ - 0.3^\circ]$, $\theta_2 = [0.75^\circ - 1.5^\circ]$, 
$\theta_3 = [20.7^\circ - 41.4^\circ]$, $\theta_\tau = [0.06^\circ - 0.3^\circ]$ for the model in 
Table \ref{tab:param}. For the alternative model $\theta_2$ is the same, $\theta_v \sim 0.06^\circ$,
$\theta_3 = [0.87^\circ - 1.74^\circ]$, $\theta_\tau = [0.08^\circ - 0.4^\circ]$. Small off-axis angles 
found for this profile are inconsistent with observed superluminal motion. Even for 
$\Gamma_{max} \sim 1000$ off-axis angles remain too small to be consistent with a superluminal motion. 
Therefore, we rule out an exponential profile.

The 3-component model provides also information about density profile of the jet. Indeed in 
agreement with simulations of jet acceleration to ultra-relativistic 
velocities~\citep{grbjetsimul}, the density in the core of the accelerated jet, where Lorentz 
factor is maximum, is much lower than its surrounding. Moreover, as described earlier, at the time 
of external shocks the jet had been already subjected to partial dissipation. Thus, a large 
fraction of its content was scattered, slowed down, and acquired significant transverse momentum. 
These processes should make side lobes - {\it the cocoon} - slower and denser, and increase gradient 
of variation in the outflow. These expectations are consistent with the model presented here. 
In addition, fraction of kinetic energy of the jet transferred to electric and magnetic fields during 
shocks in outer boundary of the outflow are expected to be smaller. C3 component which is only mildly 
relativistic satisfies these expectations. However, Table \ref{tab:param} shows that the distance of 
this component from the center at the onset of external shocks is $\sim 50\%$ larger than the other 
components. This may have multiple origins: uncertainties of simulations when $\Gamma \gtrsim 1$ 
and/or delay in the formation of shocks due to small velocity of C3 and low density of 
ISM/circum-burst material.

\subsection{Delayed brightening} \label {sec:brighten}
A significant difference between the afterglow of GW/GRB 170817A and other short bursts is its 
seemingly intrinsic faintness at early times, i.e. at $\gtrsim T+10^5$~sec, when X-ray and optical 
follow up began, and its brightening at late times. However, none of these conclusions are certain. 
Of course, as we showed in~\citep{hourigw170817}, prompt emission of this burst was intrinsically 
the faintest among all GRBs with known redshift. Therefore, its early afterglow had to be equally 
faint. Indeed, X-ray flux upper limits of $\sim 10^{-13}$~erg~sec$^{-1}$~cm$^{-2}$ from observations 
of the Swift-XRT at $\gtrsim T+1.6$~days in $0.3-10$~keV~\citep{gw170817swiftnustar} and 
$\sim 10^{-14}$~erg~sec$^{-1}$~cm$^{-2}$ in $0.3-8$~keV from Chandra observations at 
$\sim T+2.2$~days~\citep{gw170817xray} are intrinsically the most stringent among short bursts with 
redshift~\citep{gw170817grbcomp}. However, early X-ray afterglow of both short and long GRBs have 
variety of slopes~\citep{grbshortag,grbshortrev0}, see also simulations in 
Appendix \ref{app:paramvar}. Therefore, in absence of early observations, GW/GRB 170817A cannot be 
classified as an early dark burst.

Late brightening of the afterglow of this burst is not unique either. For instance, a slight 
brightening may have been observed at $\sim T+10$ days for GRB 130603B and its accompanying 
kilonova~\citep{grb130603bxray}. However, the excess found by~\citep{grb130603bxray} was with respect 
to the model fitted to the data rather than in the observed flux. In any case, no other GRB has ever 
followed up for as long as GW/GRB 170817A. Therefore, we do not know how uncommon is the brightening 
of afterglows up to $\gtrsim T+200$ days. Afterglows of GRBs up to few thousands of seconds are most 
probably a superposition of weak internal shocks in what remains from the relativistic jet after the 
main prompt shock, and emission from external shocks~\citep{xrtafterglow}. Because a highly 
relativistic jet follows closely the prompt photons, the delay between their arrival to external 
material in general would be short - of order of few tens of seconds or less for ultra-relativistic 
jets and typical distances to circum-burst material. Assuming a narrow prompt spike, the delay 
between the prompt and onset of afterglow emission is $\Delta t \sim r_e / 2c \Gamma^2$, where $c$ is 
speed of light. For the model of Table \ref{tab:param} the delay is $\sim 4$~sec. For alternative 
models discussed in Appendix \ref{app:paramvar} with shorter distances to central object this delay 
is even shorter. Therefore, the early afterglows, if they were observed, were superimposed on a 
decaying {\it tail} emission from internal shocks and time evolution of total emission depended 
on relative strength of decaying prompt tail and ascending afterglow. The absence of early 
brightening of afterglows in most GRBs means that they are dominated by {\it tail} emission of 
internal shocks. There are however exceptions. For instance, in long 
GRB 110213A~\citep{grb110213axraytail} rapid decay of the prompt X-ray made ascending and peak of 
X-ray afterglow visible. But such cases are rare. More frequent case, specially in short GRBs, 
is the presence of a plateau. Best examples are GRB 070724A~\citep{grb070724a} and 
GRB 070809~\citep{grb070809}, which have an early X-ray light curve with plateau similar to long GRBs.

Our simulations show that in the case of GW/GRB170817A 
the slow rise of the afterglow is due to low density of ISM/circum-burst material and low column 
density of the jet. The weakness of the jet was in part due to intrinsically low density and low 
Lorentz factor of polar ejecta - at least in our direction - and in part the result of large 
distance of surrounding material from center, that is $\sim 1000$~AU (see Table \ref{tab:param}) 
rather than e.g. $\sim 100$~AU for the termination shock in solar system or $\sim 200$~AU for 
recently detected NIR emitting material around the isolated neutron star 
RXJ0806.4-4123~\citep{nstarmatt}. The initial suggestions about off-axis view and brightening of 
afterglows when the jet become dissipated and its content scattered to our line of sight, is not 
consistent with relatively early break of light curves in all three energy bands after 
$\sim T+110$~days. In fact, high Lorentz factor of X-ray emitting component found here and in the 
literature means that it is emitted very close to our line of sight. Therefore, the effect of distance 
and densities may be stronger than off-axis. It is however cautious to consider that this conclusion 
may be somehow biased and correlated to spatial resolution of our simulations, because as 
Table \ref{tab:paramdef} shows, the initial/final width of synchrotron emitting region is defined 
as a fraction of initial distance of the shock from central source. Nonetheless, estimation of shock 
distance for this GRB in the literature and for other GRBs with various methods, such as cooling of 
thermal emission~\citep{grb100316d} show that the range of distances obtained from simulation of 
GRBs~\citep{hourigrbmag} with the code used here are realistic.

\subsubsection{Comparison with standard afterglow model} \label{sec:agstadard}
There is no essential difference between model of~\citep{hourigrb,hourigrbmag} and Blandford-McKee 
approach used by~\citep{emission1}. In both cases the shock is assumed to be radiative, i.e. falling 
material to the shock front (in its rest-frame) loses its kinetic energy as synchrotron radiation and 
makes the jet heavier. The only difference is that~\citep{hourigrb} formulation takes into account the 
back-reaction of this process in a systematic way and also allows evolution of induced fields, which 
cannot be formulated from first principles, in a parametric manner. 

We remind that asymptotic behaviour of external shock emission formulated in~\citep{emission1}, 
which is used in works reviews in Sec. \ref{sec:comp}, considers a uniform spherical ejecta. 
Therefore, conclusions about viewing angle of the jet is based on the value of Lorentz factor and 
beaming of emissions from a relativistic source~\citep{emissionbook}:
\be
\frac {dP_e/d\Omega}{dP_r/d\Omega} = \frac {1}{\Gamma^2 \biggl (1+\beta \cos (\theta)\biggr )}
\ee
where $P_e$ and $P_r$ are emitted and received power in the observer frame, and $\theta$ is the 
angle between the line of light and the boost direction.

Comparing simulated spectrum shown Fig. \ref{fig:spect}-a, which is up to uncertainties fully 
consistent with observations shown in the same plots, with Fig. 1-a \& b 
of~\citep{emission1}\footnote{We remind that spectra shown in Fig. \ref{fig:spect} includes an 
additional energy factor with respect to those in Fig. 1-a \& b of~\citep{emission1}.}, we find that 
the total spectrum is consistent with a {\it fast cooling} model with $\nu_c < \nu_m \sim \nu_x$. 
However, as power of electrons Lorentz factor distribution $p \sim 2$ in all the components models 
of the model, the spectrum can be also classified as {\it slow cooling} with 
$\nu_m < \nu_c \sim \nu_x$. Spectra in Fig. \ref{fig:spect}-a show that for each component the 
dominant emission band is very close to the peak of its spectrum - thus consistent with being 
dominant - and satisfies relations similar to those of the total spectrum given above. 
In particular, similarity of components spectra and their sum demonstrates why many of analysis 
in the literature reviewed in Sec. \ref{sec:comp}, specially those modelling GRB 170817A with a 
single relativistic component, find that mission in radio, optical/IR, and X-ray bands belong to 
a same power-law spectrum with a negative slope.

Synchrotron cooling frequency $\nu_c$ depends on the induced magnetic field, which in turn depends 
on the density of ISM; larger the density of accumulated (fallen) material, larger magnetic field 
and higher synchrotron characteristic frequency. It can be shown, using equations (4) and (6) 
of~\citep{emission1}, that $\nu_c \propto n^{-1} t^{-3/2}$, where $n$ is density of accelerated 
charges. Therefore, for a given $\nu_c$ lower density means slower emission variation and longer 
duration of emission because of smaller induced magnetic field. However, in~\citep{emission1} 
formulation the effect of jet column density is not explicit. We remind that the formation of a 
shock needs a significant density difference in the opposite sides of a boundary. Therefore, column 
density and density of ISM/circum-burst material must be considered together. Slow evolution of 
emissions needs relatively low ISM/circumburst material and/or low density jet. The length of the 
jet/ejecta is also important. For the same column density, a longer jet is more diluted and its 
collision with circum-burst material generates less turbulence and less energy is transferred to 
induced fields. Moreover, if densities of colliding shells are low, the faster shell must sweep a 
longer distance inside the slow shell to accumulate material and form a shock front. In model 
C1 in Table \ref{tab:param} ISM/circum-burst material density is in the middle of the range 
estimated for short bursts~\citep{grbshortrev0}. However, column density of jet is low. In the 
variant model in Fig. \ref{fig:paramvarc1}-b, which is also a good fit to data, ISM/circum-burst 
material density is lower, but a longer active region is assumed. Thus, for the ultra-relativistic 
component C1, the low density of colliding material may be the reason for long distance of external 
shocks from central source. By contrast, other components, which had larger column densities, 
have been radially more extended.

The total spectrum of the model in Fig. \ref{fig:spect}-a does not show how $\sim \nu_c$ and $\nu_m$ 
change with time. Nonetheless, evolution of spectrum discussed in Sec. \ref{sec:ag} show that, 
as expected, after the peak of emission $\sim \nu_c$ and $\nu_m$ of each component become smaller 
than what was their values before the peak. 

In summary, we conclude that the early X-ray afterglow of GW/GRB 170817A at $\lesssim T+10^5$~sec 
had to be faint and dominated by weak internal shocks. It declined quickly. A consequence of this 
conclusion, which unfortunately due to the lack of data cannot be verified, is that the claimed 
excess of UV emission at 
$\sim T+1.6$~days~\citep{gw170817bluekilonova,gw170817bluekilonovapol,gw170817bluekilonovamod} was 
indeed from kilonova rather than the afterglow of GRB 170817A. The origin of excess may be heating 
of a strong wind by neutrinos and strong magnetic field of the short lived 
HMNS~\citep{gw170817kilonovamassexc}

\subsubsection {Slope of afterglow rise} \label{sec:riseslope}
We notice that after initial fast rise the slope of light curves shown in Figs. \ref{fig:lc}, 
\ref{fig:paramvarc1}, \ref{fig:paramvarc2}, \ref{fig:paramvarc3} in the region far from the peak is 
$\sim 1 \pm 0.3$ and does not vary strongly\footnote{We thank the anonymous referee for bringing 
this point to our attention.}. To understand this behaviour we first remind that the asymptotic 
formulation of standard afterglow model by~\citep{emission1} predicts a constant rising slope of 
$\sim 0.5$ for energy bands close but on the low energy wing with respect to the spectral peak. 
As we discussed in Sec. \ref{sec:agstadard}, in the multi-component models of Table \ref{tab:param} 
and Appendix \ref{app:paramvar} the dominant emission band is close to the spectral peak. Therefore, 
having a roughly similar rising slope in these models is consistent with the asymptotic afterglow 
model. However, the latter underestimate the slope. 

One reason for difference between predictions of the two models can be the fact 
that~\citep{emission1} formulation depends only on the density of accelerated particles in the jet, 
which generate the shocked induced electromagnetic fields. The density of circum-burst material 
$N'$ on which the jet is shocked does not explicitly appear in the formulation. To see the importance 
of $N'$ we notice that in the C1 alternative model of Fig. \ref{fig:paramvarc1}-a, in which only 
$N'$ is less than that of C1 in Table \ref{tab:param}, the rise slope of the X-ray light curve is 
$\sim 0.7$ and the peak flux is smaller. As explained in Sec. \ref{sec:agstadard}, time/radius 
dependence of quantities which determine strength of the shock, such as ISM/circum-burst density 
and fraction of kinetic energy transferred to electrons and to induced magnetic field - parametrized 
by $\kappa$, $\alpha_e$ and $\alpha_B$, respectively - influence the slope of emission rise and 
the observed differences with the standard model. However, they are not the only parameters which 
influence the rise of emission. For instance, in the main C1 model the slope after initial fast rise 
is $\sim 1$. But it  increases gradually up to the peak of the X-ray light curve. 

A quantitative explanation of the rise slope in the framework of the model 
of~\citep{hourigrb,hourigrbmag} needs an analytical solution of the model, which due to its 
complexity is not available. Nonetheless, we can use some of analytical results obtained 
in~\citep{hourigrb,hourigrbmag} to understand which physical properties are involved in the rise 
of light curves and its dependence on the parameters of the model.

The phenomenological model of~\citep{hourigrb,hourigrbmag} divides the process of relativistic shocks 
and synchrotron/self-Compton to two parts: 1) kinematic of the ejecta; 2) synchrotron/self-Compton 
emission. It calculates the evolution of kinematic by taking into account the total energy 
dissipated as synchrotron/self-Compton and obtain perturbative solutions for Lorentz factor 
$\Gamma (r)$ and column density $n_c(r) = n(r) \Delta r$, where $r$ is an average distance of a 
narrow relativistic ejecta from center and $\Delta r$ is the width of shocked region - 
{\it the active region}. Under this approximation time $t$ and $r$ are not independent 
and $r (t) - r(t_0) = c \int_{t_0}^t \beta (t') dt'$. The results of this step is used to 
calculate synchrotron/self-Compton emission, from well known textbook formulation of these processes, 
with the difference that it takes into account variation of electron and baryon densities in the 
jet/fast shell, their varying Lorentz factor, and possible radial density inhomogeneities of the slow 
shell/ISM. Time/radius dependence of these quantities connect kinematical and dynamical parts of 
the model. Additionally, in the case of internal shocks a total boost of colliding shells must be 
take into account. However, this phenomenological formulation is unable to determine how the 
width of active region $\Delta r$ changes with time/radius. Nonetheless considering formation and 
decay of a shock front, its initial width of active region should be zero. It should grow gradually 
to a maximum and decline to zero at the end of shocks. A series of phenomenological models 
presenting such evolution is defined in Appendix \ref{app:def}. 

Under these approximations the synchrotron energy flux can be written as:
\be
dP/(\omega d\omega) ~ \propto ~ r^2\Delta r / \Gamma (r) \int_{\gamma_m}^\infty d\gamma_e n'_e (\gamma_e) 
f (\gamma_e, r) \label{synchflux}
\ee
where $n'_e$ is the density of electrons and $f$ is a function of $r$ and electrons Lorentz factor 
$\gamma_e$, see e.g. equation (62) in~\citep{hourigrb} for full expression. The $r^2$ factor changes 
densities to total emission from volume of the active region. It reflects the fact that for the 
same density, column density, and jet opening angle, the total amount of emitting material is 
larger if the average radius of the shell is larger. Therefore, in addition to quantities related 
to micro-physics of the ejecta and environment, geometrical setup, namely the initial distance of 
external shocks $r_0$, extension of the ejecta and evolution of the thickness of active region 
$\Delta r$ are crucial for the rise and peak amplitude of emissions, see equation 30 and 
Appendix A in~\citep{hourigrb} for technical details. In particular, for the same densities and 
other parameters, longer distance to center leads to longer rise time and higher and later peak 
emission.

The term $n'_e \Delta r$ in (\ref{synchflux}) can be considered as column density of electrons with 
Lorentz factor $\gamma_e$. Normalization factor of the distribution of electrons Lorentz factor is 
proportional to density of active regions $n'= n'_c / \Delta r$ and inversely proportional to minimum 
Lorentz factor of electrons $\gamma_m \propto \epsilon_e \propto (r/r_0)^{\alpha_e}$. However, due to 
non-linear term $f (\gamma_e, r)$ factors of $\Delta r$ in (\ref{synchflux}) do not cancel each others. 
Evolution of $\Gamma (r)$ is also dependent on the phenomenological exponent of $\delta$, $\alpha_e$, 
$\alpha_B$ and $\kappa$, see equation (19) in~\citep{hourigrbmag}. As the values of these indices are 
the same in all the models discussed here - except the model shown 
in Fig. \ref{fig:paramvarc1}-e in which $\kappa = 0$ - it is not a surprise that light curves have 
roughly the same rise slope. Nonetheless, due to the nonlinearity of evolution equations the slope 
in not exactly the same in all the models. In addition, as mentioned earlier, it changes with time. 
This fact reflects the influence of complicated dynamics, which cannot be 
estimated by a simple asymptotic power-law. For instance, the model in Fig. \ref{fig:paramvarc1}-e 
(dotted line) has a faster rise than C1 in Table \ref{tab:param} because $\kappa = 0$, meaning that 
the density of circum-burst material in this model does not decline at large distances from the 
merger. By contrast, in the model of Fig. \ref{fig:paramvarc1}-a, which has the same $\kappa$ as C1, 
the rise of flux is slower because the density of circum-burst material is much smaller than in C1. 
This shows how initial conditions and constant quantities in the model influence the dynamics of 
the emission.

\subsection{Material surrounding BNS}
In the phenomenological model of~\citep{hourigrb,hourigrbmag} the density of circum-burst material 
and its variation with distance are defined by parameters $n'$ and $\kappa$, respectively. 
Table \ref{tab:param} shows that $\kappa \neq 0$ and $n'$ is not the same for all components of 
the model. This is probably an evidence that circum-burst material was not only the ISM, which a 
priori should be independent of the merger and approximately uniform. Thus, additional material should 
have been present.

The origin of circum-burst material and its properties can be traced back to the evolution of progenitor 
neutron stars. In young neutron stars and pulsars the distance to wind Termination Shock (TS) is 
$R_{TS} = \sqrt {\dot {E} / (4\pi \eta P)}$ where $\dot {E}$ is the change in the rotational kinetic 
energy and $P$ is average pressure in the wind nebula surrounding the neutron star~\citep{nssheath}. 
After initial expansion of nebula and establishment of an approximately steady state condition 
pressure inside the nebula is balanced with the ISM pressure and extension of pulsar nebula is 
stabilized to $R_{TS} \sim \mathcal {O}(0.1)$~pc~\citep{nssheath}. However, in old neutron stars 
reduction of glitching activities and dissipation of magnetic field gradually decreases 
$\dot {E}$ and may reduce $R_{TS}$. Recent observation of thermal material at a relatively short 
distance of $\sim 200$~AU$~\sim 3 \times 10^{15}$~cm around the isolated neutron star RXJ0806.4-4123 
with an age of $\sim 10$~Myr~\citep{nstarmatt} is an example of such cases. On the other hand, 
during BNS formation and merger if neutron stars were initially at a distance $\gtrsim R_{TS}$ 
and if there were still a remnant of their wind nebula around them, its content as well as any 
other material would be disrupted. Moreover, during early stages of inspiral glitching activities 
and mass ejection might have been resumed due to perturbation of neutron stars crust and might have 
replenished the nebula. These processes can explain the putative additional surrounding material and 
its anisotropic distribution according to the model presented here. 

To estimate column density of this material we can use distribution used in the model i.e. 
$N'(r) = N'(r_0) (r/r_0)^-\kappa$. For model C1 this estimation gives a column density of 
$\sim 4 \times 10^{14}$~cm$^{-2}$, i.e. much smaller than column density of C1 which has the smallest 
column density among components of the model. This is smaller than swept material in the first 
$\sim 3 \times 10^5$~sec after the onset of the external shocks and is completely negligible. See also 
Fig. \ref{fig:paramvarc1}-d \& e for simulation of shocks on the ISM/circum-burst material with the 
same density as C1 but at low lower distances from center . On the other hand, if we consider much 
denser circum-burst material at shorter distances, much higher X-ray flux generated by the shocks 
violates upper limits at $\sim T+2$~days, see Fig. \ref{fig:paramvarc1}-f for an example with 
$N' = 4$~cm$^{-3}$, i.e. 2 orders of magnitude larger than C1 in Table \ref{tab:param}, in which the 
early X-ray flux is by about 2 orders of magnitude larger than upper limits. In such a model the column 
density of material between sites of internal and external shocks would be 
$\sim 4 \times 10^{16}$~cm$^{-2}$. Using these simulations and estimations, we conclude that 
the column density of material inside $r_e \sim 10^{16}$ was 
$< \mathcal{O} (1)\times 10^{15}$~cm$^{-2}$ or equivalently its average density was $< 0.4$~cm$^{-3}$.

In conclusion, diversity of short GRB afterglows reflect age, history, and environmental 
differences of progenitor neutron stars. In the case of GW/GRB 170817A we notice that $N'$ is 
much smaller than typical values predicted for young neutron stars/pulsars~\citep{nssheath}. 
This means that the progenitor stars were most probably old and had lost most of material they had 
ejected during their youth. This conclusion is consistent with analysis of the prompt gamma-ray of 
this transient~\citep{hourigw170817}. The relatively large distance of circum-burst material may 
have several reasons. But with available information about progenitors it is not possible to pin 
down the dominant cause.

\section{Outline} \label{sec:outline}
In this work we used the same  phenomenological formulation which had been used to model 
the prompt gamma-ray emission of GW/GRB 170817A to analyze its afterglows. We found 
a 3-component model presenting a structured relativistic jet which its collision with circum-burst 
material at a distance of $\sim 1000$~AU generated observed afterglows. It reproduces 
radio and X-ray light curves and photometric spectrum in these bands. Its optical emission 
is consistent with the dominance of the kilonova emission in this band up to 
$\gtrsim 200-300$ days after the merger event, where this range covers degenerate models 
studied here.

Additionally, this analysis helps understand physical conditions around the progenitor BNS 
before and after their merger. In particular, it shows that a small fraction of the prompt 
ultra-relativistic jet had survived internal shocks, and despite dissipation and lateral expansion 
it had preserved in some extent its internal coherence up to a distance of $\sim 10^{16}$~cm from 
central source, where it collided with surrounding material. Another result of the model studied here 
is that despite oblique view of the jet, our line of sight passed through an ultra-relativistic region. 
This conclusion is consistent with works in the literature which fit observations up to $200$ days 
post merger and beyond. It seems that an ultra-relativistic component is indispensable for a 
proper fit of the whole data. Other components of the model, which their emissions are dominantly 
in low energies, are interpreted as approximately presenting side lobes/cocoon of a structured 
jet at the time of external shocks. 

The line of sight angle is model dependent and correlated to the jet profile and its maximum - 
on-axis - Lorentz factor. In the models studied here it can be as small as $\sim 5^\circ$ for a 
maximum Lorentz factor of $\sim 250$ or as large as $\sim 18^\circ$ if on-axis Lorentz factor 
approached $1000$. Considering observed orbital inclination of the BNS and superluminal motion of 
the radio counterpart, and marginalizing on the models and their parameters, the line of sight 
angle can be estimated as $5^\circ - 18^\circ$. 

We identified low density of circum-burst material from merger, its low density and low density of 
the jet as factors leading to slow rise of afterglows. Relatively long distance of the circum-burst 
materials from center might have been in part the reason for extended expansion of the jet up to 
longer distances than in typical short bursts and its dilation, which in addition to slow rise of 
afterglows, pushed the peak of emission to $\gtrsim 110$~days, rather than a couple of days seen in 
short GRBs with a X-ray plateau. Unfortunately, in absence of long duration follow up of short GRBs 
we are not able to access whether late shocks and brightening is an exception in this burst or a 
common behaviour of many short GRBs, i.e. those for which an early X-ray plateau is not observed.

\paragraph*{Acknowledgment:}
The author thanks Hans Krimm for providing the unpublished upper limit of the Swift-BAT on the 
late gamma-ray emission from GW/GRB 170817, and Amy Lien and Scott Barthelmy from Swift Science 
Team for their help to acquire this data.

\appendix
\section{Definition of parameters and evolution of active region} \label{app:def}
Note that in the formulation model in~\citep{hourigrb,hourigrbmag} $\Gamma$ is the Lorentz 
factor slow shell with respect to a far observer and $\gamma'_0$ indicates the relative 
Lorentz factor fast and slow shells. In external shocks on the ISM or circumburst material, 
their velocity is negligible and $\Gamma = 1$. For this reason, here we have replaced 
$\gamma'_0$ with $Gamma$ which is usually used in the literature and indicates the Lorentz 
factor of outflow at the beginning of external shocks.
\begin{table}
\begin{center}
\caption{Parameters of the phenomenological relativistic shock model \label{tab:paramdef}}
\vspace{0.5cm}
\begin{tabular}{| p{2.5cm} | p{12cm} |}
\hline
Model (mod.) & Model for evolution of active region with distance from central engine. \\
$r_0$ (cm) & Initial distance of shock front from central engine. \\
$\Delta r_0$ & Initial (or final, depending on the model) thickness of active region. \\
$p$ & Slope of power-law spectrum for accelerated electrons; See eq. (3.8) of~\citep{hourigrbmag}. \\
$p_1,~p_2$ & Slopes of double power-law spectrum for accelerated electrons; See eq. (3.14) 
of~\citep{hourigrbmag}. \\
$\gamma_{cut}$ & Cut-off Lorentz factor in power-law with exponential cutoff spectrum for 
accelerated electrons; See eq. (3.11) of~\citep{hourigrbmag}. \\
$\Gamma$ & Lorentz factor of jet with respect to far observer.\\
$\delta$ & Index in the model defined in eq. (3.29) of~\citep{hourigrbmag}. \\
$Y_e$ & Electron yield defined as the ratio of electron (or proton) number density to baryon number 
density. \\
$\epsilon_e$ & Fraction of the kinetic energy of falling baryons of fast shell transferred to 
leptons in the slow shell (defined in the slow shell frame). \\
$\alpha_e$ & Power index of $\epsilon_e$ as a function of $r$. \\
$\epsilon_B$ & Fraction of baryons kinetic energy transferred to induced magnetic field in the 
active region. \\
$\alpha_B$ & Power index of $\epsilon_B$ as a function of $r$. \\
$N'$ & Baryon number density of slow shell. \\
$\kappa$ & Power-law index for N' dependence on $r'$. \\
$n'_c$ & Column density of fast shell at $r'_0$. \\
\hline
\end{tabular}
\end{center}
{\small
\begin{description}
\item{$\star$} The phenomenological model discussed in~\citep{hourigrb} and its 
simulation~\citep{hourigrbmag} depends only on the combination $Y_e\epsilon_e$. For this reason 
only the value of this combination is given for simulations.
\item{$\star$} The model neglects variation of physical properties along the jet or active region. 
They only depend on the average distance from center $r$, that is $r-r_0 \propto t-t_0$.
\item{$\star$} Quantities with prime are defined with respect to rest frame of slow shell, and 
without prime with respect to central object, which is assumed to be at rest with respect to 
a far observer. Power indices do not follow this rule.
\end{description}
}
\end{table}

As explained in Sec. \ref{sec:riseslope} the evolution of $\Delta r'(r')$ cannot be 
determined from first principles. For this reason we consider the following phenomenological 
models:
\bea
&& \Delta r' = \Delta r'_0 \biggl (\frac {\gamma'_0 \beta'}{\beta'_0 \gamma'} 
\biggr )^{\tau}\Theta (r'-r'_0) \quad \text {dynamical model, Model = 0} \label {drdyn} \\
&& \Delta r' = \Delta r'_{\infty} \bigg [1-\biggl (\frac{r'}{r'_0} \biggr )^
{-\delta}\biggr ] \Theta (r'-r'_0) \quad \text {Steady state model, Model = 1} \label {drquasi} \\
&& \Delta r' = \Delta r'_0 \biggl (\frac{r'}{r'_0} \biggr )^{-\delta} 
\Theta (r'-r'_0) \quad \text {Power-law model, Model = 2} \label {drquasiend} \\
&& \Delta r' = \Delta r_{\infty} \bigg [1- \exp (- \frac{\delta(r'-r'_0)}{r'_0}) \biggr ] 
\Theta (r-r'_0) \quad \text {Exponential model, Model = 3} \label {expon} \\
&& \Delta r' = \Delta r'_0 \exp \biggl (-\delta\frac{r'}{r'_0} \biggr )
\Theta (r'-r'_0) \quad \text {Exponential decay model, Model = 4} \label {expodecay}
\eea
The column called Model in Table \ref{tab:paramdef} refers to these evolution models for 
$\Delta r'(r')$. The initial width $\Delta r'(r'_0)$ in Model = 1 \& 3 is zero. Therefore, they are 
suitable for description of initial formation of an active region in internal or external shocks. 
Other models are suitable for describing more moderate growth or decline of the active region. 
In Table \ref{tab:param} the column $mod.$ indicates which evolution rule is used in a simulation 
regime - as defined in the foot notes of this table - using model number given in 
\ref{drdyn}-\ref{expodecay}.

\section{Alternative models for the structured jet of GW/GRB 170817A} \label{app:paramvar}
As mentioned in Sec. \ref{sec:ag} the large number of parameters in the phenomenological formulation 
used here to model afterglows of GW/GRB 170817A does not allow to perform a systematic 
search in the parameters space to optimize selected models. Moreover, based on physical arguments 
there must be correlation between parameters. Unfortunately in absence of a first-principle formalism 
they cannot be easily removed and induce degeneracies in the model.

To investigate how degeneracies affect the model of afterglows presented here and whether they 
can alter our conclusions, Figs. \ref{fig:paramvarc1}, \ref{fig:paramvarc2}, \ref{fig:paramvarc3} 
show some variants of the model presented in Table \ref {tab:param}. In most of these simulations 
only distance of shock from merger, Lorentz factor, column density of the jet, thickness of 
shocked (active) region, and density of ISM/circumburst material, are changed and other parameters 
are kept the same as ones shown in Table \ref {tab:param}. In some models index of energy 
distribution of electrons $p$ and fraction of energy transferred to electric field $\epsilon_e$ are 
also changed. 

Although some of these models fit the data as good as the model of Table \ref {tab:param}, they have 
some issues which we will discuss case by case. For instance, a variant of C1 with larger 
$\Delta r_0/r_0$ and smaller ISM/circum-burst density presented in Fig. \ref{fig:paramvarc1}-b is 
degenerate with C1 in Table \ref {tab:param}, but future observations of X-ray afterglow can 
distinguish between them. A variant of C2 with larger $\Delta r_0/r_0$ and smaller $\epsilon_b$ shown 
in Fig. \ref{fig:paramvarc2}-d has a slightly better fit to the data. But it is not exactly a 
replacement of C2 and can be considered as a finer division of the structured jet to components. 

In the following subsections we summarize properties of variant models separately for each component.

\subsection{Variants of C1} \label{sec:c1var}
\begin{center}
\begin{figure}
\begin{tabular}{p{6cm}p{6cm}p{6cm}}
a) & \hspace{-1cm} b) & \hspace{-2cm} c) \\
\includegraphics[width=7cm]{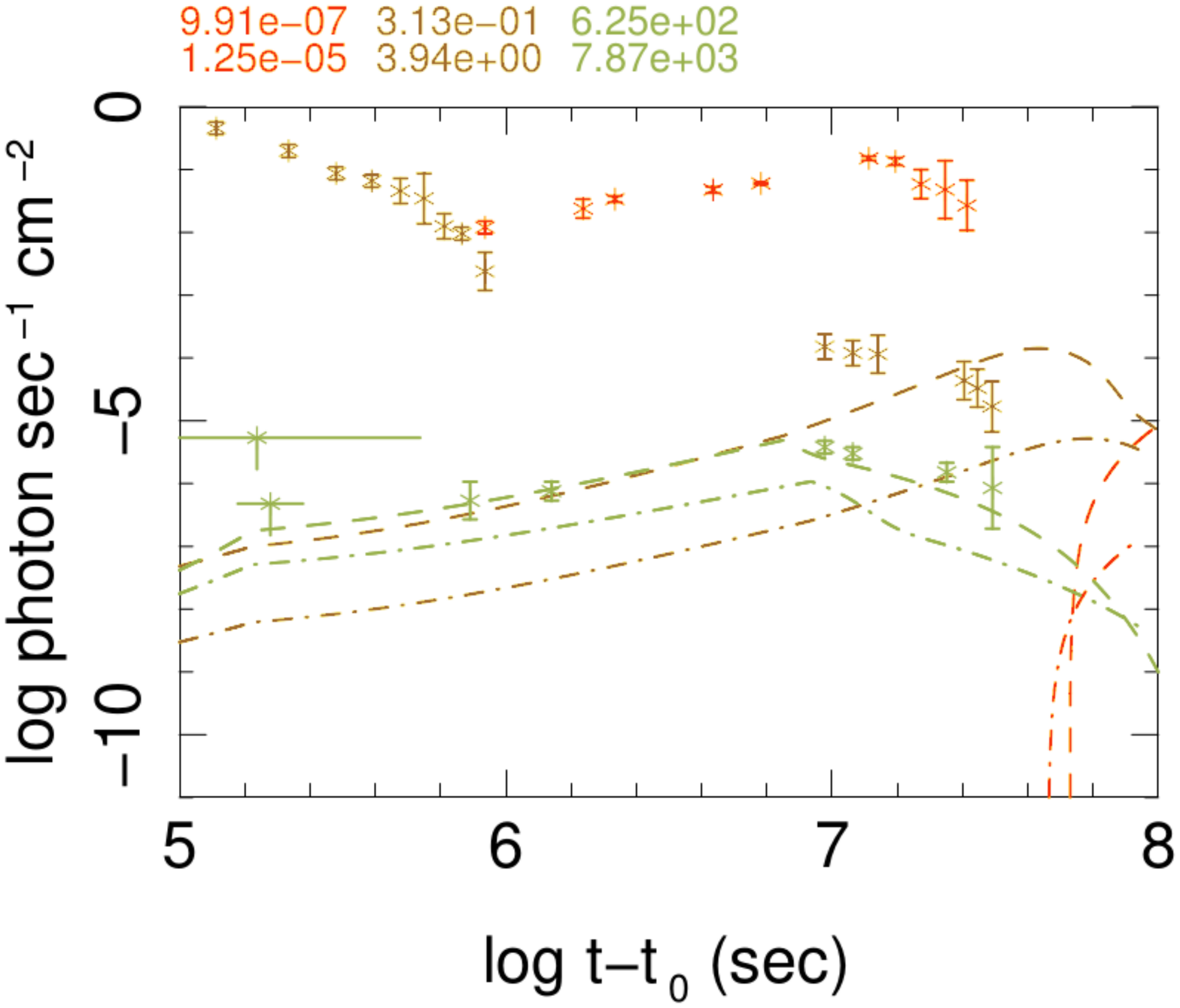} & 
\hspace{-1cm}\includegraphics[width=7cm]{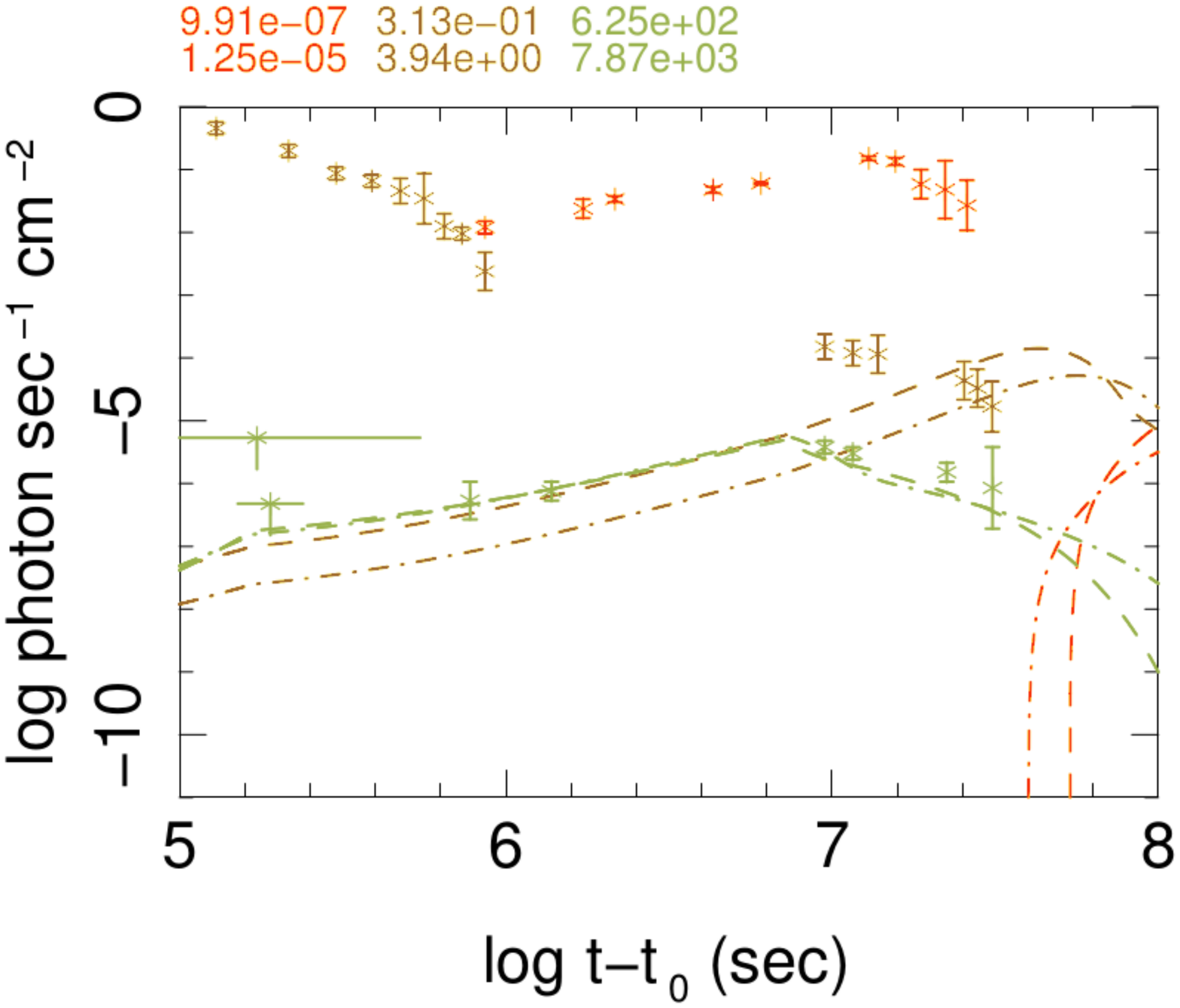} & 
\hspace{-2cm}\includegraphics[width=7cm]{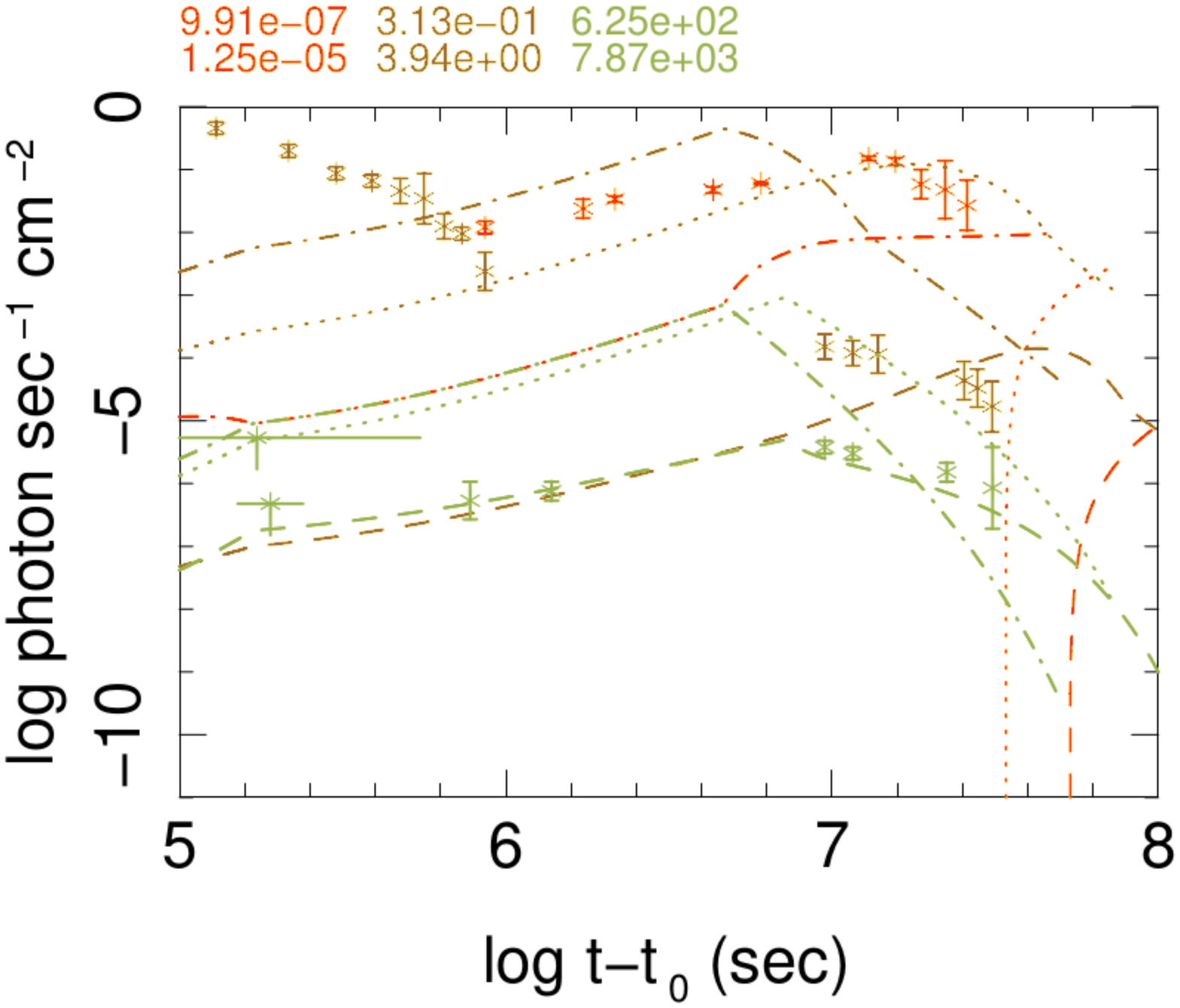} \\
d) & \hspace{-1cm} e) & \hspace{-2cm} f) \\
\includegraphics[width=7cm]{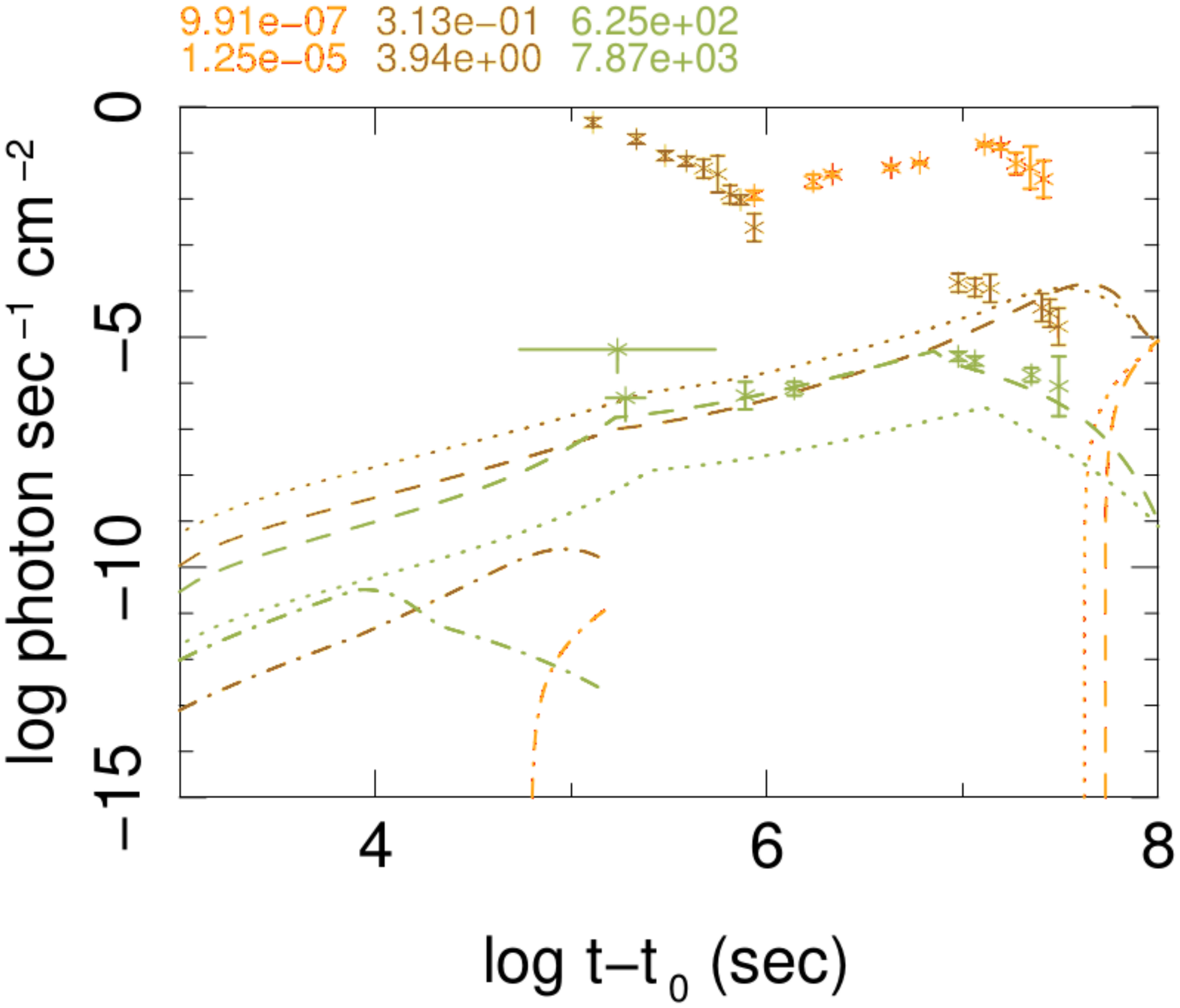} & 
\hspace{-1cm}\includegraphics[width=7cm]{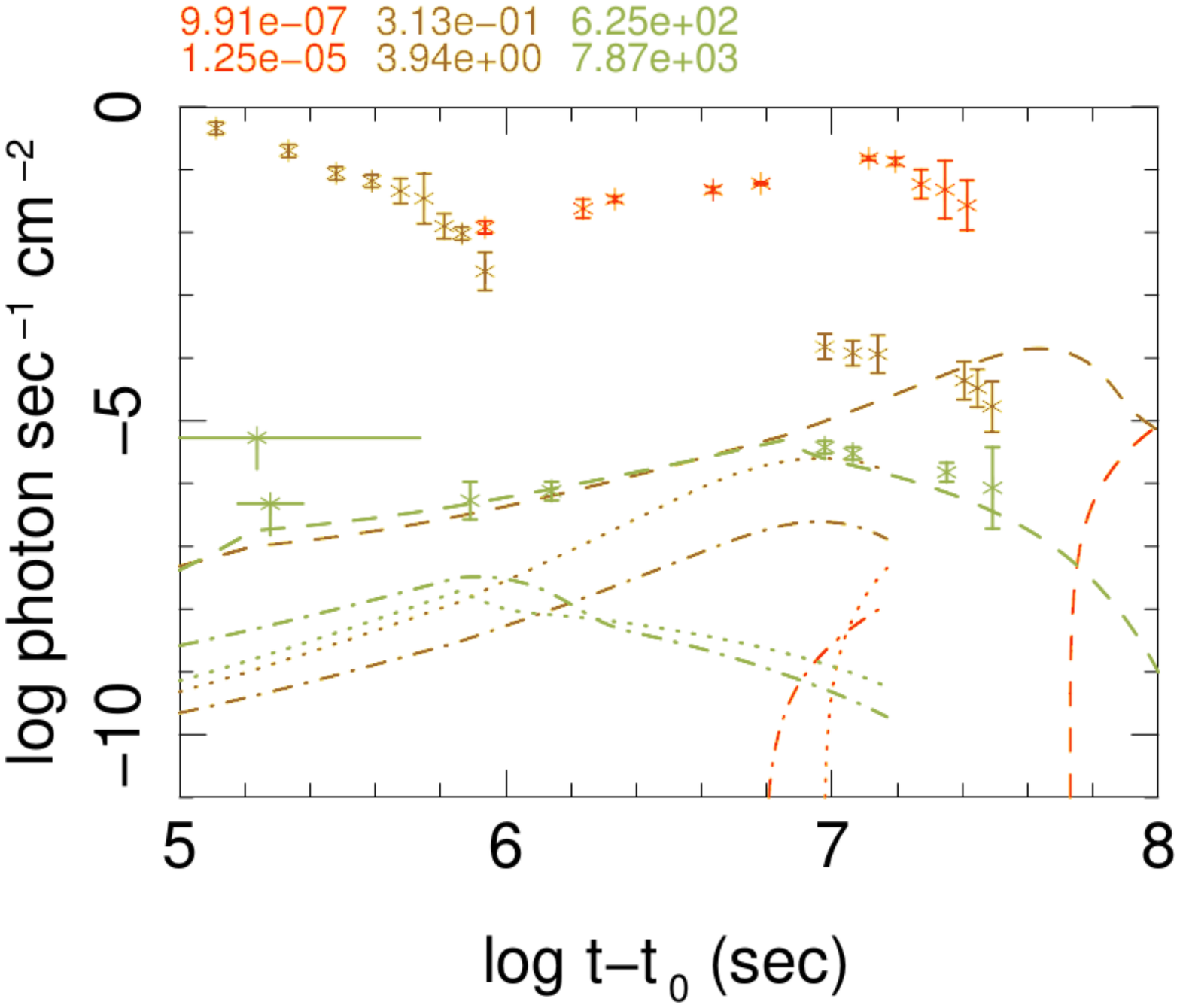} & 
\hspace{-2cm}\includegraphics[width=7cm]{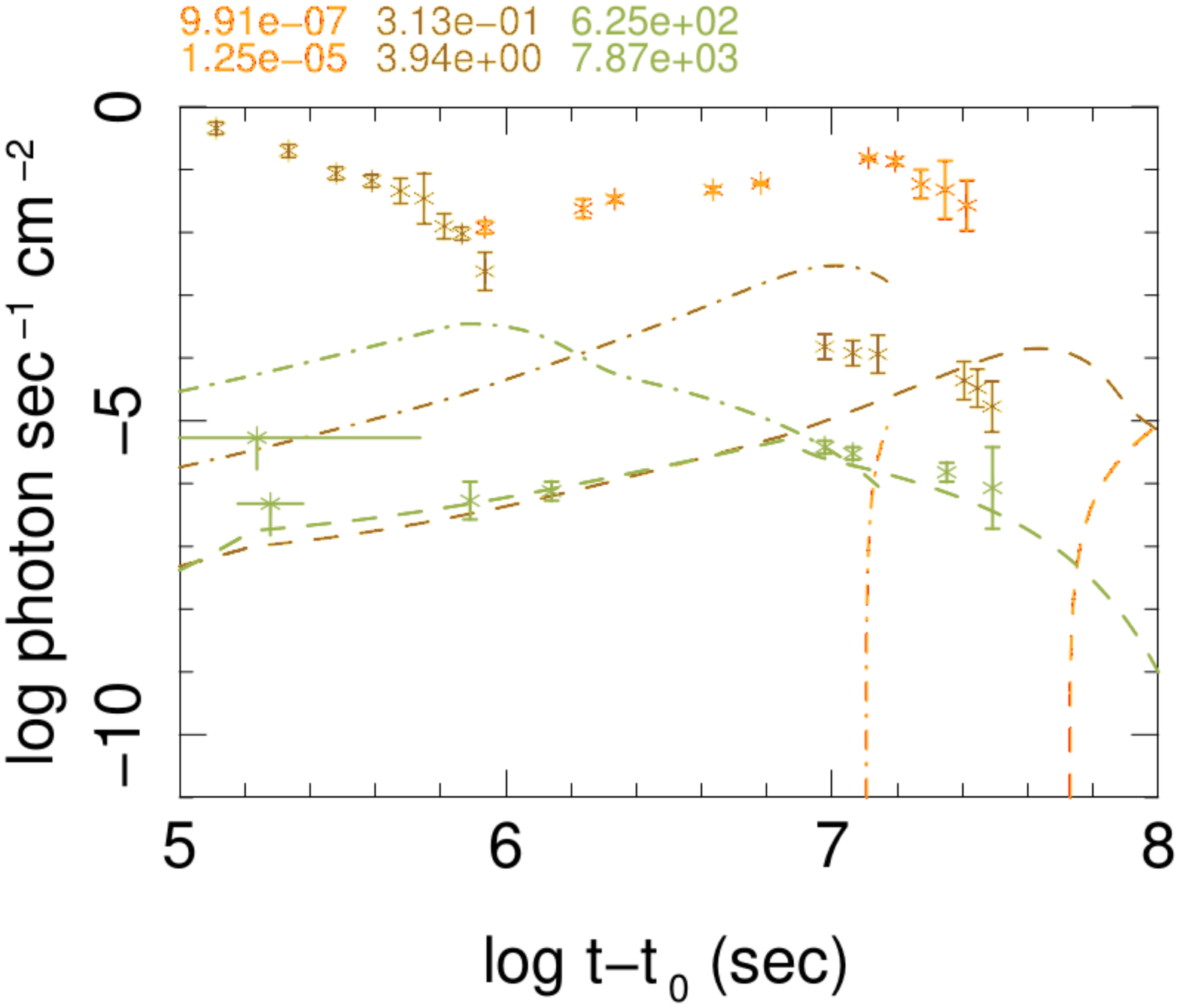}
\end{tabular}
\caption{Models of component C1 with varied parameters: 
a) $N' = 0.008$ (cm$^{-3}$ (dash-dot); 
b) $\Delta r_0/r_0 = 10^{-6}$, $N' = 0.004$~cm$^{-3}$ (dash-dot); 
c) $\gamma'_0 = \Gamma = 8$; $\Delta r_0/r_0 = 5 \times 10^{-3}$, $p = 2.1$, $N' = 0.03$~cm$^{-3}$, 
$n'_c = 5 \times 10^{23}$~cm$^{-2}$ (dash-dot), $\gamma'_0 = \Gamma = 80$; $\Delta r_0/r_0 = 3 
\times 10^{-4}$, $p = 2.1$, $N' = 0.03$~cm$^{-3}$, $n'_c = 5 \times 10^{22}$~cm$^{-2}$ (dotted); 
d) $r_0 = 10^{13}$~cm, $\Delta r_0/r_0 = 10^{-4}$, $n'_c = 10^{23}$~cm$^{-2}$ (dash-dot); 
$r_0 = 1.5 \times 10^{16}$~cm, $\Delta r_0/r_0 = 10^{-6}$, $p = 2.1$, $\epsilon_e = 0.02$, 
$N' = 0.008$~cm$^{-3}$, $n'_c = 5 \times 10^{23}$~cm$^{-2}$ (dotted); 
e) $r_0 = 10^{15}$~cm, $n'_c = 5 \times 10^{23}$~cm$^{-2}$ (dash-dot); $r_0 = 10^{15}$~cm, 
$n'_c = 5 \times 10^{24}$~cm$^{-2}$, $\kappa = 0$, i.e. a uniform ISM/circum-burst density (dotted)
f) $r_0 = 10^{15}$~cm, $n'_c = 5 \times 10^{24}$~cm$^{-2}$, $N' = 4$~cm$^{-3}$ (dash-dot);
Other parameters of these models are the same as model C1 in Table \ref {tab:param}. In all plots 
dash line corresponds to component C1 in this table. Note that simulated optical light curves have 
a broader width than in Fig. \ref{fig:lc}. \label{fig:paramvarc1}}
\end{figure}
\end{center}

Fig. \ref{fig:paramvarc1} shows variants of component C1. In model a) density of 
ISM/circum-burst material is smaller and consequently its X-ray emission is not sufficient. 
Model b) has larger $\Delta r_0/r_0$ but smaller ISM/circum-burst density, which falls 
in the lower half of the range of ISM/circum-burst material density estimated for other short 
GRBs~\citep{grbshortrev0}. This model is an example of a model degenerate with C1 in 
Table \ref{tab:param}. However, it has flatter X-ray light curve at later times, i.e. after 
$T + 4 \times 10^7$~sec~$\sim T + 470$~days. Thus, future observations may distinguish between this 
model and C1 in Table \ref {tab:param}. None of models in Fig. \ref{fig:paramvarc1}-a or -b can 
explain the radio afterglow. Fig. \ref{fig:paramvarc1}-c includes two models with smaller Lorentz 
factors, larger $\Delta r_0/r_0$, and larger ISM/circumburst and/or jet column densities. Both models 
have inconsistent X-ray and optical light curves and insufficient radio emission. We also notice 
that although $r_0$ in these models is the same as C1 in Table \ref{tab:param}, the time of peaks 
in the same bands are not the same in the 3 models.

Fig. \ref{fig:paramvarc1}-d presents two models, one with a shorter initial radius $r_0$ for external 
shocks and the other with a longer distance. The most important difference of these models with 
C1 in Table \ref {tab:param} is the time of peak emission. Although the amplitude of X-ray light curves 
of these models are low, it can be adjusted by increasing $\Delta r_0/r_0$ and/or column density. 
Therefore, the essential problem of these models is the position of light curves peaks. 
Model e) has $r_0$ less than C1 but larger than the first model in 
Fig. \ref{fig:paramvarc1}-d and a jet column density one order of magnitude larger than C1. Similar 
to models in Fig. \ref{fig:paramvarc1}-d the time of peak emission and its amplitude are 
inconsistent with data. Finally, model f) has similar distance as the model in 
Fig. \ref{fig:paramvarc1}-e but higher jet column density and ISM/circum-burst material density. It is 
too bright in X-ray and optical and peaks are too early.

\subsection{Variants of C2} \label{sec:c3var}
As mentioned in Sec. \ref{sec:ag}, optical and IR emissions are dominantly from kilonova and only 
during the latest observations at $> T + 200$~days contribution of the GRB might have become 
significant. For this reason optical data should be used as an upper limit for GRB contribution. It 
is also more difficult to select the best model for C2 and quantify characteristics of the 
component/section of the jet which emitted dominantly in optical bands.

\begin{center}
\begin{figure}
\begin{tabular}{p{6cm}p{6cm}p{6cm}p{6cm}}
a) & \hspace{-2cm} b) & \hspace{-4cm} c) & \hspace{-6cm} d) \\
\includegraphics[width=6cm]{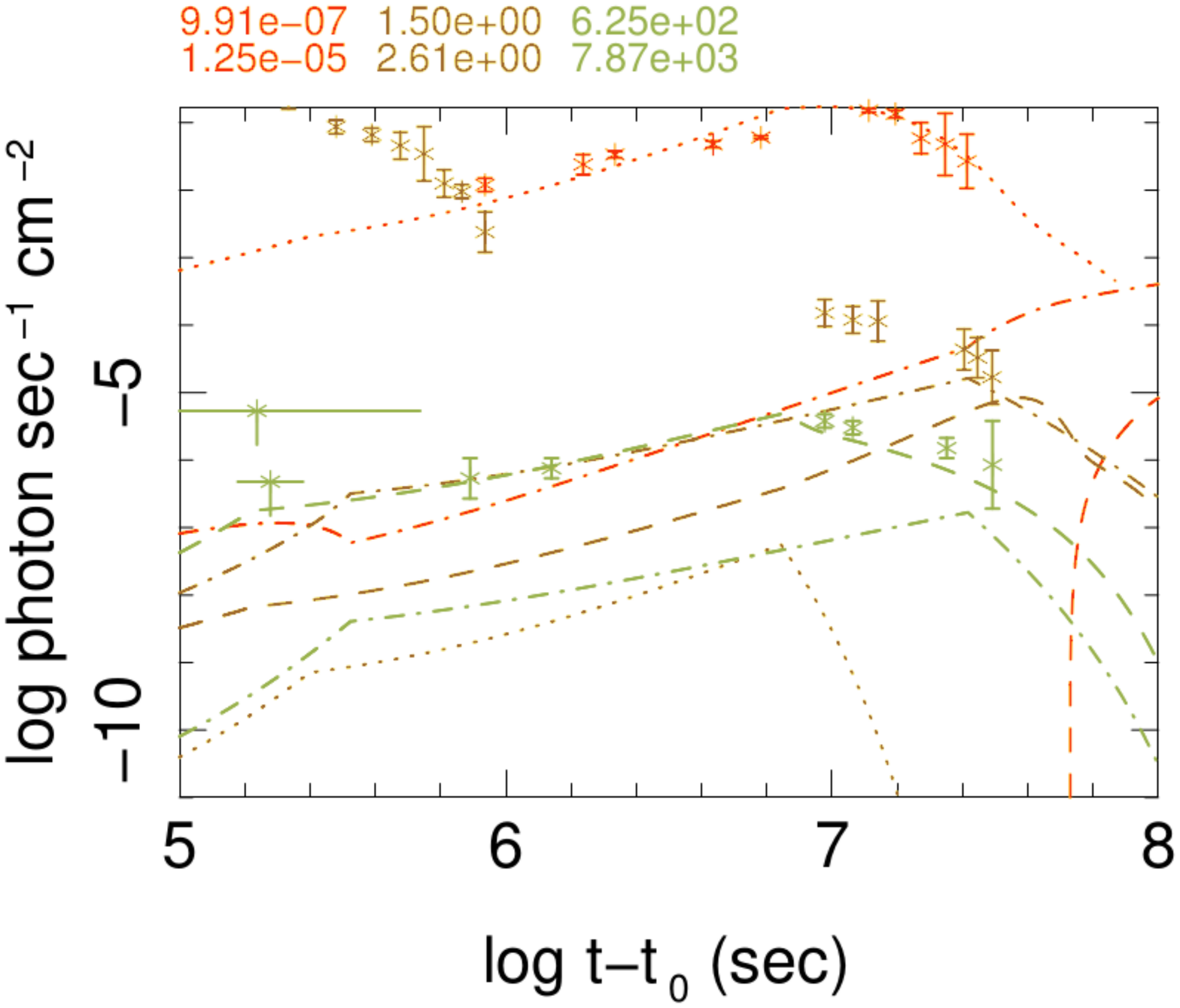} & 
\hspace{-2cm}\includegraphics[width=6cm]{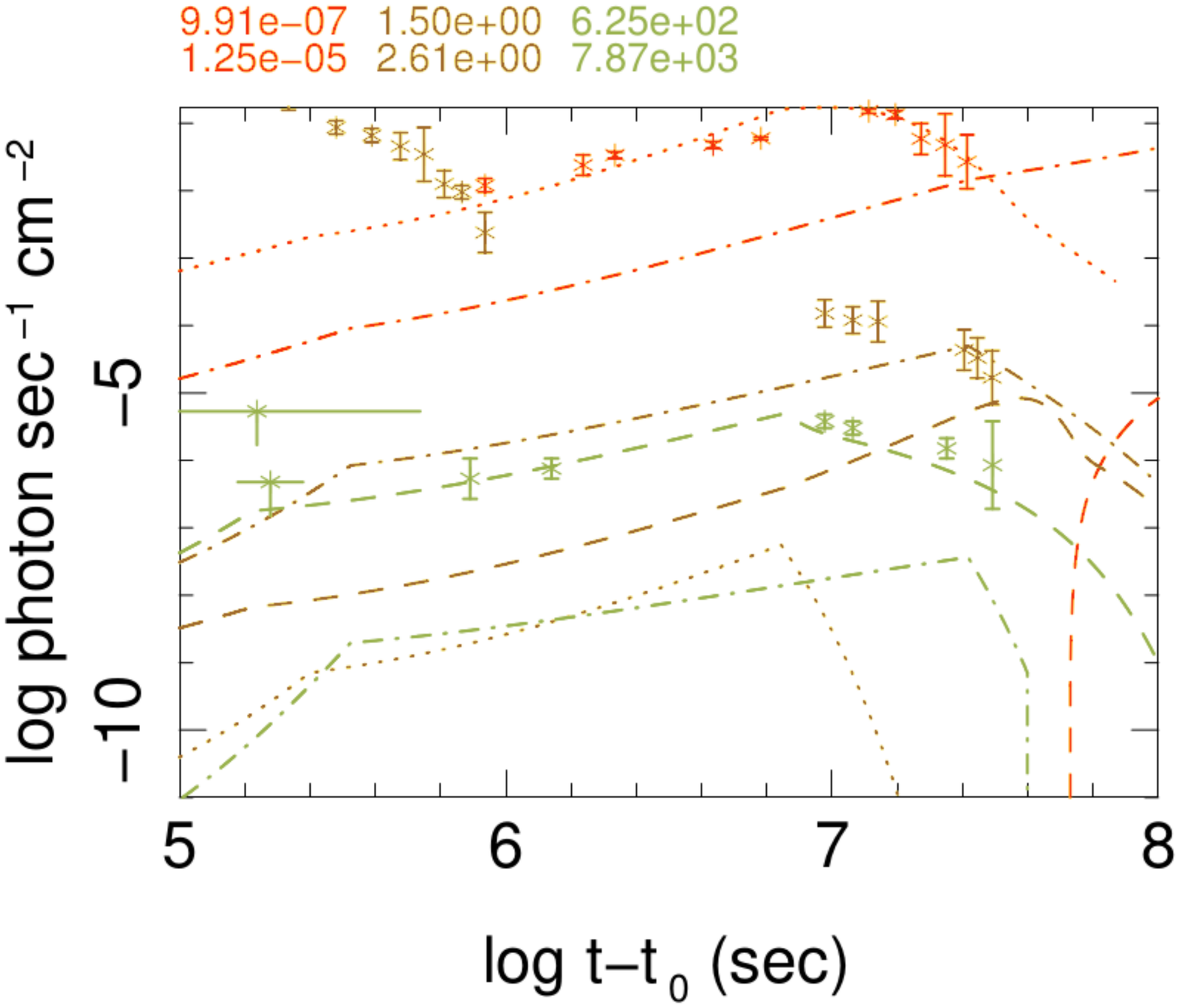} & 
\hspace{-4cm}\includegraphics[width=6cm]{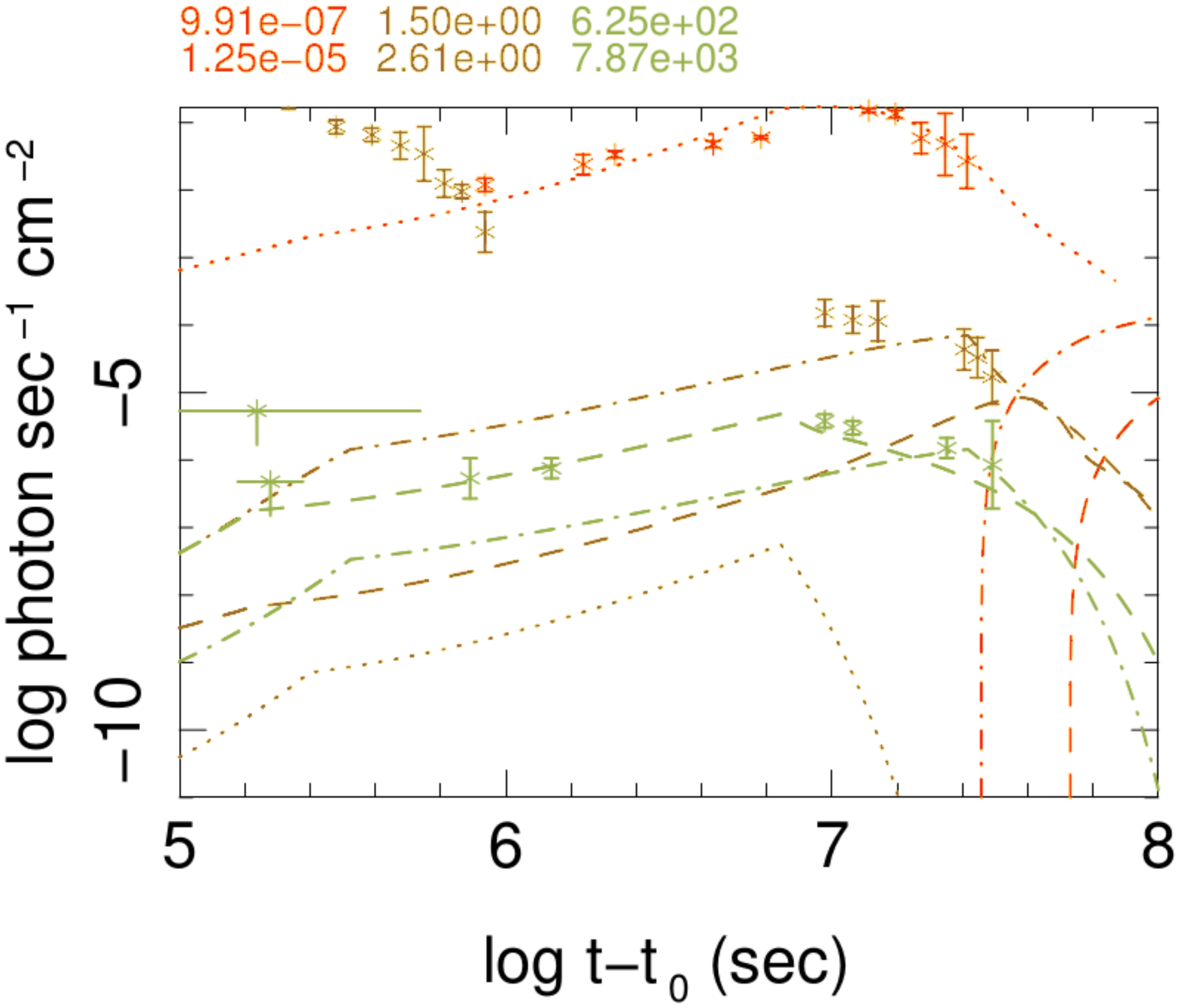} &
\hspace{-6cm}\includegraphics[width=6cm]{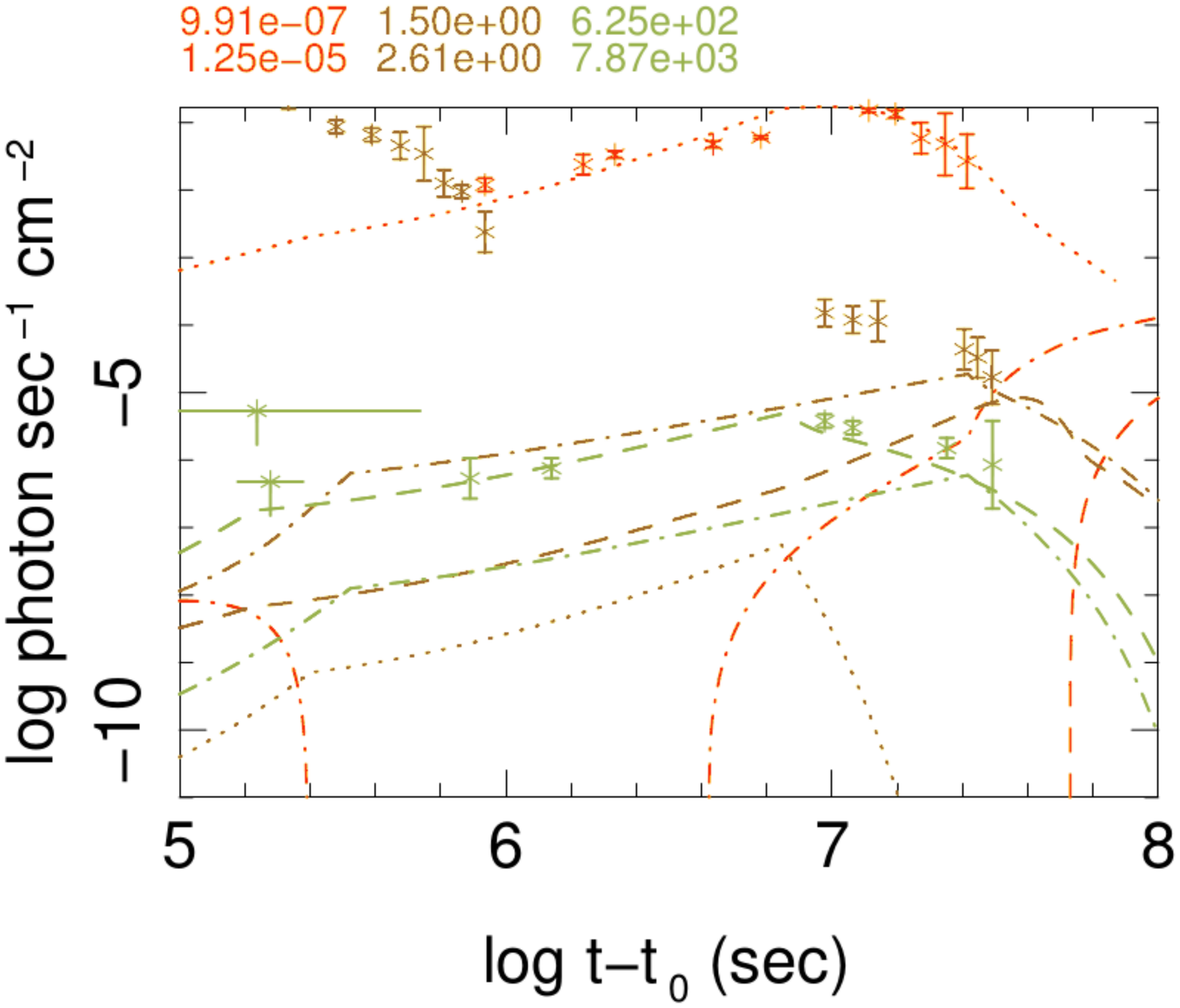} \\
\includegraphics[width=6cm]{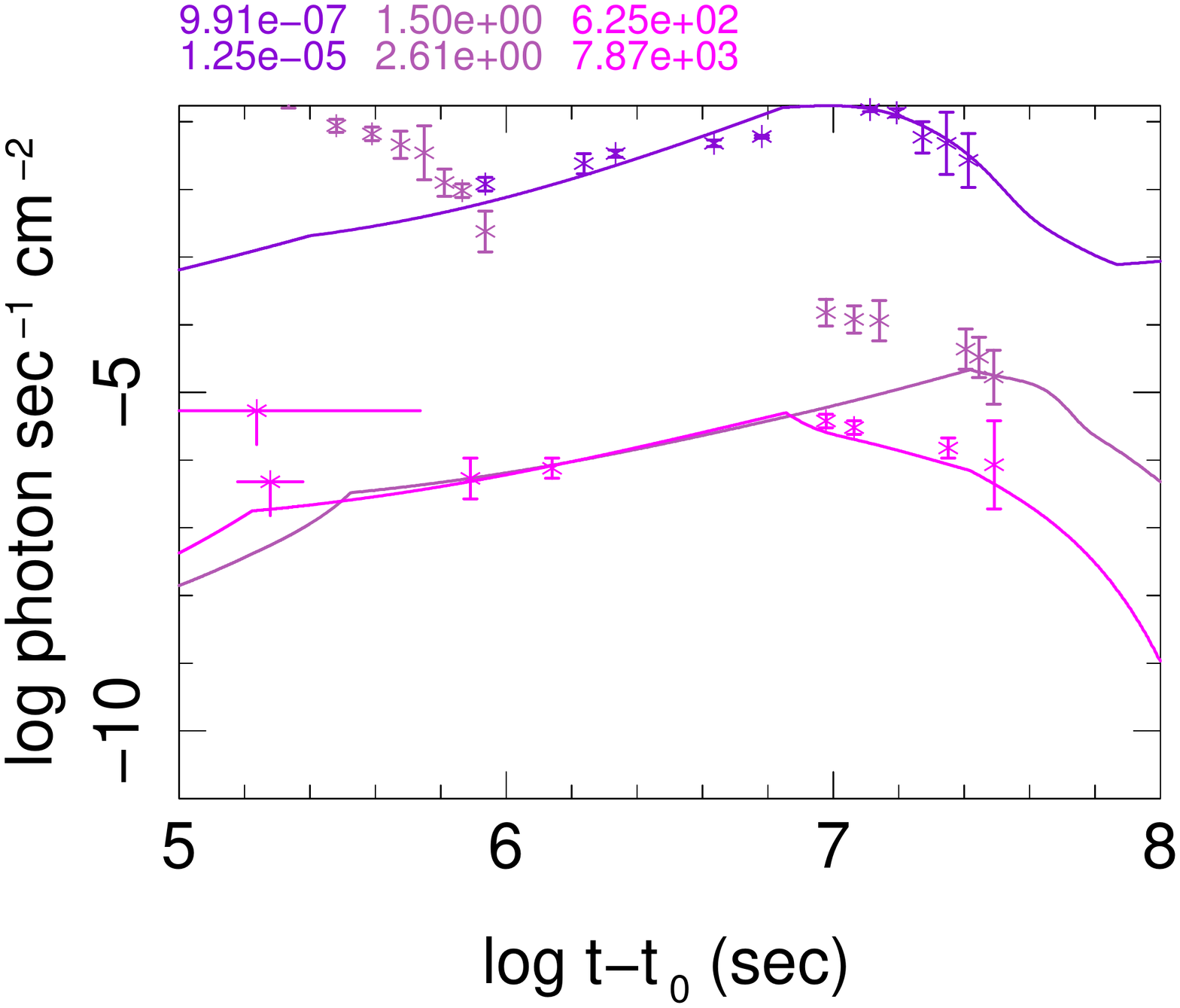} &
\hspace{-2cm}\includegraphics[width=6cm]{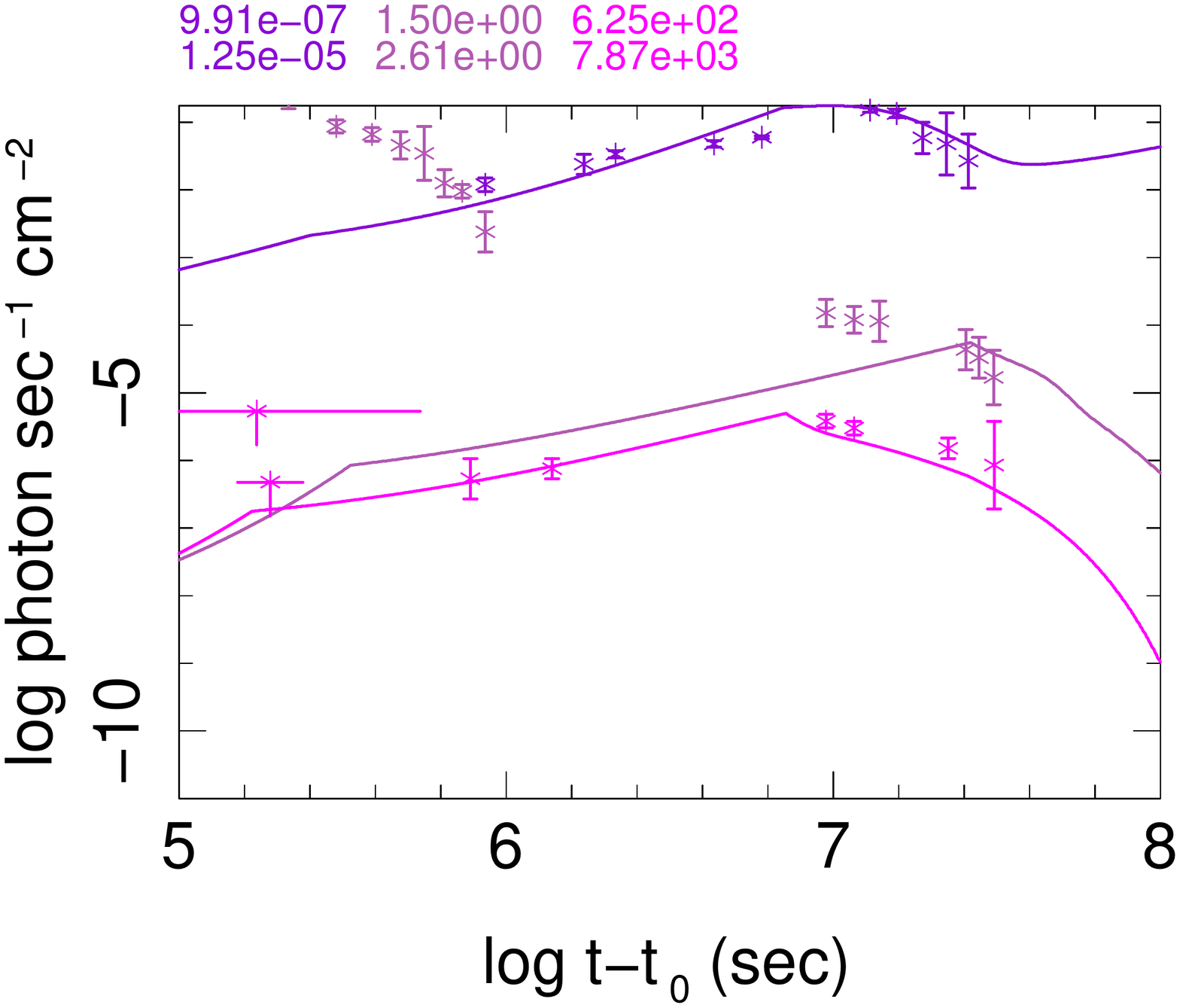} &
\hspace{-4cm}\includegraphics[width=6cm]{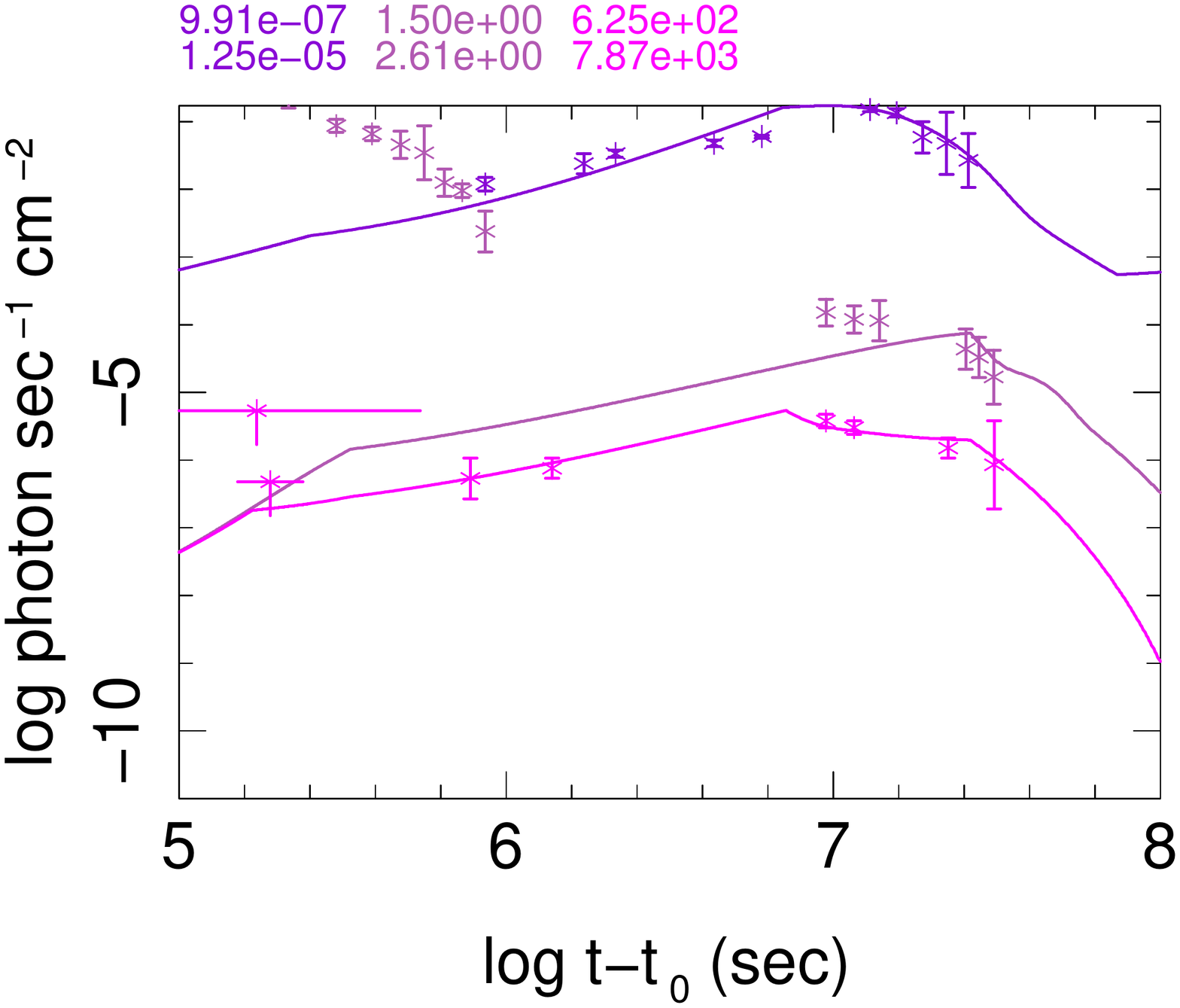} &
\hspace{-6cm}\includegraphics[width=6cm]{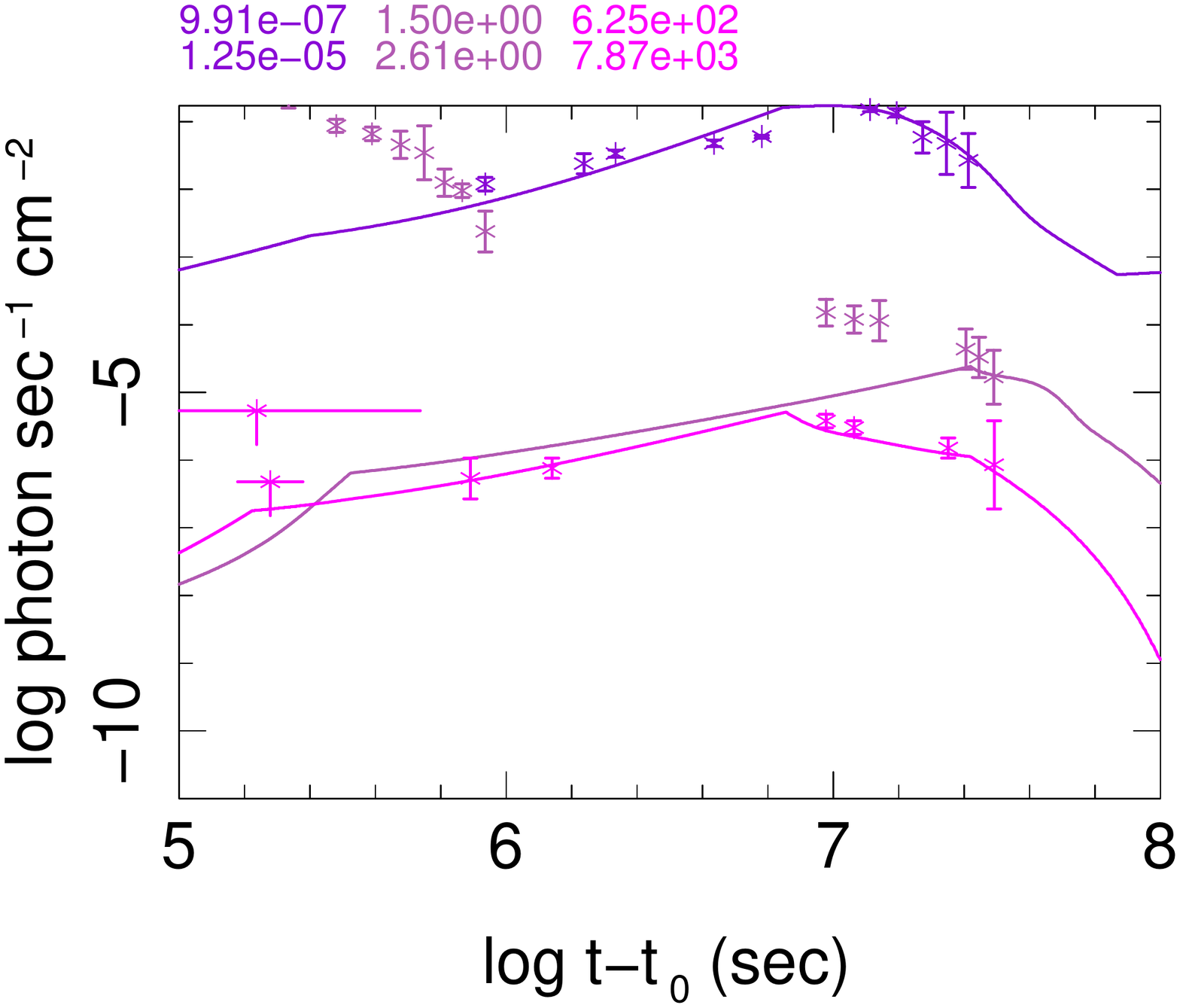}
\end{tabular}
\caption{Light curves of 3-component model with variant component C2. Upper row: Light curves of 
components; Lower row: Sum of light curves of 3 components. Variant 
of C2 have following parameters different from C2 in Table \ref {tab:param}:
a) $\Delta r_0/r_0 = 10^{-5}$; 
b) $\Delta r_0/r_0 = 5 \times 10^{-4}$, $\epsilon_e = 0.03$, $N' = 0.02$~cm$^{-3}$;
c) $\gamma'_0 = \Gamma = 30$, $\Delta r_0/r_0 = 10^{-6}$, $n'_c = 10^{22}$~cm$^{-2}$;
d) $\gamma'_0 = \Gamma = 30$, $\Delta r_0/r_0 = 10^{-5}$, $n'_c = 10^{22}$~cm$^{-2}$, 
$N' = 0.008$~cm$^{-3}$. C1 and C3 correspond to models shown in Table \ref {tab:param}. Definition 
of light curves is the same as in Fig. \ref {fig:lc}. \label{fig:paramvarc2}}
\end{figure}
\end{center}

Fig. \ref{fig:paramvarc2} shows light curves of components and their sum for 4 variant models to C2 in 
Table \ref{tab:param}. Model a) has a smaller $\Delta r_0/r_0$ and predicts a lower contribution from 
GRB in the optical/IR emission than C2. Model b) has a larger $\Delta r_0/r_0$ than C2 but smaller 
ISM/circum-burst density and fraction of kinetic energy transferred to accelerated electrons. This 
model fits the last 3 optical/IR observations and has a slightly better fit - $\chi^2 \approx 1.57$ 
for C2 against $\chi^2 \approx 0.05$ for this model (only the last 3 data points are considered). 
However, the model in Fig. \ref{fig:paramvarc2}-b may be also interpreted as slightly over-estimating 
optical emission at the epoch of latest 3 observations.

Models c) and d) are simulated with $\Gamma = 30$, i.e. similar to the second prompt gamma-ray 
peak at the end of internal shocks~\citep{hourigw170817}. The purpose for this choice is to 
see whether C2 can be a remnant of shells or sector of jet creating the second prompt peak. 
Interestingly, they fit X-ray light curve sightly better than model C1 in Table \ref{tab:param} 
because they have a significant X-ray mission at late times. But c) slightly over-estimates optical 
emission. This inconsistency can be resolved by e.g. slightly reducing the width of active region or 
column density. However, such adjustments decrease its X-ray contribution too and neutralizes advantage 
of this model with respect to the model of Table \ref{tab:param}. Model d) has both X-ray 
and optical well consistent with the data. However, the density of ISM/circum-burst material in this 
model is 5 folds less than in C1 component. Giving the fact that radiation from models c) and d) 
should come from an azimuthal angle $\lesssim 2^\circ$ further from C1, such a large variation of 
ISM/circum-burst density seems unrealistic. By contrast, if in place of C1 we use its variant in 
Fig. \ref {fig:paramvarc1}-b, which has much smaller ISM/circum-burst density closer to that of model 
d) here, then together with d) for C2 they make an overall consistent model. In any case, column 
density of c) and d) models is much smaller than other components. In a picture in which the 
3-component model crudely presents profile of the jet, a component with a tiny column density can 
be included to more prominent ones and neglected. As mentioned in the footnote \ref{foot:shortcom}, 
a better model of the jet should include continuous variation of column density and Lorentz factor 
with azimuthal angle.

\subsection{Variants of C3} \label{sec:c3var}
\begin{center}
\begin{figure}
\begin{tabular}{p{6cm}p{6cm}p{6cm}}
a) & \hspace{-1cm} b) & \hspace{-2cm} c) \\
\includegraphics[width=7cm]{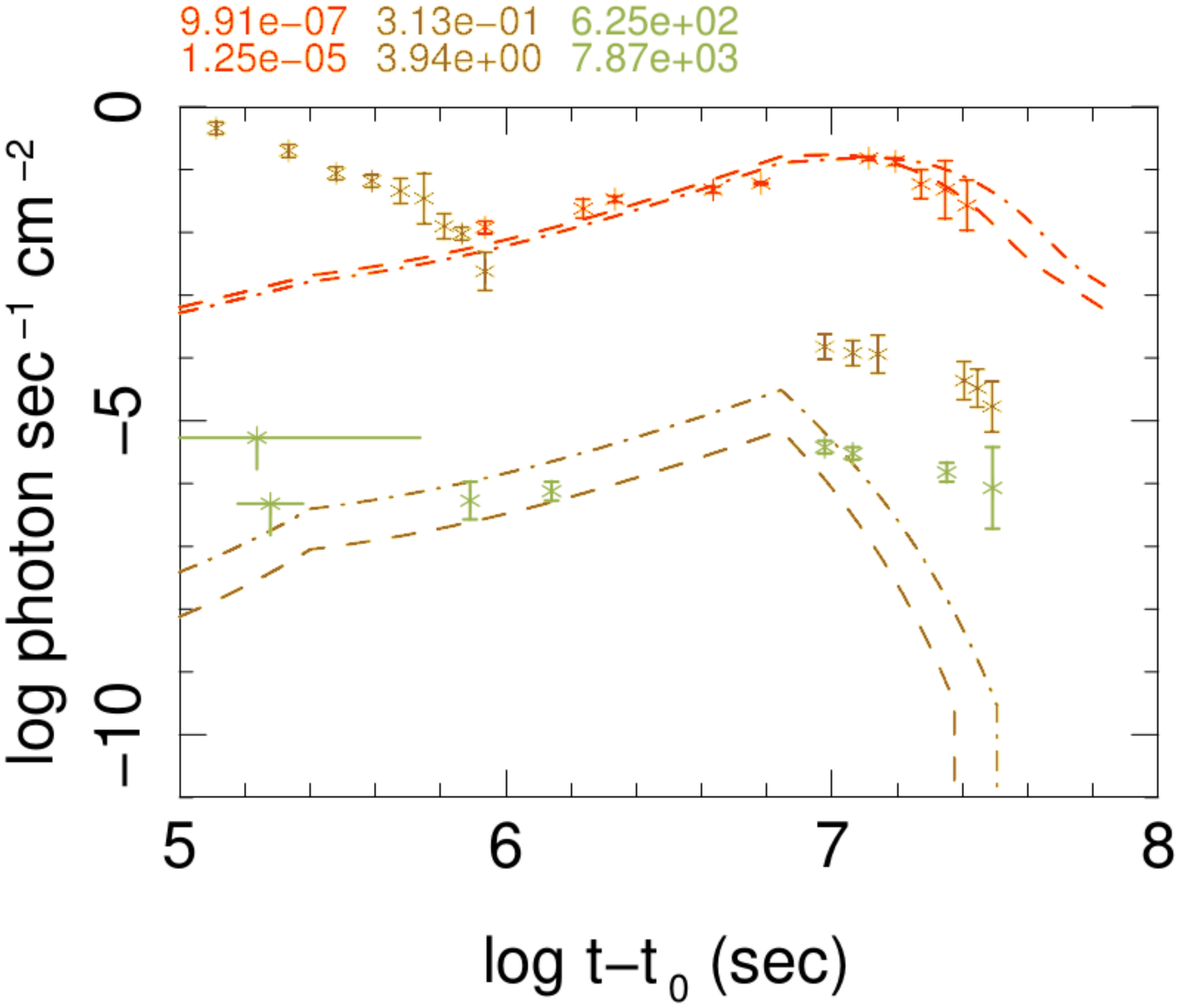} &
\hspace{-1cm}\includegraphics[width=7cm]{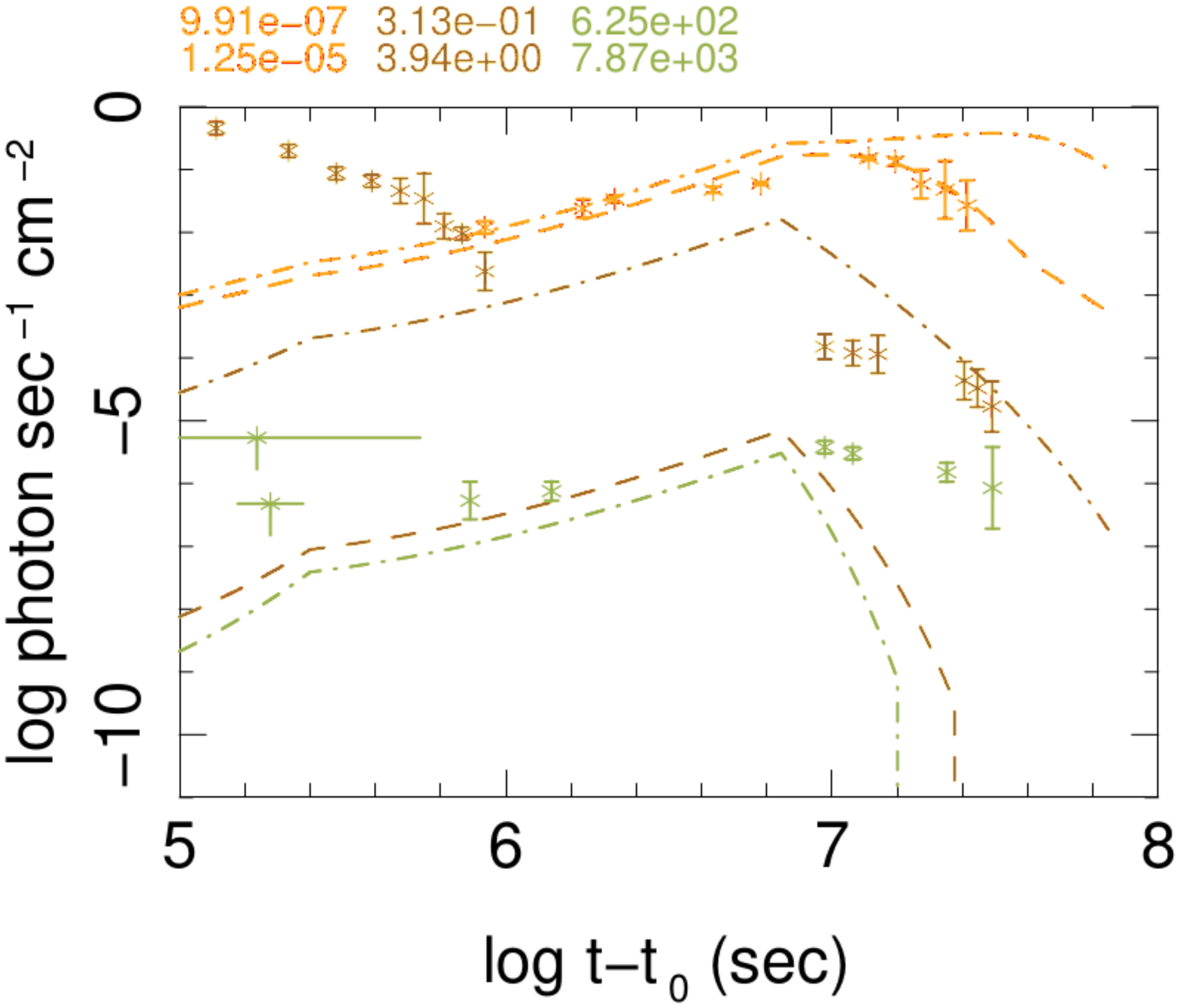} &
\hspace{-2cm}\includegraphics[width=7cm]{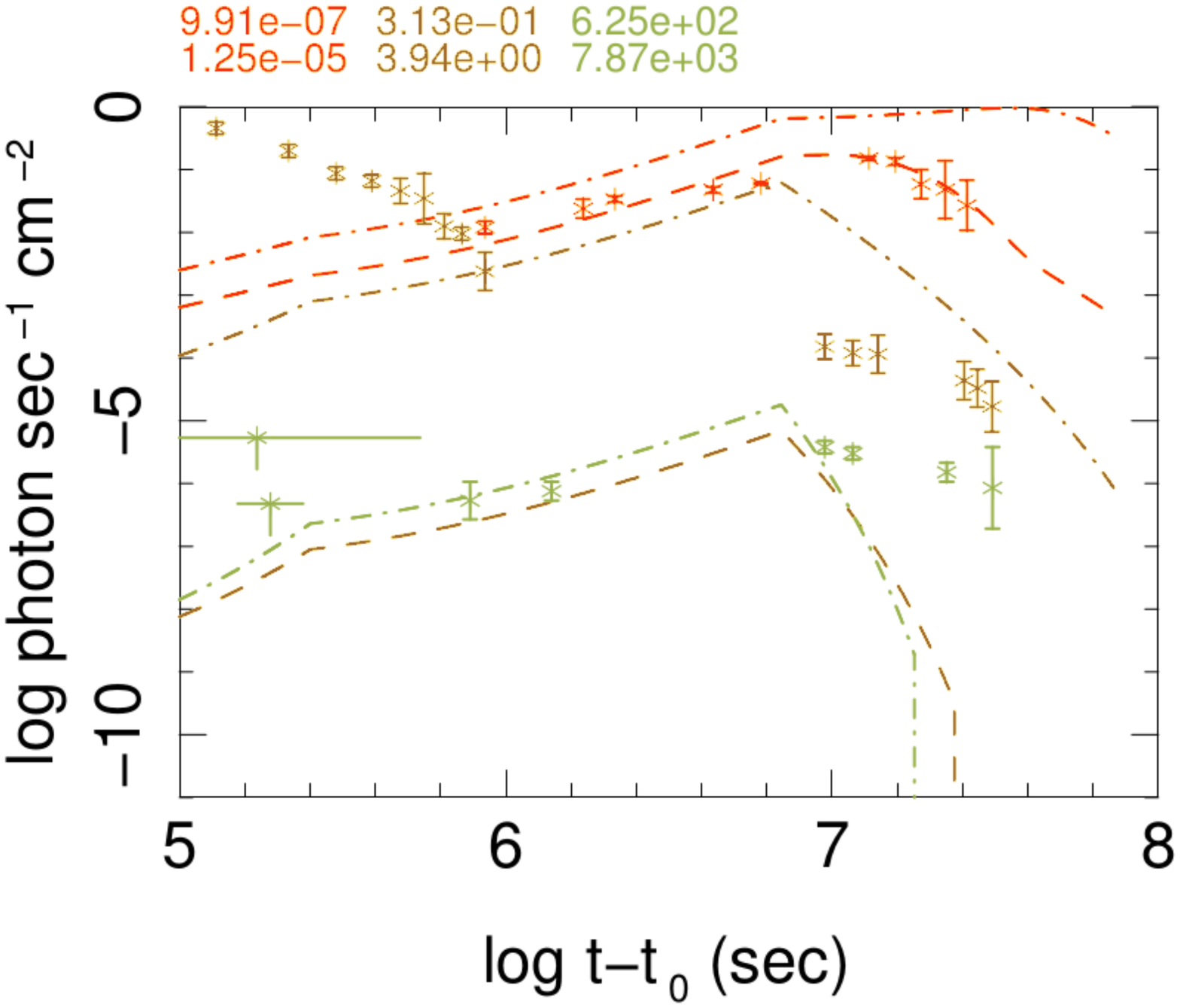} \\
d) & \hspace{-1cm} e) & \hspace{-2cm} f) \\
\includegraphics[width=7cm]{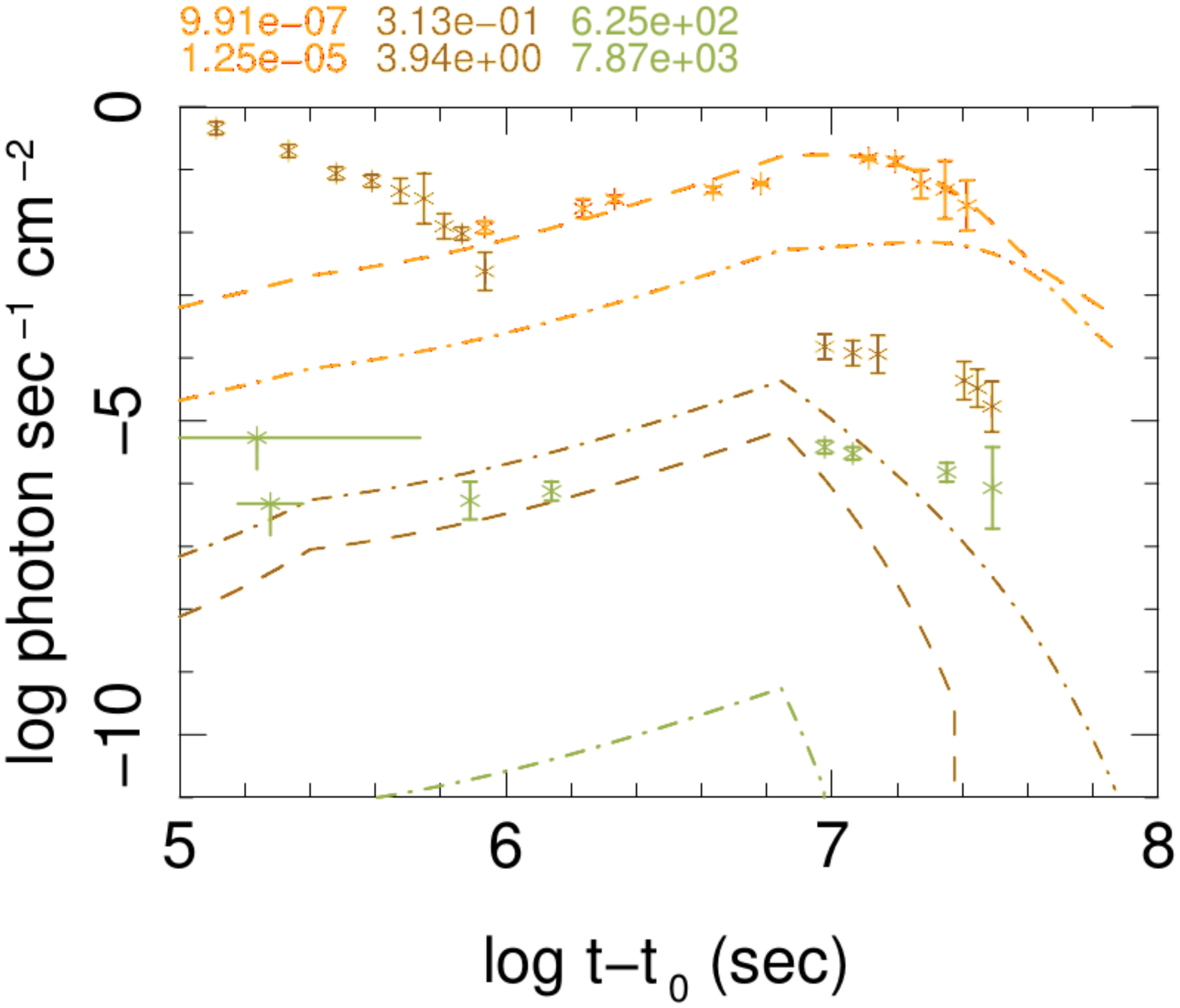} &
\hspace{-1cm}\includegraphics[width=7cm]{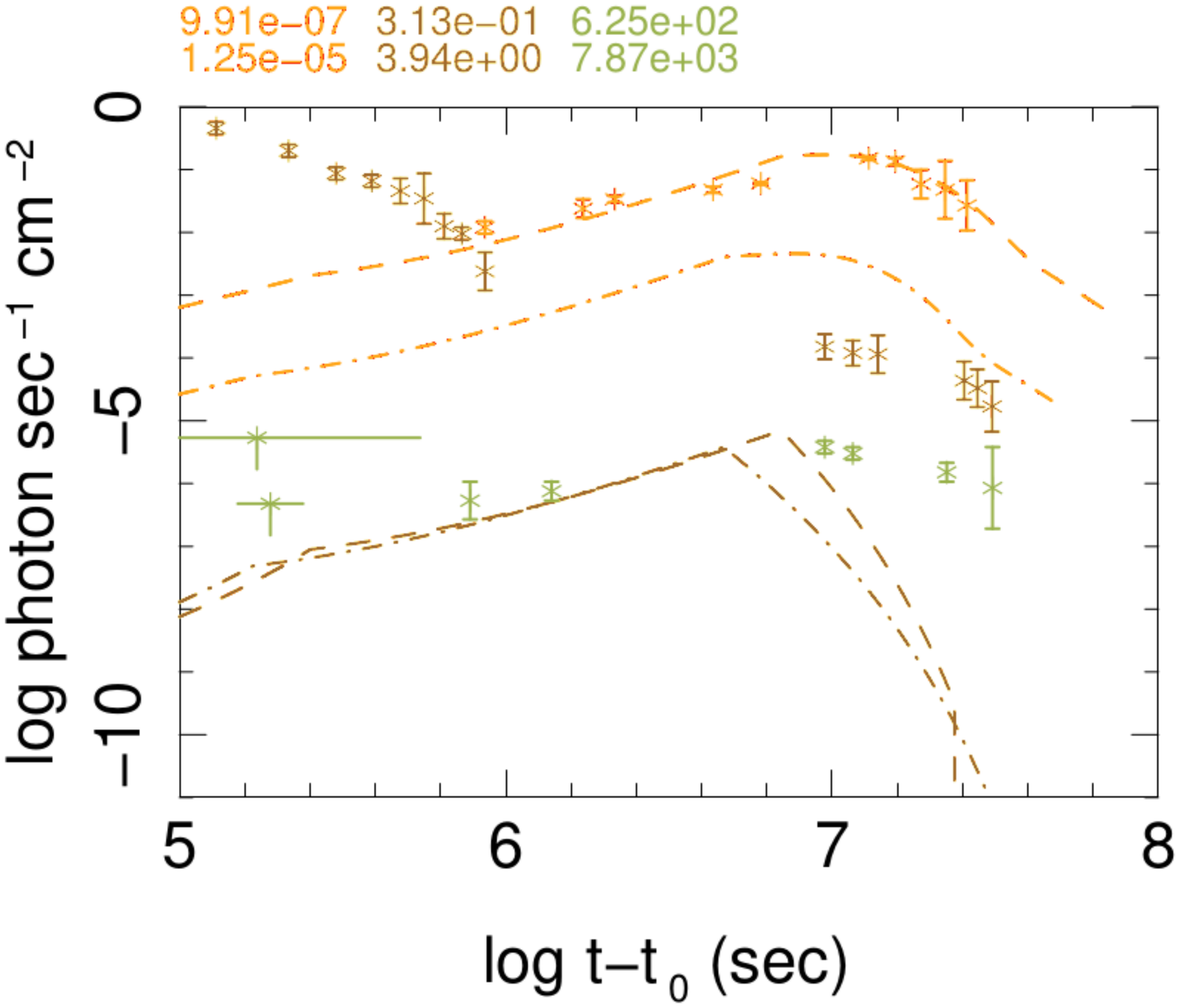} &
\hspace{-2cm}\includegraphics[width=7cm]{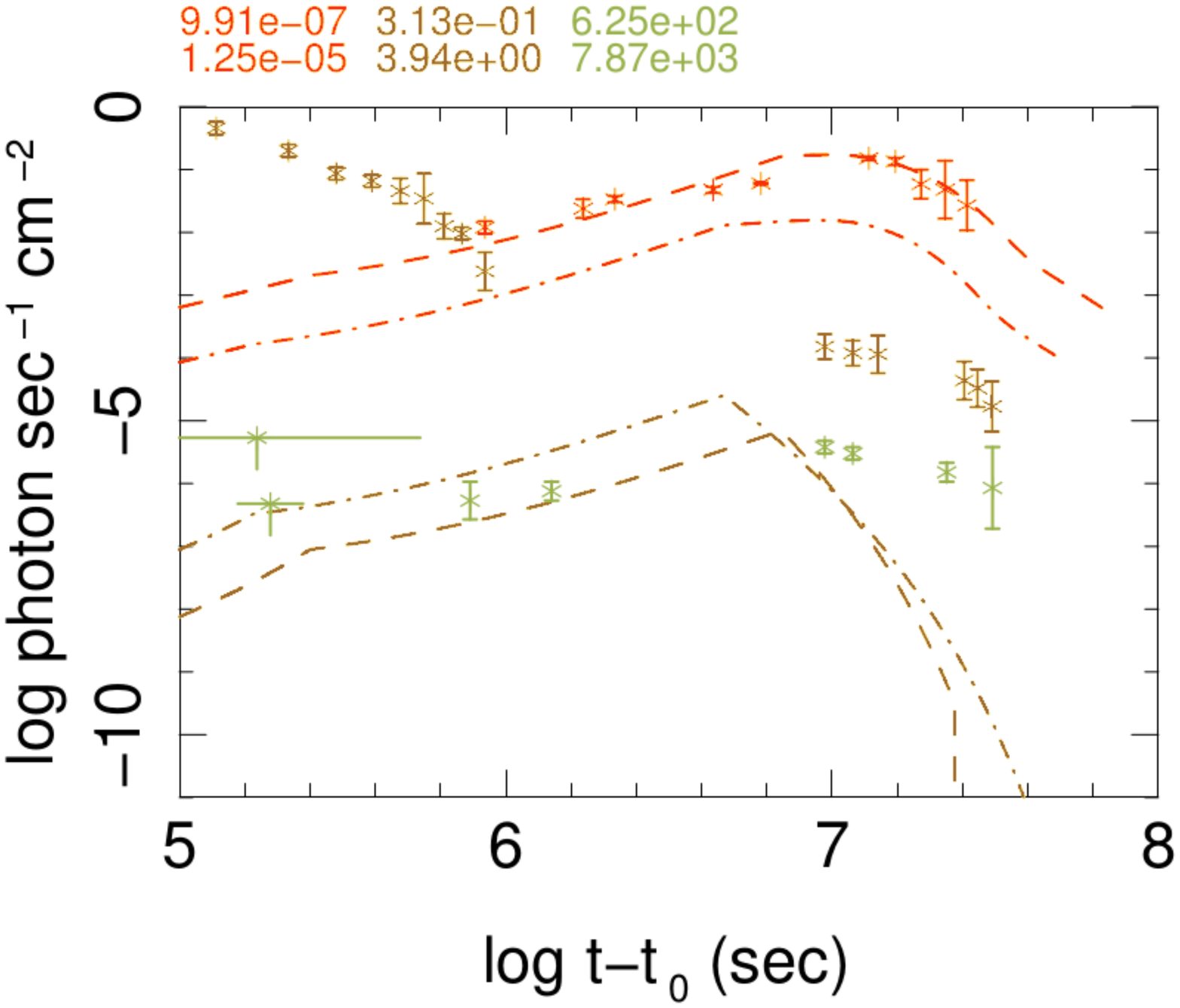}
\end{tabular}
\caption{Models of component C3 with varied parameters: 
a) $\gamma'_0 = \Gamma = 4$, $r_0 = 10^{16}$~cm, $\Delta r_0/r_0 = 10^{-3}$, $N' = 0.001$~cm$^{-3}$, 
$n'_c = 10^{25}$~cm$^{-2}$ (dash-dot); 
b) $\gamma'_0 = \Gamma = 4$, $r_0 = 10^{16}$~cm, $\Delta r_0/r_0 = 10^{-3}$, $n'_c = 10^{25}$~cm$^{-2}$ 
(dash-dot); 
c) $\gamma'_0 = \Gamma = 4$, $r_0 = 10^{16}$~cm, $\Delta r_0/r_0 = 10^{-3}$, $N' = 0.04$~cm$^{-3}$, 
$n'_c = 10^{25}$~cm$^{-2}$ (dash-dot);  
d) $\gamma'_0 = \Gamma = 1.6$~cm$^{-2}$, $r_0 = 10^{16}$~cm (dash-dot);
e) $\gamma'_0 = \Gamma = 1.6$, $r_0 = 10^{16}$~cm, $\Delta r_0/r_0 = 10^{-3}$ (dash-dot); 
f) $\gamma'_0 = \Gamma = 1.6$, $r_0 = 10^{16}$~cm, $\Delta r_0/r_0 = 10^{-3}$, $n'_c = 10^{25}$~cm$^{-2}$
(dash-dot). 
Other parameters of these models are the same as model C3 in Table \ref {tab:param}. In all plots 
dash lines corresponds to component C3 in Table \ref {tab:param}. Note that optical band in these plots 
have a broader width than in Fig. \ref{fig:lc}. \label{fig:paramvarc3}}
\end{figure}
\end{center}

Fig. \ref{fig:paramvarc3} shows variant models for component C3. Models \ref{fig:paramvarc3}-a, -b, -c 
have a Lorentz factor of 4, i.e. similar to estimation of~\citep{gw170817lateradiosuprlum} from 
observation of superluminal movement of radio counterpart. Model a) has a shorter initial distance 
$r_0$ from center, thinner active region, and lower ISM/circum-burst material density, and denser jet. 
Radio light curve of this model is consistent with data up to $\sim T + 2\times 10^7$~sec~$\sim 230$~
days. This date corresponds to the second epoch of observations reported 
in~\citep{gw170817lateradiosuprlum}, which was used to measure apparent superluminal displacement of 
the source. However, C3 model in Table \ref{tab:param} fits later observations better. Nonetheless, 
considering uncertainties of the data, this model a priori remains an acceptable alternative. On the 
other hand, a column density of $\sim 10^{25}$~cm$^{-2}$ in a) seems too large to be realistic. Indeed 
this is equal to the column density of the ultra-relativistic component 
at much closer distances to the merger and before prompt internal shocks.  Energy dissipation in the 
outflow during its propagation from site of internal shocks up to location of external shocks should 
have reduced column density unless it had kept its coherence and collimation, and thereby its kinetic 
energy and Lorentz factor. But reduction of $\Gamma$ from 130 to just 4 is inconsistent with these 
assumptions. Another way to reconcile model a) with the overall picture is to assume that it is 
dominated by less accelerated, initially unshocked material ejected by the merger at azimuthal angles 
between ultra-relativistic component and slow kilonova disk/torus. A component with these properties 
may be present in some GRMHD simulations, for instance those 
by~\citep{nsmergerrprocsimulout,nstarbhmergsimul1}. In any case, replacing C3 in Table \ref{tab:param} 
with a) does not significantly modify conclusions of Sec. \ref{sec:interpret}. 

Models b) and c) have larger ISM/circum-burst density and over-produce both optical and radio 
emissions. Models d), e) and f) have a Lorentz factor slightly larger than C3 in Table \ref{tab:param} 
and $r_0$ shorter by $50\%$. None of them produces enough radio emission even with a column 
density $n'_c$ as large as $10^{25}$~cm$^{-2}$. These models show the interplay between Lorentz factor,  
densities, and thickness of active synchrotron emitting region.

\begin{thebibliography}{99}
\bibitem [\protect\citeauthoryear{Alexander, \etal}{2017}] {gw170817earlyradio1} \Name{K.D.}{Alexander}, \Name{E.}{Berger}, \Name{W.}{Fong}, \Name{P.K.G.}{Williams}, \Name{C.}{Guidorzi}, \Name{R.}{Margutti}, \Name{B.D.}{Metzger}, \Name{J.}{Annis}, \etal, \Journal{\APJL}{848}{2017}{L21} \href{https://arxiv.org/abs/1710.05457}{[arXiv:1710.05457]}.
\bibitem [\protect\citeauthoryear{Alexander, \etal}{2018}] {gw170817latexoptradio1} \Name{K.D.}{Alexander}, \Name{R.}{Margutti}, \Name{P.K.}{Blanchard}, \Name{W.}{Fong}, \Name{E.}{Berger}, \Name{A.}{Hajela}, \Name{T.}{Eftekhari}, \Name{R.}{Chornock}, \etal, \Journal{\APJL}{863}{2018}{L18} \href{https://arxiv.org/abs/1805.02870}{[arXiv:1805.02870]}.
\bibitem [\protect\citeauthoryear{Beuermann, \etal}{1999}] {synchradiofit} \Name{K.}{Beuermann}, \Name{F.V.}{Hessman}, \Name{K.}{Reinsch}, \Name{H.}{Nicklas}, \Name{P.M.}{Vreeswijk}, \Name{T.J.}{Galama}, \Name{E.}{Rol}, \Name{J.}{van Paradijs}, \Name{C.}{Kouveliotou}, \Name{F.}{Frontera}, \Name{N.}{Masetti}, \Name{E.}{Palazzi}, \Name{E.}{Pian}, \Journal{\AA}{352}{1999}{L26} \href{https://arxiv.org/abs/astro-ph/9909043}{astro-ph/9909043}.
\bibitem [\protect\citeauthoryear{Chatterjee \& Cordes}{2004}] {nstarbowshock} \Name{S.}{Chatterjee}, \Name{J.M.}{Cordes}, \Journal{\APJL}{600}{2004}{L51} 
\bibitem [\protect\citeauthoryear{Covino, \etal}{2017}] {gw170817bluekilonovapol} \Name{S.}{Covino}, \Name{K.}{Wiersema}, \Name{Y.}{Z.}{Fan}, \Name{K.}{Toma}, \Name{A.}{B.}{Higgins}, \Name{A.}{Melandri}, \Name{P.}{D'Avanzo}, \Name{C.}{G.}{Mundell}, \etal, \Journal {\NATA}{1}{2017}{791} \href{https://arxiv.org/abs/1710.05849}{[arXiv:1710.05849]}.
\bibitem [\protect\citeauthoryear{Cowperthwaite, \etal}{2017}] {gw170817bluekilonovamod} \Name{P.S.}{Cowperthwaite}, \Name{E.}{Berger}, \Name{V.A.}{Villar}, \Name{B.D.}{Metzger}, \Name{M.}{Nicholl}, \Name{R.}{Chornock}, \Name{P.K.}{Blanchard}, \Name{W.}{Fong}, \etal, \Journal{\APJL}{848}{2017}{17} \href{https://arxiv.org/abs/1710.05840}{[arXiv:1710.05840]}.
\bibitem [\protect\citeauthoryear{D'Avanzo, \etal}{2018}] {gw170817latedecline} \Name{P.}{D'Avanzo}, \Name{S.}{Campana}, \Name{G.}{Ghisellini}, \Name{A.}{Melandri}, \Name{M.G.}{Bernardini}, \Name{S.}{Covino}, \Name{V.}{D'Elia}, \Name{L.}{Nava}, \etal, \Journal{\AA}{613}{2018}{L1} \href{https://arxiv.org/abs/1801.06164}{[arXiv:1801.06164]}. 
\bibitem [\protect\citeauthoryear{D'Elia, \etal}{2011}] {grb110213axraytail} \Name{V.}{D'Elia}, \Name{N.}{Gehrels}, \Name{J.M.}{Gelbord}, \Name{S.T.}{Holland}, \Name{H.A.}{Krimm}, \Name{C.B.}{Markwardt}, \Name{D.M.}{Palmer}, \Name{M.H.}{Siegel}, \etal, (2011) \cir{\href{https://gcn.gsfc.nasa.gov/gcn3/11705.gcn3}{11705}}.
\bibitem [\protect\citeauthoryear{Dingus}{1995}] {fermi} \Name{B.L.}{Dingus} \Journal{\SSC}{231}{1995}{187}.
\bibitem [\protect\citeauthoryear{Dionysopoulou, \etal}{2015}] {nsmergerrprocsimulout} \Name{K.}{Dionysopoulou}, \Name{D.}{Alic}, \Name{L.}{Rezzolla}, \Journal {\PRD}{92}{2015}{084064} \href{https://arxiv.org/abs/1502.02021}{[arXiv:1502.02021]}. 
\bibitem [\protect\citeauthoryear{Dobie, \etal}{2018}] {lateradio} \Name{D.}{Dobie}, \Name{D.L.}{Kaplan}, \Name{T.}{Murphy}, \Name{E.}{Lenc}, \Name{K.P.}{Mooley}, \Name{C.}{Lynch}, \Name{A.}{Corsi}, \Name{D.}{Frail}, \Name{M.}{Kasliwal}, \Name{G.}{Hallinan}, \Journal{\APJ}{858}{2018}{L15} \href{https://arxiv.org/abs/1803.06853}{[arXiv:1803.06853]}.
\bibitem [\protect\citeauthoryear{Duffell \& MacFadyen}{2013}] {mhdjetsimul} \Name{P.C.}{Duffell}, \Name{A.I.}{MacFadyen}, \Journal{\APJ}{775}{2013}{87} \href{https://arxiv.org/abs/1302.7306}{arXiv:1302.7306}.
\bibitem [\protect\citeauthoryear{Duffell, \etal}{2015}] {mhdjetsimul0} \Name{P.C.}{Duffell}, \Name{E.}{Quataert}, \Name{A.I.}{MacFadyen}, \Journal{\APJ}{813}{2015}{64} \href{https://arxiv.org/abs/1505.05538}{arXiv:1505.05538}. 
\bibitem [\protect\citeauthoryear{Evans, \etal}{2017}] {gw170817swiftnustar} \Name{P.A.}{Evans}, \Name{S.B.}{Cenko}, \Name{J.A.}{Kennea}, \Name{S.W.K.}{Emery}, \Name{N.P.M.}{Kuin}, \Name{O.}{Korobkin}, \Name{R.T.}{Wollaeger}, \Name{C.L.}{Fryer}, \etal, \Journal{\SCI}{358}{2017}{1565} \href{https://arxiv.org/abs/1710.05437}{[arXiv:1710.05437]}.
\bibitem [\protect\citeauthoryear{Fong, \etal}{2013}] {grb130603bxray} \Name{W.F.}{Fong}, \Name{E.}{Berger}, \Name{B.D.}{Metzger}, \Name{R.}{Margutti}, \Name{R.}{Chornock}, \Name{G.}{Migliori}, \Name{R.J.}{Foley}, \Name{B.A.}{Zauderer}, \etal, \Journal{\APJ}{780}{2013}{118} \href{https://arxiv.org/abs/1309.7479}{[arXiv:1309.7479]}.
\bibitem [\protect\citeauthoryear{Fong, \etal}{2015}] {grbshortrev0} \Name{W.F.}{Fong}, \Name{E.}{Berger}, \Name{R.}{Margutti}, \Name{B.A.}{Zauderer}, \Journal{\APJ}{815}{2015}{102} \href{https://arxiv.org/abs/1509.02922}{[arXiv:1509.02922]}.
\bibitem [\protect\citeauthoryear{Fong, \etal}{2018}] {gw170817grbcomp} \Name{W.F.}{Fong}, \Name{E.}{Berger}, \Name{P.K.}{Blanchard}, \Name{R.}{Margutti}, \Name{P.S.}{Cowperthwaite}, \Name{R.}{Chornock}, \Name{K.D.}{Alexander}, \Name{B.D.}{Metzger}, \Name{V.A.}{Villar}, \etal, \Journal{\APJL}{848}{2017}{L29} \href{https://arxiv.org/abs/1710.05438}{[arXiv:1710.05438]}.
\bibitem [\protect\citeauthoryear{Goldstein, \etal}{2017}] {gw170817fermi} \Name{A.}{Goldstein}, \Name{P.}{Veres}, \Name{E.}{Burns}, \Name{M.S.}{Briggs}, \Name{R.}{Hamburg}, \Name{D.}{Kocevski}, \Name{C.A.}{Wilson-Hodge}, \Name{R.D.}{Preece}, \Journal{\APJL}{848}{2017}{L14} \href{https://arxiv.org/abs/1710.05446}{[arXiv:1710.05446]}.
\bibitem [\protect\citeauthoryear{Gottlieb, \etal}{2017}] {gw170817cocoon0} \Name{O.}{Gottlieb}, \Name{E.}{Nakar}, \Name{T.}{Piran}, \Name{K.}{Hotokezaka}, (2017) \href{https://arxiv.org/abs/1710.05896}{[arXiv:1710.05896]}.
\bibitem [\protect\citeauthoryear{Grano \& Piran}{2012}] {jetlateralexpan} \Name{J.}{Grano}, \Name{T.}{Piran}, \Journal{\MRA}{421}{2012}{570} \href{https://arxiv.org/abs/1109.6468}{[arXiv:1109.6468]}.
\bibitem [\protect\citeauthoryear{Ghirlanda, \etal}{2019}] {gw170817lateradiosuprlum0} \Name{G.}{Ghirlanda}, \Name{O.S.}{Salafia}, \Name{Z.}{Paragi}, \Name{M.}{Giroletti}, \Name{J.}{Yang}, \Name{B.}{Marcote}, \Name{J.}{Blanchard}, \Name{I.}{Agudo}, \etal, \Journal{\SCI}{363}{2019}{968} \href{https://arxiv.org/abs/1808.00469}{[arXiv:1808.00469]}.
\bibitem [\protect\citeauthoryear{Hallinan, \etal}{2017}] {gw170817earlyradio} \Name{G.}{Hallinan}, \Name{A.}{Corsi}, \Name{K.}{P.}{Mooley}, \Name{K.}{Hotokezaka}, \Name{E.}{Nakar}, \Name{M.M.}{Kasliwal}, \Name{D.L.}{Kaplan}, \Name{D.A.}{Frail}, \etal, \Journal{\SCI}{358}{2017}{1579} \href{https://arxiv.org/abs/1710.05435}{[arXiv:1710.05435]}.
\bibitem [\protect\citeauthoryear{Haggard, \etal}{2018}] {gw170817chandraxray358} \Name {D.}{Haggard}, \Name{M}{Nynka}, \Name{J.J.}{Ruan}, (2018) \cir{\href{https://gcn.gsfc.nasa.gov/gcn3/23137.gcn3}{23137}}, Correction: \cir{\href{https://gcn.gsfc.nasa.gov/gcn3/23140.gcn3}{23140}}.
\bibitem [\protect\citeauthoryear{Hajela, \etal}{2018}] {gw170817chandraxray260} \Name{A.}{Hajela}, \Name{K.D.}{Alexander}, \Name{T.}{Eftekhari}, \Name{R.}{Margutti}, \Name{W.}{Fong}, \Name{E.}{Berger}, (2018) \cir{\href{https://gcn.gsfc.nasa.gov/gcn3/22692.gcn3}{22692}}.
\bibitem [\protect\citeauthoryear{Hotokezaka, \etal}{2018}]{gw170817latefasttail} \Name{K.}{Hotokezaka}, \Name{K.}{Kiuchi}, \Name{M.}{Shibata}, \Name{E.}{Nakar}, \Name{T.}{Piran}, (2018) \href{https://arxiv.org/abs/1803.00599}{[arXiv:1803.00599]}.
\bibitem [\protect\citeauthoryear{Izzo, \etal}{2017}]{grb171205asn2017uk} \Name{L.}{Izzo}, \Name{A.}{de Ugarte Postigo}, \Name{K.}{Maeda}, \Name{C.C.}{Th\"one}, \Name{D.A.}{Kann}, \Name{M.}{Della Valle}, \Name{A.}{Sagues Carracedo}, \Name{M.J.}{Michałowski}, \etal, \Journal{\NAT}{565}{2019}{324} \href{https://arxiv.org/abs/1901.05500}{[arXiv:1901.05500]}.
\bibitem [\protect\citeauthoryear{Kann}{2012}]{grbshortag} \Name{D.A.}{Kann} \Journal{\EAS}{61}{2013}{309} \href{https://arxiv.org/abs/1212.0040}{[arXiv:1212.0040]}.
\bibitem [\protect\citeauthoryear{Kasen, \etal}{2017}] {gw170817optkilonovath} \Name{D.}{Kasen}, \Name{B.}{Metzger}, \Name{J.}{Barnes}, \Name{E.}{Quataert}, \Name{E.}{Ramirez-Ruiz}, \Journal{\NAT}{551}{2017}{80} \href{https://arxiv.org/abs/1710.05463}{[arXiv:1710.05463]}.
\bibitem [\protect\citeauthoryear{Kasen \& Barnes}{2019}] {gw170817kilonovafaint0} \Name{D.}{Kasen}, \Name{J.}{Barnes}, \Journal{\APJ}{876}{2019}{128} \href{https://arxiv.org/abs/1807.03319}{[arXiv:1807.03319]}.
\bibitem [\protect\citeauthoryear{Kasliwal, \etal}{2017}] {gw170817cocoon} \Name{M.M.}{Kasliwal}, \Name{E.}{Nakar}, \Name{L.P.}{Singer}, \Name{D.L.}{Kaplan}, \Name{D.O.}{Cook}, \Name{A.}{Van Sistine}, \Name{R.M.}{Lau}, \etal, \Journal {\SCI}{358}{2017}{1559} \href{https://arxiv.org/abs/1710.05436}{[arXiv:1710.05436]}.
\bibitem [\protect\citeauthoryear{Kiuchi, \etal}{2015}] {nstarbhmergsimul1} \Name{K.}{Kiuchi}, \Name{Y.}{Sekiguchi}, \Name{K.}{Kyutoku}, \Name{M.}{Shibata}, \Name{K.}{Taniguchi}, \Name{T.}{Wada}, \Journal {\PRD}{92}{2015}{064034} \href{https://arxiv.org/abs/1506.06811}{[arXiv:1506.06811]}.
\bibitem [\protect\citeauthoryear{Komissarov, \etal}{2009}] {grbjetsimul} \Name{S.}{Komissarov}, \Name{N.}{Vlahakis}, \Name{A.}{Konigl}, \Name{M.}{Barkov}, \Journal {\MRA}{394}{2009}{1182} \href{https://arxiv.org/abs/0811.1467}{[arXiv:0811.1467]}.
\bibitem [\protect\citeauthoryear{Krimm}{2018}] {batgammaupper} \Name{H.A.}{Krimm}, Swift Science Team, private communication (2018).
\bibitem [\protect\citeauthoryear{Lamb \& Kobayashi}{2017}] {structuredjet} \Name{G.P.}{Lamb}, \Name{S.}{Kobayashi}, \Journal{\MRA}{472}{2017}{4953} \href{https://arxiv.org/abs/1706.03000}{[arXiv:1706.03000]}.
\bibitem [\protect\citeauthoryear{Lamb, \etal}{2019}]{gw170817lateopthstir} \Name{G.P.}{Lamb}, \Name{J.D.}{Lyman}, \Name{A.J.}{Levan}, \Name{N.R.}{Tanvir}, \Name{T.}{Kangas}, \Name{A.S.}{Fruchter}, \Name{B.}{Gompertz}, \Name{J.}{Hjorth}, \Name{I.}{Mandel}, \Name{S.R.}{Oates}, \Journal{\APJL}{870}{2019}{L15} \href{https://arxiv.org/abs/1811.11491}{[arXiv:1811.11491]}.
\bibitem [\protect\citeauthoryear{Lazzati, \etal}{2016}] {grboffaxis} \Name{D.}{Lazzati}, \Name{A.}{Deich}, \Name{B.J.}{Morsony}, \Name{J.C.}{Workman}, \Journal {\MRA}{471}{2017}{1652} \href{https://arxiv.org/abs/1610.01157}{[arXiv:1610.01157]}.
\bibitem [\protect\citeauthoryear{Lazzati, \etal}{2018}] {gw170817latexraystructjet} \Name{D.}{Lazzati}, \Name{R.}{Perna}, \Name{B.J.}{Morsony}, \Name{D.}{L\'opez-Cámara}, \Name{M.}{Cantiello}, \Name{R.}{Ciolfi}, \Name{B.}{Giacomazzo}, \Name{J.C.}{Workman}, \Journal {\PRL}{120}{2018}{241103} \href{https://arxiv.org/abs/1712.03237}{[arXiv:1712.03237]}.
\bibitem [\protect\citeauthoryear{LIGO, \etal}{2017a}] {gw170817multimess} \Name{LIGO Scientific Collaboration}, \Name{Virgo Collaboration}, \Name{Fermi GBM}, \Name{INTEGRAL}, \Name{IceCube Collaboration}, \Name{AstroSat Cadmium Zinc Telluride Imager Team}, \Name{IPN Collaboration}, \Name{The Insight-Hxmt Collaboration}, \etal, \Journal{\APJL}{848}{2017a}{L12} \href{https://arxiv.org/abs/1710.05833}{[arXiv:1710.05833]}.
\bibitem [\protect\citeauthoryear{LIGO, \etal}{2017b}] {gw170817fermimulti} \Name{LIGO Scientific Collaboration}, \Name{Virgo Collaboration}, \Name{Fermi Gamma-Ray Burst Monitor Collaboration}, \Name{INTEGRAL Collaboration}, \Journal{\APJL}{848}{2017b}{L13} \href{https://arxiv.org/abs/1710.05834}{[arXiv:1710.05834]}.
\bibitem [\protect\citeauthoryear{Lind \& Blandford}{1985}] {jetbeaming} \Name{K.R.}{Lind}, \Name{R.D.}{Blandford}, \Journal{\APJ}{295}{1985}{358}.
\bibitem [\protect\citeauthoryear{Lyman, \etal}{2018}] {gw170817lateopt} \Name{J.D.}{Lyman}, \Name{G.P.}{Lamb}, \Name{A.J.}{Levan}, \Name{I.}{Mandel}, \Name{N.R.}{Tanvir}, \Name{S.}{Kobayashi}, \Name{B.}{Gompertz}, \Name{J.}{Hjorth, \etal}, (2017) \href{https://arxiv.org/abs/1801.02669}{[arXiv:1801.02669]}.
\bibitem [\protect\citeauthoryear{Mandel}{2018}]{gw170817decline} \Name {I.}{Mandel}, \Journal{\APJ}{853}{2018}{L12} \href{https://arxiv.org/abs/1712.03958}{[arXiv:1712.03958]}.
\bibitem [\protect\citeauthoryear{Marshall, \etal}{2007}]{grb070809} \Name {F.E.}{Marshall}, \Name {D.N.}{Burrows}, \Name {M.M.}{Chester}, \Name {J.R.}{Cummings}, \Name {P.A.}{Evans}, \Name {N.}{Gehrels}, \Name {C.}{Guidorzi}, \Name {S.T.}{Holland}, \etal, (2007) \cir{\href{https://gcn.gsfc.nasa.gov/gcn3/6728.gcn3}{6728}}.
\bibitem [\protect\citeauthoryear{Margutti, \etal}{2017}] {gw170817cxc2day} \Name{R.}{Margutti}, \Name{E.}{Berger}, \Name{W.}{Fong}, \Name{C.}{Guidorzi}, \Name{K.D.}{Alexander}, \Name{B.D.}{Metzger}, \Name{P.K.}{Blanchard}, \Name{P.S.}{Cowperthwaite}, \etal, \Journal {\APJL}{848}{2017}{L20} \href{https://arxiv.org/abs/1710.05431}{[arXiv:1710.05431]}.
\bibitem [\protect\citeauthoryear{Margutti, \etal}{2018}] {gw170817latexary} \Name{R.}{Margutti}, \Name{K.D.}{Alexander}, \Name{X.}{Xie}, \Name{L.}{Sironi}, \Name{B.D.}{Metzger}, \Name{A.}{Kathirgamaraju}, \Name{W.}{Fong}, \Name{P.K.}{Blanchard}, \Name{E.}{Berger, \etal}, (2017) \href{https://arxiv.org/abs/1801.03531}{[arXiv:1801.03531]}.
\bibitem [\protect\citeauthoryear{Metzger, \etal}{2018}] {gw170817kilonovamassexc} \Name{B.D.}{Metzger}, \Name{T.A.}{Thompson}, \Name{E.}{Quataert}, \Journal{\APJ}{856}{2018}{101} \href{https://arxiv.org/abs/1801.04286}{[arXiv:1801.04286]}.
\bibitem [\protect\citeauthoryear{Montanari, \etal}{2005}] {grbshorttailbepposax} \Name{E.}{Montanari}, {F.}{Frontera}, \Name{C.}{Guidorzi}, \Name{M.}{Rapisarda}, \Journal{\APJ}{625}{2005}{L17} \href{https://arxiv.org/abs/astro-ph/0504199}{[astro-ph/0504199]}.
\bibitem [\protect\citeauthoryear{Mooley, \etal}{2017}] {gw170817lateradio} \Name{K.P.}{Mooley}, \Name{E.}{Nakar}, \Name{K.}{Hotokezaka}, \Name{G.}{Hallinan}, \Name{A.}{Corsi}, \Name{D.A.}{Frail}, \Name{A.}{Horesh}, \Name{T.}{Murphy}, \Name{E.}{Lenc}, \etal, \Journal{\NAT}{554}{2018}{207} \href{https://arxiv.org/abs/1711.11573}{[arXiv:1711.11573]}.
\bibitem [\protect\citeauthoryear{Mooley, \etal}{2018a}] {gw170817lateradiosuprlum} \Name{K.P.}{Mooley}, \Name{A.T.}{Deller}, \Name{O.}{Gottlieb}, \Name{E.}{Nakar}, \Name{G.}{Hallinan}, \Name{S.}{Bourke}, \Name{D.A.}{Frail}, \Name{A.}{Horesh}, \etal, \Journal{\NAT}{561}{2018a}{355} \href{https://arxiv.org/abs/1806.09693}{[arXiv:1806.09693]}
\bibitem [\protect\citeauthoryear{Mooley, \etal}{2018b}] {gw170817lateradio300} \Name{K.P.}{Mooley}, \Name{D.A.}{Frail}, \Name{D.}{Dobie}, \Name{E.}{Lenc}, \Name{A.}{Corsi}, \Name{K.}{De}, \Name{A.J.}{Nayana}, \Name{S.}{Makhathini}, \Name{I.}{Heywood}, \etal, \Journal{\APJL}{868}{2018b}{L11} \href{https://arxiv.org/abs/1810.12927}{[arXiv:1810.12927]}.
\bibitem [\protect\citeauthoryear{Murguia-Berthier, \etal}{2017}] {grboffaxprobab0} \Name{A.}{Murguia-Berthier}, \Name{E.}{Ramirez-Ruiz}, \Name{C.D.}{Kilpatrick}, \Name{R.J.}{Foley}, \Name{D.}{Kasen}, \Name{W.H.}{Lee}, \Name{A.L.}{Piro}, \Name{D.A.}{Coulter}, \etal, \Journal {\APJL}{848}{2017}{L34} \href{https://arxiv.org/abs/arXiv:1710.05453}{[arXiv:1710.05453]}. 
\bibitem [\protect\citeauthoryear{Nakar, \etal}{2018}] {gw170817cocoonsimul} \Name{E.}{Nakar}, \Name{O.}{Gottlieb}, \Name{T.}{Piran}, \Name{M.M.}{Kasliwal}, \Name{G.}{Hallinan} (2018) \href{https://arxiv.org/abs/1803.07595}{[arXiv:1803.07595]}.
\bibitem [\protect\citeauthoryear{Nicholl, \etal}{2017}] {gw170817kilonovaspeed} \Name{M.}{Nicholl}, \Name{E.}{Berger}, \Name{D.}{Kasen}, \Name{B.}{D.}{Metzger}, \Name{J.}{Elias}, \Name{C.}{Briceno}, \Name{K.}{D.}{Alexander}, \Name{P.}{K.}{Blanchard}, \etal, \Journal{\APJL}{848}{2017}{18} \href{https://arxiv.org/abs/1710.05456}{[arXiv:1710.05456]}.
\bibitem [\protect\citeauthoryear{Nishikawa, \etal}{2016}] {jetpicsimul} \Name{K.I.}{Nishikawa}, \Name{J.T.}{Frederiksen}, \Name{A.}{Nordlund}, \Name{Y.}{Mizuno}, \Name{P.E.}{Hardee}, \Name{J.}{Niemiec}, \Name{J.L.}{Gomez}, \Name{A.}{Pe'er}, \etal, \Journal{\APJ}{820}{2016}{94} \href{https://arxiv.org/abs/1511.03581}{[arXiv:1511.03581]}.
\bibitem [\protect\citeauthoryear{Norris, \etal}{2010}] {grbshorttailbat} \Name{J.P.} Norris, \Name{N.}{Gehrels}, \Name{J.D.}{Scargle} \Journal{\APJ}{717}{2010}{411} \href{https://arxiv.org/abs/0910.2456}{[arXiv:0910.2456]}.
\bibitem [\protect\citeauthoryear{Nynka, \etal}{2018}] {gw170817xraycxc260} \Name{M.}{Nynka}, \Name{J.J.}{Ruan}, \Name{D.}{Haggard}, (2018)  \href{https://arxiv.org/abs/1805.04093}{arXiv:1805.04093}
\bibitem [\protect\citeauthoryear{Olivares, \etal}{2012}] {grb100316d} \Name{F.}{Olivares}, \Name{J.}{Greiner}, \Name{P.}{Schady}, \Name{A.}{Rau}, \Name{S.}{Klose}, \Name{T.}{Krühler}, \Name{P.M.J.}{Afonso}, \Name{A.C.}{Updike}, \etal, \Journal{\AA}{539}{2012}{A76} \href{https://arxiv.org/abs/1110.4109}{[arXiv:1110.4109]}.
\bibitem [\protect\citeauthoryear{Pian, \etal}{2017}] {gw170817rprocess} \Name{E.}{Pian}, \Name{P.}{D'Avanzo}, \Name{S.}{Benetti}, \Name{M.}{Branchesi}, \Name{E.}{Brocato}, \Name{S.}{Campana}, \Name{E.}{Cappellaro}, \Name{S.}{Covino}, \etal, \Journal{\NAT}{551}{2017}{67} \href{https://arxiv.org/abs/1710.05858}{[arXiv:1710.05858]}.
\bibitem [\protect\citeauthoryear{Piro \& Kollmeier}{2017}]{gw170817cocoonevid} \Name{L.}{Piro}, \Name{J.}{Kollmeier}, \Journal{\APJ}{855}{2018}{103} \href{https://arxiv.org/abs/1710.05822}{[arXiv:1710.05822]}.
\bibitem [\protect\citeauthoryear{Posselt, \etal}{2018}] {nstarmatt} \Name{B.}{Posselt}, \Name{G.G.}{Pavlov}, \Name{Ü.}{Ertan}, \Name{S.}{Çalışkan}, \Name{K.L.}{Luhman}, \Name{C.C.}{Williams}, \Journal{\APJ}{865}{2018}{1} \href{https://arxiv.org/abs/1809.08107}{[arXiv:1809.08107]}. 
\bibitem [\protect\citeauthoryear{Rossi, \etal}{2004}] {grbjetoffaxis} \Name{E.M.}{Rossi}, \Name{D.}{Lazzati}, \Name{J.D.}{Salmonson}, \Name{G.}{Ghisellini}, \Journal{\MRA}{354}{2004}{86} \href{https://arxiv.org/abs/astro-ph/0401124}{[astro-ph/]}.
\bibitem [\protect\citeauthoryear{Rossi, \etal}{2018}] {gw170817lateopt160} \Name{A.}{Rossi}, \Name{M.}{Cantiello}, \Name{V.}{Testa}, \Name{D.}{Paris}, \Name{A.}{Melandri}, \Name{S.}{Covino}, \Name{O.S.}{Salafia}, \Name{P.}{D'Avanzo}, \etal, (2018) \cir{\href{https://gcn.gsfc.nasa.gov/gcn3/22763.gcn3}{22763}}. 
\bibitem [\protect\citeauthoryear{Rybicki \& Lightman}{2004}] {emissionbook} \Name{G.B.}{Rybicki}, \Name{A.P.}{Lightman}, 2004, "Radiative Processes in Astrophysics'', Wieley-VCH verlag GmbH \& Co.KGaA, Weinheim.
\bibitem [\protect\citeauthoryear{Sari \etal}{1998}]{emission1} Sari, R., Piran, T., \& Narayan, R., \Journal{\APJ}{497}{1998}{17} \href{https://arxiv.org/abs/astro-ph/9712005}{astro-ph/9712005}.
\bibitem [\protect\citeauthoryear{Savchenko, \etal}{2017}] {gw170817integral} \Name{V.}{Savchenko}, \Name{C.}{Ferrigno}, \Name{E.}{Kuulkers}, \Name{A.}{Bazzano}, \Name{E.}{Bozzo}, \Name{S.}{Brandt}, \Name{J.}{Chenevez}, \Name{T.J.-L.}{Courvoisier}, \etal, \Journal{\APJL}{848}{2017}{L15} \href{https://arxiv.org/abs/1710.05449}{[arXiv:1710.05449]}.
\bibitem [\protect\citeauthoryear{Slane}{2017}] {nssheath} \Name{P.}{Slane} in ``Handbook of Supernovae'' (2017), 2159, Eds. A.W. Alsabti, P. Murdin, Springer, Cham \href{https://arxiv.org/abs/1703.09311}{[arXiv:1703.09311]}.
\bibitem [\protect\citeauthoryear{Smartt, \etal}{2017}] {gw170817bluekilonova} \Name{S.J.}{Smartt}, \Name{T.W.}{Chen}, \Name{A.}{Jerkstrand}, \Name{M.}{Coughlin}, \Name{E.}{Kankare}, \Name{S.A.}{Sim}, \Name{M.}{Fraser}, \Name{C.}{Inserra}, \etal, \Journal{\NAT}{551}{2017}{75} \href{https://arxiv.org/abs/1710.05841}{[arXiv:1710.05841]}.
\bibitem [\protect\citeauthoryear{Soares-Santos, \etal}{2017}] {gw170817optdes} \Name{M.}{Soares-Santos}, \Name{D.E.}{Holz}, \Name{J.}{Annis}, \Name{R.}{Chornock}, \Name{K.}{Herner}, \Name{E.}{Berger}, \Name{D.}{Brout}, \Name{H.}{Chen}, \etal, \Journal{\APJL}{848}{2017}{L16} \href{https://arxiv.org/abs/1710.05459}{[arXiv:1710.05459]}.
\bibitem [\protect\citeauthoryear{Spitkovsky}{2008}] {fermiaccspec} \Name{A.}{Spitkovsky}, \Journal{\APJ}{682}{2008}{5} \href{https://arxiv.org/abs/0802.3216}{[arXiv:0802.3216]}.
\bibitem [\protect\citeauthoryear{Takami, \etal}{2007}] {grbtailsidejet} \Name{K.}{Takami}, \Name{R.}{Yamazaki}, \Name{T.}{Sakamoto}, \Name{G.}{Sato}, \Journal{\APJ}{663}{2007}{1118} \href{https://arxiv.org/abs/0704.1055}{[arXiv:0704.1055]}.
\bibitem [\protect\citeauthoryear{Troja, \etal}{2017}] {gw170817xray} \Name{E.}{Troja}, \Name{L.}{Piro}, \Name{H.}{van Eerten}, \Name{R.T.}{Wollaeger}, \Name{M.}{Im}, \Name{O.D.}{Fox}, \Name{N.R.}{Butler}, \Name{S.B.}{Cenko}, \etal, \Journal{\NAT}{551}{2017}{71} \href{https://arxiv.org/abs/1710.05433}{[arXiv:1710.05433]}.
\bibitem [\protect\citeauthoryear{Troja, \etal}{2018a}] {gw170817chandraxray260a} \Name{E.}{Troja}, \Name{L.}{Piro}, \Name{G.}{Ryan}, (2018) \cir{\href{https://gcn.gsfc.nasa.gov/gcn3/22693.gcn3}{22693}}.
\bibitem [\protect\citeauthoryear{Troja, \etal}{2018b}] {gw170817latebroad} \Name{E.}{Troja}, \Name{L.}{Piro}, \Name{G.}{Ryan}, \Name{H.}{van Eerten}, \Name{R.}{Ricci}, \Name{M.}{Wieringa}, \Name{S.}{Lotti}, \Name{T.}{Sakamoto}, \Name{S.B.}{Cenko}, (2018) \href{https://arxiv.org/abs/1801.06516}{[arXiv:1801.06516]}.
\bibitem [\protect\citeauthoryear{Troja, \etal}{2018c}] {gw170817kilonovadec} \Name{E.}{Troja}, \Name{H.}{van Eerten}, \Name{G.}{Ryan}, \Name{R.}{Ricci}, \Name{J.M.}{Burgess}, \Name{M.}{Wieringa}, \Name{L.}{Piro}, \Name{S.B.}{Cenko}, \Name{T.}{Sakamoto}, (2018) \href{https://arxiv.org/abs/1808.06617}{[arXiv:1808.06617]}.
\bibitem [\protect\citeauthoryear{Tunnicliffe, \etal}{2014}] {grb111020aredshift} \Name{R.L.}{Tunnicliffe}, \Name{A.J.}{Levan}, \Name{N.R.}{Tanvir}, \Name{A.}{Rowlinson}, \Name{D.A.}{Perley}, \Name{J.S.}{Bloom}, \Name{S.B.}{Cenko}, \Name{P.T.}{O'Brien}, \etal, \Journal {\MRA}{437}{2014}{1495} \href{https://arxiv.org/abs/1402.0766}{[arXiv:1402.0766]}.
\bibitem [\protect\citeauthoryear{Waxman, \etal}{2019}]{gw170817kilonovafaint} \Name{E.}{Waxman}, \Name{E.O.}{Ofek}, \Name{D.}{Kushnir}, \Journal{\APJ}{878}{2019}{93} \href{https://arxiv.org/abs/1902.01197}{[arXiv:1902.01197]}.
\bibitem [\protect\citeauthoryear{Willingale, \etal}{2006}]{xrtafterglow} \Name{R.}{Willingale}, \Name{P.}{O'Brien}, \Name{J.P.}{Osborne}, \Name{O.}{Godet}, \Name{K.L.}{Page}, \Name{M.R.}{Goad}, \Name{D.N.}{Burrows}, \Name{B.}{Zhang}, \Name{E.}{Rol}, \Name{N.}{Gehrels}, \Name{G.}{Chincarini}, \Journal{\APJ}{662}{2007}{1093} [astro-ph/0612031].
\bibitem [\protect\citeauthoryear{Winkler, \etal}{2003}] {integral} \Name{C.}{Winkler}, \Name{T.J.L.}{Couvoisier}, \Name{G.}{Di Cocco}, \Name{N.}{Gehrels}, \Name{A.}{Giménez}, \Name{S.}{Grebenev}, \Name{W.}{Hermsen}, \Name{J.M.}{Mas-Hesse}, \etal, \Journal{\AA}{411}{2003}{L1}.
\bibitem [\protect\citeauthoryear{Wu \& MacFadyen}{2019}] {gw170817optfit} \Name{Y.}{Wu}, \Name{A.}{MacFadyen}, \Journal{\APJ}{869}{2019}{55} \href{https://arxiv.org/abs/1809.06843}{[arXiv:1809.06843]}.
\bibitem [\protect\citeauthoryear{Xie, \etal}{2018}] {gw170817aglowlorentz} \Name{X.}{Xie}, \Name{J.}{Zrake}, \Name{A.}{MacFadyen}, \Journal{\APJ}{863}{2018}{58} \href{https://arxiv.org/abs/1804.09345}{[arXiv:1804.09345]}.
\bibitem [\protect\citeauthoryear{Ziaeepour, \etal}{2007}] {grb070724a} \Name{H.}{Ziaeepour}, \Name{S.D.}{Barthelmy}, \Name{A.}{Parsons}, \Name{K.L.}{Page}, \Name{M.}{De Pasquale}, \Name{P.}{Schady}, \etal, (2007) \rep{74.2}. 
\bibitem [\protect\citeauthoryear{Ziaeepour}{2009}] {hourigrb} \Name{H.}{Ziaeepour}, \Journal{\MRA}{397}{2009}{361} \href{https://arxiv.org/abs/0812.3277}{[arXiv:0812.3277]}.
\bibitem [\protect\citeauthoryear{Ziaeepour \& Gardner}{2011}] {hourigrbmag} \Name{H.}{Ziaeepour}, \Name{B.}{Gardner}, \Journal {\JCA}{12}{2011}{001} \href{https://arxiv.org/abs/1101.3909}{[arXiv:1101.3909]}.
\bibitem [\protect\citeauthoryear{Ziaeepour}{2018a}] {hourigw170817} \Name{H.}{Ziaeepour} \Journal{\MRA}{478}{2018a}{3233} \href{https://arxiv.org/abs/1801.06124}{[arXiv:1801.06124]}.
\bibitem [\protect\citeauthoryear{Ziaeepour}{2018b}] {hourigw170817ag} \Name{H.}{Ziaeepour} (2018b) \href{https://arxiv.org/abs/1806.11161}{[arXiv:1806.11161]}.

\end{thebibliography}
\end{document}